\documentclass[letterpaper,12pt]{article}
\usepackage{epsfig,rotating,setspace,latexsym,amsmath,epsf,amssymb,bm}
\usepackage{cite,graphicx,authblk,multirow,color,caption,subcaption}
\usepackage[title]{appendix}

\newtheorem{Theo}{Theorem}

\newtheorem{Lem}{Lemma}

\def \ypn{Y_{pn}^n}
\def \ynp{Y_{np}^n}
\def \znp{Z_{np}^n}
\def \zpn{Z_{pn}^n}
\def \ynn{Y_{nn}^n}
\def \ydp{Y_{dp}^n}
\def \ypd{Y_{pd}^n}
\def \zpp{Z_{pp}^n}
\def \zdp{Z_{dp}^n}
\def \zpd{Z_{pd}^n}
\def \zdd{Z_{dd}^n}
\def \ypp{Y_{pp}^n}
\def \ztilde{\tilde{Z}}
\def \ytilde{\tilde{Y}}
\def \ztildepn{\ztilde_{pn}^n}

\def \ytildenp{\ytilde_{np}^n}
\def \ytildedp{\ytilde_{dp}^n}

\def \ztildepd{\ztilde_{pd}^n}

\newcommand{\CSIT}{\text{CSIT}}
\newcommand{\CSI}{\text{CSI}}
\newcommand{\PD}{\mathsf{PD}}
\newcommand{\DP}{\mathsf{DP}}
\newcommand{\DD}{\mathsf{DD}}
\newcommand{\PP}{\mathsf{PP}}
\newcommand{\NP}{\mathsf{NP}}
\newcommand{\PN}{\mathsf{PN}}
\newcommand{\DN}{\mathsf{DN}}
\newcommand{\ND}{\mathsf{ND}}
\newcommand{\NN}{\mathsf{NN}}
\newcommand{\ubar}[1]{\underline{\mathbf{#1}}}
\allowdisplaybreaks
\title{Secure Degrees of Freedom Region of the Two-User MISO Broadcast Channel with Alternating CSIT\thanks{This work was supported by NSF Grants CNS 13-14733, CCF 14-22111, CCF 14-22129 and CCF 14-22090, and presented in part at IEEE ISIT 2014 and to be presented in part at IEEE ICC 2015.} 
}
\author[1]{Pritam Mukherjee}
\author[2]{Ravi Tandon}
\author[1]{Sennur Ulukus}
\affil[1]{Department of ECE, University of Maryland, College Park, MD}
\affil[2]{Discovery Analytics Center and Department of CS, Virginia Tech, VA}

\begin{document}
\maketitle
\thispagestyle{empty}

\begin{abstract}
The two user multiple-input single-output (MISO) broadcast channel with confidential messages (BCCM) is studied in which the nature of channel state information at the transmitter (CSIT) from each user can be of the form $I_{i}$, $i=1,2$ where $I_{1}, I_{2}\in \{\mathsf{P}, \mathsf{D}, \mathsf{N}\}$, and the forms $\mathsf{P}$, $\mathsf{D}$ and $\mathsf{N}$ correspond to perfect and instantaneous, completely delayed,  and no CSIT, respectively. Thus, the overall CSIT can alternate between $9$ possible states corresponding to all possible values of $I_{1}I_{2}$, with each state occurring for $\lambda_{I_{1}I_{2}}$ fraction of the total duration.  The main contribution of this paper is to establish the secure degrees of freedom (s.d.o.f.) region of the MISO BCCM with alternating CSIT with the symmetry assumption, where $\lambda_{I_{1} I_{2}}=\lambda_{I_{2}I_{1}}$. 

The main technical contributions include developing a) novel achievable schemes for MISO BCCM with alternating CSIT with security constraints which also highlight the synergistic benefits of inter-state coding for secrecy, b) new converse proofs via local statistical equivalence and channel enhancement; and  c) showing the interplay between various aspects of channel knowledge and their impact on s.d.o.f. \end{abstract}

\section{Introduction}
Wireless systems are particularly vulnerable to security attacks because of the inherent openness of the transmission medium. With the widespread adoption of multiple-input multiple-output (MIMO) systems, there has been a significant recent interest in information theoretic physical layer security, the main premise of which is to exploit the difference in the wireless channels between different users. Information theoretic security has been investigated for a variety of channel models ranging from fading channels \cite{liang,LiYatesTrappechapter,gopala,PMukherjee_Ulukus2013}, MIMO wiretap channels \cite{2-2-1_ulukus_journal, khisti_mimome, mimo_wiretap, liu_shamai_mimo_wiretap}, multiple access channels \cite{tekin-yener-it2,tekin_yener_mac2008,ersen_ulukus_mac_2008,bassily_ergodic_align, jianwei_ulukus_mac_2013}, multi-receiver wiretap channels \cite{multireceiver_mimo_wiretap,ersen_eurasip,khandani_broadcast}, broadcast channels with confidential messages \cite{liu_maric_yates_BCCM_2008, liu_poor_BCCM_2009, liu_poor_shamai_BCCM_MIMO_2010}, wiretap channels with helpers \cite{tang_helper,jianwei_ulukus_helper_2012}, interference channels with confidential messages \cite{koyluoglu_gamal_lai_poor2011,ozan_intererence,he_yener_iccm_2011,jianwei_ulukus_interference_2013}, X-channels with confidential messages \cite{jafar_x_channel,skoglund_x_channel}, relay eavesdropper channels \cite{lai_elgamal_relay,yuksel_erkip_relay,bloch_relay,he_yener_relay,ersen_relay}, etc.

\begin{figure}[t]
\centering{\includegraphics[width=0.5\linewidth]{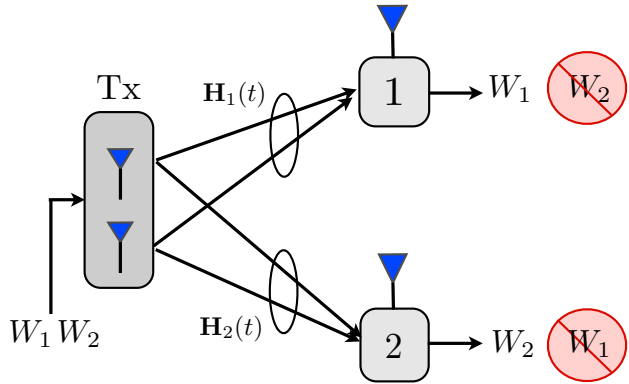}}
\caption{MISO broadcast channel with confidential messages (BCCM).}
\label{fig:system_model}
\end{figure}

The focus of this paper is on the secure degrees of freedom (s.d.o.f.) region of the fading two-user multiple-input single-output (MISO) broadcast channel with confidential messages (BCCM), in which the transmitter with two antennas has two confidential messages, one for each of the single antenna users (see Fig.~\ref{fig:system_model}). The secrecy capacity region of the MISO broadcast channel for the case of perfect and instantaneous CSI at all terminals (transmitter and the receivers) has been characterized in \cite{liu_poor_BCCM_2009, liu_poor_shamai_BCCM_MIMO_2010}. Using these results, it follows that for the two-user MISO BCCM, the sum s.d.o.f.~is $2$ with perfect and instantaneous channel state information at the transmitter (CSIT). In practice, the assumption of perfect and instantaneous CSIT may be too optimistic as $\CSIT$ may be delayed, imprecise or may not even be available at all. 

The impact of relaxing such assumptions on the d.o.f.~(secure or otherwise) has been widely studied in the literature. With perfect $\CSIT$ ($\mathsf{P}$), the sum d.o.f.~for the two-user MISO broadcast channel is $2$. With no $\CSIT$ ($\mathsf{N}$) however, reference\cite{vaze_varanasi_mimo_broadcast_noCSIT} showed that the sum d.o.f.\footnote{We refer to sum d.o.f.~as the sum degrees of freedom for a network without any confidentiality constraints (e.g., MISO BC); and sum s.d.o.f.~as the  sum secure degrees of freedom for the same network with confidential messages (e.g., MISO BCCM).} collapses to $1$. With delayed\footnote{By delayed $\CSIT$, we refer to the standard assumption as in \cite{maddah_ali_tse} in which the delay in acquiring $\CSIT$ is larger than the channel coherence time.} $\CSIT$ ($\mathsf{D}$), it is shown in \cite{maddah_ali_tse} that the sum d.o.f.~for the two-user MISO BC increases to $\frac{4}{3}$. \cite{maddah_ali_tse}  also presents novel results for the more general setting of $K$-user MISO BC, for $K\geq 2$. With delayed CSI, \cite{vaze_varanasi_delayed_csi} established the d.o.f.~region for the two-user MIMO BC. Other channel models besides the BC has also been investigated. Reference \cite{ravi_interference_delayed} provided the d.o.f.~region of the MIMO interference channel with delayed CSIT and output feedback. For the X-channel, references \cite{jafar_shamai_x_channel,maddah_ali_x_channel} showed that the optimal sum d.o.f.~is $\frac{4}{3}$ with perfect channel knowledge. With delayed CSIT the optimal sum d.o.f.~of the X-channel remains unknown in general. However, with a restriction of the transmission policies to linear schemes, reference \cite{x_channel_linear_dof} determined the sum d.o.f.~of the channel to be $\frac{6}{5}$ (also see \cite{mimo_x_channel_linear_dof, x-channel-delayed-Khandani} and the references therein). With global feedback, where each transmitter receives output feedback from every receiver, \cite{ravi_x_channel_delayed} showed that the sum d.o.f.~of the two-user X-channel with delayed CSIT is the same as that of the two-user MISO broadcast channel with 2 antennas at the transmitter; thus, all the transmitters can cooperate and behave like a single 2-antenna MISO system, and the optimal sum d.o.f.~is $\frac{4}{3}$. 

When security constraints are introduced, the s.d.o.f.~is known for several scenarios of delayed or no $\CSIT$. For the two-user MISO BCCM with no $\CSIT$, the sum s.d.o.f.~is zero as the two users are statistically equivalent and hence no secrecy is possible. On the other hand, with completely outdated $\CSIT$ from both users, \cite{kobayashi_delayed_csit} showed that the sum s.d.o.f.~increases to $1$. For the two-user SISO X-channel with confidential messages and global output feedback, \cite{mimo_x_channel_feedback_delayedcsi} showed that the optimal s.d.o.f.~is $1$; thus, two distributed transmitters with one antenna each are as good as a single transmitter with $2$ antennas and the X-channel behaves like a two-user MISO BCCM.
The aforementioned literature primarily deals with homogeneous CSIT scenarios in which the nature of channel knowledge supplied by every receiver is of the same form. In practice, however,  the nature of $\CSIT$ can vary across users. 
This observation naturally leads to the setting  of heterogeneous (or hybrid) CSIT which models the variability in the quality/delay of channel knowledge supplied by different users. In contrast to homogeneous CSIT, the setting of heterogeneous CSIT 
is much less understood. To the best of our knowledge, the complete characterization of the d.o.f.~of all fixed heterogeneous CSIT configurations is only known for the two-user MISO broadcast channel: see \cite{jafar_retrospective_alignment, Ravi-maddah-ali} for state $\PD$ for which the optimal sum d.o.f.~is shown to be $3/2$; and\cite{aligned_image_sets_jafar} which recently settled the states $\PN$ and $\DN$ through a novel converse proof and showed that the optimal sum d.o.f.~is given by $1$.  Beyond these results, partial results are available for the three-user MISO BC with hybrid CSIT in\cite{ravi_3_user_bc_2014,k_user_bc_hybrid} but by and large the problem of  heterogeneous CSIT even without secrecy constraints remains open.

Besides exhibiting heterogeneity across users, the nature of channel knowledge may also vary over time/frequency. Such variability can arise either naturally (due to the time variation in tolerable feedback overhead from a user) or it can be artificially induced (by deliberately altering the channel feedback mechanism over time/frequency).  For example, instead of requiring perfect $\CSIT$ from one user and delayed $\CSIT$ from the other user throughout the duration of communication, one may require that for half of the time, the first user provide perfect $\CSIT$ while the second user provide delayed $\CSIT$ (state $\PD$), and the roles of the users are reversed for the remaining half of the time (state $\DP$), the total network feedback overhead being the same in both cases. This leads naturally to the setting of alternating $\CSIT$ in which multiple $\CSIT$ states, for instance, $\PD$ and $\DP$ in the above example, arise over time. The alternating $\CSIT$ framework was introduced in \cite{ravi_alternating} where the d.o.f.~region was characterized for the two-user MISO BC. It was shown that synergistic gains in d.o.f.~are possible by jointly coding across these states. It was observed in \cite{ravi_alternating} for the two-user case that the final d.o.f.~region depends only on the marginal fractions of perfect, delayed and no CSIT, that is, the fractions of the time a user provides perfect, delayed and no CSIT. Given these results, several natural questions arise: a) do such synergistic gains still exist with additional confidentiality constraints on the messages, b) if yes, what is the optimal s.d.o.f.~region and how to achieve it, c) what is the penalty for incorporating confidentiality in contrast to \cite{ravi_alternating} and d) the fundamental impact of the variability of channel knowledge on secrecy.


In this paper, we consider the two-user MISO BCCM with alternating $\CSIT$ with all $9$ possible $\CSIT$ states: $\PP$, $\PD$, $\PN$, $\DP$, $\NP$, $\DD$, $\DN$, $\ND$, and $\NN$. We assume that these states occur for arbitrary fractions of time, except for a mild condition of symmetry, which is that states $I_1I_2$ and $I_2I_1$ occur for equal fractions of the time if $I_1\neq I_2$. The main contribution of this paper is the characterization of the optimal s.d.o.f.~region for this general model\footnote{In our preliminary work \cite{pmukherjee_ravi_ulukus_isit2014}, we considered the problem with only two states, $\PD$ and $\DP$ and established the optimal s.d.o.f.~region for this  specific problem. Reference \cite{sezgin_alternating_2013} considered another special case with four states: $\PP$, $\PD$, $\DP$ and $\DD$, but provided only an inner bound for the s.d.o.f.~region.}. With $9$ states, each occurring for arbitrary fractions of the time, it is not immediately clear how to optimally code across the states and the achievability of the s.d.o.f.~region is highly non-trivial. To this end, we first develop several key constituent schemes, where each scheme uses a subset of the $9$ states to achieve a particular s.d.o.f.~value.  We present all the constituent schemes in Section \ref{constituent_schemes}. Now given an arbitrary\footnote{Arbitrary subject to mild symmetry, i.e., $\lambda_{I_1I_2} = \lambda_{I_2I_1}$} probability mass function (pmf) on the $9$ $\CSIT$ states, we need to judiciously time share between the constituent schemes to achieve the optimal s.d.o.f.~region. It is not immediately clear how this should be done. Thus, we consider different sub-cases based on the relative proportions of the various states and explicitly characterize how the constituent schemes should be time shared to obtain the optimal s.d.o.f.~region in each sub-case. This characterization is done in Section~\ref{achievability}. 

Next, we provide a matching converse for the full region. We first generalize the \emph{local statistical equivalence} property introduced in \cite{pmukherjee_ravi_ulukus_isit2014}. The idea behind the converse is to first enhance the channel by providing more $\CSIT$ to obtain a new channel with fewer number of states but at least as large secrecy capacity as the original channel. Outer bounds on the s.d.o.f.~region for the enhanced channel give us the desired outer bounds for the original channel. 

Thus, the main contributions of this paper can be summarized as follows: a) We obtain the full s.d.o.f.~region with all possible  $9$ states occurring for arbitrary fractions of time constrained only by the requirement of symmetry, which is that states $I_1I_2$ and $I_2I_1$ occur for equal fractions of the time if $I_1\neq I_2$. b) To achieve this region, we provide several new optimal achievable schemes for different alternating $\CSIT$ scenarios. c) In addition, we provide an explicit method of combining the various achievable schemes judiciously to achieve the region. d) We provide a matching converse for the full region using channel enhancement and generalizing the \emph{local statistical equivalence} property introduced in \cite{pmukherjee_ravi_ulukus_isit2014}. e) We establish the s.d.o.f.~regions of the MISO BCCM under two heterogeneous CSIT settings: $\PD$ and $\DN$ states alone. These results completely settle the problem of characterizing the s.d.o.f.~regions of all individual heterogenous CSIT states: $\PD$, $\PN$, $\DN$. f) We show synergistic benefits of coding across the different alternating states even under security constraints.

\section{System Model}

We consider a two-user MISO BC, shown in Fig.~\ref{fig:system_model}, where the transmitter Tx, equipped with 2 antennas, wishes to send independent confidential messages to two single antenna receivers 1 and 2. The input-output relations at time $t$ are given by, 
\begin{align}
Y(t)&= \mathbf{H}_{1}(t)\mathbf{X}(t) + N_{1}(t)\label{eq:channel_model_a}\\
Z(t)&= \mathbf{H}_{2}(t)\mathbf{X}(t) + N_{2}(t), \label{eq:channel_model_b}
\end{align}
where $Y(t)$ and $Z(t)$ are the channel outputs of receivers $1$ and $2$, respectively. The $2 \times 1$ channel input $\mathbf{X}(t)$ is power constrained as $\mathbb{E}[||\mathbf{X}(t)||^{2}]\leq P$, and $N_{1}(t)$ and $N_{2}(t)$ are circularly symmetric complex white Gaussian noises with zero-mean and unit-variance. The $1\times 2$ channel vectors $\mathbf{H}_{1}(t)$ and $\mathbf{H}_{2}(t)$ of receivers 1 and 2, respectively, are independent and identically distributed (i.i.d.) with continuous distributions, and are also i.i.d.~over time. We denote $\mathbf{H}(t)=\{\mathbf{H}_{1}(t), \mathbf{H}_{2}(t)\}$ as the collective channel vectors at time $t$ and $\mathbf{H}^{n}= \{\mathbf{H}(1), \ldots, \mathbf{H}(n)\}$ as the sequence of channel vectors up until and including time $n$.

In practice, the receivers estimate the channel coefficients and feed them back to the transmitter. In general, the receiver can choose to send not only the current measurements, but rather any function of all the channel measurements it has taken upto that time. The CSIT at time $t$ can thus be any function of the measured channel coefficients upto time $t$. There are two key aspects to the CSIT: precision and delay. Precision captures the fact that the measurements made at the receivers and sent to the transmitter are imprecise (usually, quantized) and noisy. Delay is introduced since making measurements and feeding them back to the transmitter takes time. We will focus on the delay aspect of CSIT, and assume that the CSIT when available, has infinite precision. 

In order to model the delay in CSIT, we assume that at each time $t$, there are  3 possible CSIT states for each user:     
\begin{itemize}
\item \textit{Perfect CSIT}  ($\mathsf{P}$): This denotes the availability of precise and instantaneous CSI of a user at the transmitter. In this state, the transmitter has precise channel knowledge before the start of the communication.
\item \textit{Delayed CSIT} ($\mathsf{D}$): In this state, the transmitter does not have the CSI at the beginning of the communication. In slot $t$, the receiver may send any function of all the channel coefficients upto and including time $t$ as CSI to the transmitter. However, the CSIT becomes available only after a delay such that the CSI is completely outdated, that is, independent of the current channel realization.
\item \textit{No CSIT} ($\mathsf{N}$): In this state, there is no CSI of the user available at the transmitter. 
\end{itemize}
Denote the CSIT of user 1 by $I_1$ and the CSIT of user 2 by $I_2$. Then,
\begin{align}
I_1,I_2 \in\left\lbrace  \mathsf{P},\mathsf{D},\mathsf{N}\right\rbrace.
\end{align}
Thus, for the two-user MISO BC, we have 9 CSIT states, namely $\PP$, $\DD$, $\NN$, $\PD$, $\DP$, $\PN$, $\NP$, $\DN$, and $\ND$. Let $\lambda_{I_1I_2}$ be the fraction of the time the state $I_1I_2$ occurs. Then,
\begin{align}
\sum_{I_1,I_2} \lambda_{I_1I_2} =1.
\end{align}
We also assume symmetry: $\lambda_{I_1I_2} = \lambda_{I_2I_1}$ for every $I_1I_2$. Specifically,
\begin{align}
\lambda_{PD} &= \lambda_{DP}\label{eq:symmetry1}\\
\lambda_{DN} &= \lambda_{ND}\label{eq:symmetry2}\\
\lambda_{PN} &= \lambda_{NP}\label{eq:symmetry3}.
\end{align}
Further, we assume that perfect and global CSI is available at both receivers.
 
A secure rate pair $(R_1,R_2)$ is achievable if there exists a sequence of codes which satisfy the reliability constraints at the receivers, namely, $\mbox{Pr}\left[ W_{i}\neq \hat{W}_{i}\right] \leq \epsilon_{n}$, for $i=1,2$, and the confidentiality constraints, namely,
\begin{align}
 \frac{1}{n} I(W_1;Z^n,\mathbf{H}^n) \leq \epsilon_n,\qquad
 \frac{1}{n} I(W_2;Y^n,\mathbf{H}^n)\leq \epsilon_n, \label{eq:confidentiality}
\end{align}
where $\epsilon_n \rightarrow 0$ as $n \rightarrow \infty$. Informally, the constraints in \eqref{eq:confidentiality} ensure that the information leakage, per channel use, of the first receiver's message at the second receiver should be arbitrarily small, and vice versa. A s.d.o.f.~pair $(d_{1}, d_{2})$ is achievable, if there exists an achievable rate pair $(R_1,R_2)$ such that
\begin{align}
d_1 = \lim\limits_{P\rightarrow \infty} \frac{R_1}{\log P}, \qquad
d_2 =\lim\limits_{P\rightarrow \infty} \frac{R_2}{\log P}.
\end{align}

Let us define the following:
\begin{align}\label{Lambda-defn}
\lambda_{P}&\triangleq \lambda_{PP}+ \lambda_{PD}+ \lambda_{PN}\\
\lambda_{D}&\triangleq \lambda_{PD}+ \lambda_{DD}+ \lambda_{DN}\\
\lambda_{N}&\triangleq \lambda_{PN}+ \lambda_{DN}+ \lambda_{NN}.
\end{align}
Using these definitions, it is easy to verify that
\begin{align}\label{Lambda-cons}
\lambda_{P}+ \lambda_{D}+ \lambda_{N}=1.
\end{align}
Here, we can interpret these three quantities as follows:
\begin{itemize}
\item $\lambda_P$: represents the total fraction of time the CSIT of a user is in the $\mathsf{P}$ state.
\item $\lambda_D$: represents the total fraction of time the CSIT of a user is delayed, that is, the state $\mathsf{D}$.
\item $\lambda_N$: represents the total fraction of time a user supplies no CSIT.
\end{itemize}

Given the probability mass function (pmf), $\lambda_{I_1I_2}$, our goal is to characterize the s.d.o.f.~region of the two-user MISO BCCM.
\section{Main Result and Discussion}

\begin{Theo}
The s.d.o.f.~region for the two-user MISO BCCM with alternating $\CSIT$, $\mathcal{D}(\lambda_{I_1I_2})$, is the set of all non-negative pairs $(d_1,d_2)$ satisfying,
\begin{align}
d_{1}&\leq \min\left(\frac{2+2\lambda_P-\lambda_{PP}}{3},1-\lambda_{NN}\right)\label{eq:single_user_rate1}\\
d_{2}&\leq \min \left(\frac{2+2\lambda_P-\lambda_{PP}}{3},1-\lambda_{NN}\right)\label{eq:single_user_rate2} \\
3d_1 + d_2 &\leq 2 + 2\lambda_{P}\label{eq:3d1+d2_bound1}\\
d_1 + 3d_2 &\leq 2 + 2\lambda_{P} \label{eq:3d1+d2_bound2}\\
d_{1}+ d_{2}&\leq 2(\lambda_{P}+ \lambda_{D}). \label{eq:sum_rate}
\end{align}
\end{Theo}

A proof for the achievability of this region will be provided in Section \ref{achievability} using constituent schemes presented in Section~\ref{constituent_schemes}. A converse is provided in Section \ref{converse}.

\begin{figure}
\centering
\includegraphics[width =0.5\linewidth]{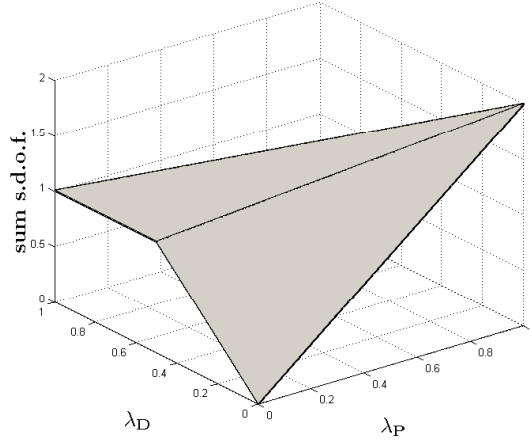}
\caption{The sum s.d.o.f.~as a function of $\lambda_P$ and $\lambda_D$.}
\label{fig:sum_dof_plot}
\end{figure}

We next make a series of remarks highlighting the consequences and interesting aspects of this theorem.
\subsubsection*{Remark 1. [Sum s.d.o.f.: $\max (d_{1}+d_{2})$]}
From the region stated in \eqref{eq:single_user_rate1}-\eqref{eq:sum_rate}, it is clear that the sum s.d.o.f.~is given by,
\begin{align}
\mbox{sum s.d.o.f.}= \min\left(2\left(\frac{2+2\lambda_P-\lambda_{PP}}{3}\right),2(1-\lambda_{NN}), 2(\lambda_P+\lambda_{D}), 1+\lambda_{P}\right).  \label{eq:sum}
\end{align}
The sum s.d.o.f.~expression in \eqref{eq:sum} can be significantly simplified by noting that the first two terms in the minimum are inactive due to the inequalities  $1+\lambda_P \leq 2\left(\frac{2+2\lambda_P-\lambda_{PP}}{3}\right)$, and $2(\lambda_P+\lambda_{D}) =  2(1-\lambda_{N})
\leq 2(1-\lambda_{NN})$. These inequalities follow directly  from (\ref{Lambda-defn})-(\ref{Lambda-cons}). Using these inequalities, the sum s.d.o.f.~expression above is equivalent to 
\begin{align}
\mbox{sum s.d.o.f.}&= \min\left( 2(\lambda_P+\lambda_{D}), 1+\lambda_{P}\right) \\
&= \min\left( 2(\lambda_P+\lambda_{D}), 2\lambda_{P}+ \lambda_{D}+\lambda_{N}\right)\\
&= 2\lambda_P+\lambda_{D} +\min(\lambda_D,\lambda_N).\label{eq:sum_rate_compact}
\end{align} 
Fig.~\ref{fig:sum_dof_plot} shows the sum s.d.o.f.~as a function of $\lambda_P$ and $\lambda_D$.

\subsubsection*{Remark 2. [Same marginals property]} From \eqref{eq:sum_rate_compact}, we notice that the marginal probabilities $\lambda_P$, $\lambda_{D}$ and $\lambda_{N}$ are sufficient to determine the sum s.d.o.f. \textit{Thus, for any given pmf $\lambda_{I_1I_2}$, satisfying the symmetry conditions \eqref{eq:symmetry1}-\eqref{eq:symmetry3}, there exists an \textbf{equivalent} alternating $\CSIT$ problem having only three states: $\PP$, $\DD$ and $\NN$ occurring for $\lambda_P$, $\lambda_{D}$ and $\lambda_{N}$ fractions of the time, respectively, that has the same sum s.d.o.f.} This observation is similar to the case when there is no secrecy \cite{ravi_alternating}. \textit{However unlike in \cite{ravi_alternating}, the s.d.o.f.~region \textbf{does not} have the same property in general} as we can see the explicit dependence of the s.d.o.f.~region in \eqref{eq:single_user_rate1}-\eqref{eq:sum_rate} on $\lambda_{PP}$ and $\lambda_{NN}$.

\begin{figure}[t]
\centering
\includegraphics[width =0.5\linewidth]{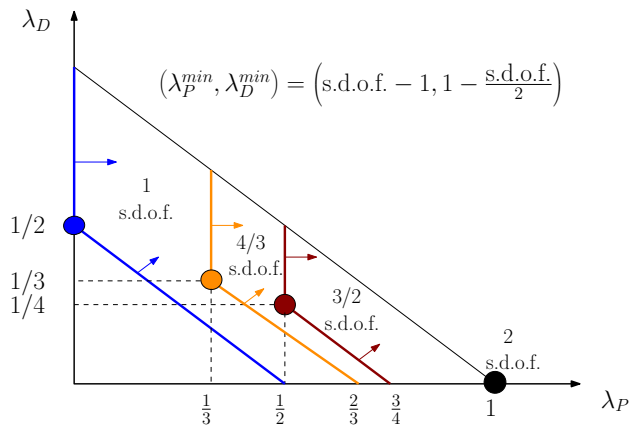}
\caption{Trade-off between delayed and perfect $\CSIT$.}
\label{fig:trade-off_plot}
\end{figure}
\subsubsection*{Remark 3. [Channel knowledge equivalence]} We next highlight an interesting property which shows that from the sum s.d.o.f.~perspective, no CSIT is equivalent to delayed CSIT when $\lambda_D\geq \lambda_N$, and delayed CSIT is equivalent to perfect CSIT when $\lambda_D < \lambda_N$.

\emph{Equivalence of delayed and no CSIT when $\lambda_D\geq\lambda_N$}: From a sum s.d.o.f.~perspective, we see that when $\lambda_D \geq \lambda_N$, the sum s.d.o.f.~depends only on $\lambda_{P}$. Hence, as long as $\lambda_D \geq \lambda_N$ holds, the $\mathsf{N}$ states behave as $\mathsf{D}$ states in the sense that, if the $\mathsf{N}$ states were enhanced to $\mathsf{D}$ states, the sum s.d.o.f.~would not increase. Essentially, the $\mathsf{N}$ states can be combined with various $\mathsf{D}$ states and we obtain the same sum s.d.o.f.~as if every $\mathsf{N}$ state were replaced by a $\mathsf{D}$ state. Consider an example, where the states $\PD$, $\DP$ and $\NN$ occur for $\frac{2}{5}$th, $\frac{2}{5}$th and $\frac{1}{5}$th fractions of the time, respectively. Note that $\lambda_D =\frac{2}{5}>\lambda_N =\frac{1}{5}$ in this case. The sum s.d.o.f., from \eqref{eq:sum_rate_compact}, is $2\lambda_P+\lambda_D+\lambda_N = \frac{7}{5}$. Now, if we enhance the $\mathsf{N}$ states to $\mathsf{D}$ states, we get the states $\PD$, $\DP$ and $\DD$ occur for $\frac{2}{5}$th, $\frac{2}{5}$th and $\frac{1}{5}$th of the time, respectively. The sum s.d.o.f.~of this enhanced system is still $\frac{7}{5}$.

\emph{Equivalence of delayed and perfect CSIT when $\lambda_D\leq\lambda_N$}: From a sum s.d.o.f.~perspective, we see that when $\lambda_D\le \lambda_N$, the sum s.d.o.f.~depends only on $\lambda_N$. Hence, in this case, if $\lambda_{D} \leq \lambda_{N}$, the delayed $\CSIT$ is as good as perfect $\CSIT$, that is, every $\mathsf{D}$ state can be enhanced to a $\mathsf{P}$ state without any increase in the sum s.d.o.f. For example, consider a system  where the states $\PD$, $\DP$ and $\NN$ occur for $\frac{1}{5}$th, $\frac{1}{5}$th and $\frac{3}{5}$th fractions of the time, respectively. Note that $\lambda_D =\frac{1}{5}<\lambda_N =\frac{3}{5}$ in this case. The sum s.d.o.f.~for this system is $\frac{4}{5}$, from \eqref{eq:sum_rate_compact}. By enhancing the $\mathsf{D}$ states to $\mathsf{P}$ states, we get a system, where the states $\PP$ and $\NN$ occur for $\frac{2}{5}$th and $\frac{3}{5}$th fractions of the time, respectively. The sum s.d.o.f.~in for this enhanced system is still $\frac{4}{5}$. 

\subsubsection*{Remark 4. [Minimum CSIT required for a sum s.d.o.f.~value]} 
Fig.~\ref{fig:trade-off_plot} shows the trade-off between $\lambda_P$ and $\lambda_D$ for a given value of sum s.d.o.f. The highlighted corner point in each curve shows the most \emph{efficient} point in terms of $\CSIT$ requirement. \emph{Any other feasible point either involves redundant $\CSIT$ or unnecessary instantaneous $\CSIT$ where delayed $\CSIT$ would have sufficed}. For example, following are the minimum $\CSIT$ requirements for various sum s.d.o.f.~values:
\begin{align}
\mbox{sum s.d.o.f.} = 2 \,:&\; (\lambda_P,\lambda_D)_{\min} = \left(1,0\right)\\
\mbox{sum s.d.o.f.} = \frac{3}{2} :&\; (\lambda_P,\lambda_D)_{\min} = \left(\frac{1}{2},\frac{1}{4}\right)\\
\mbox{sum s.d.o.f.} = \frac{4}{3} :&\; (\lambda_P,\lambda_D)_{\min} = \left(\frac{1}{3},\frac{1}{3}\right)\\
\mbox{sum s.d.o.f.} = 1 \,:&\; (\lambda_P,\lambda_D)_{\min} = \left(0,\frac{1}{2}\right).
\end{align} 
In general, for a given value of sum s.d.o.f.~$=s$, the minimum $\CSIT$ requirements are given by:
\begin{align}
\left(\lambda_P,\lambda_D\right)_{\min} = \begin{cases}
\left(s-1,1-\frac{s}{2}\right), & \mbox{if } 1\leq s\leq 2\\
\left(0,\frac{s}{2}\right), & \mbox{if } 0\leq s\leq 1.
\end{cases}
\end{align}  
\subsubsection*{Remark 5. [Cost of security]} 
\begin{figure}[t]
\centering
\includegraphics[width =0.5\linewidth]{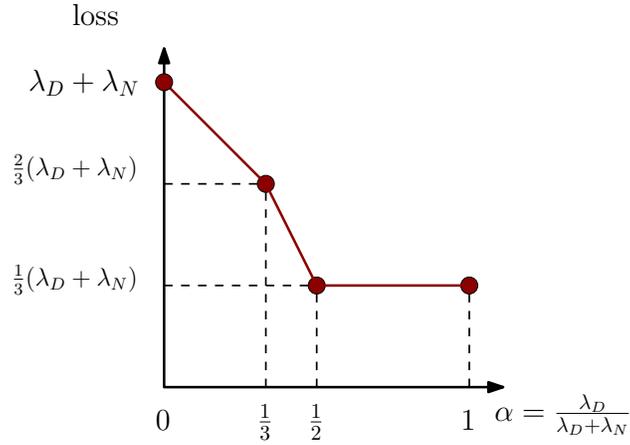}
\caption{Cost of security.}
\label{fig:cost_of_security}
\end{figure}

We recall that in the case with no security \cite{ravi_alternating}, the sum d.o.f.~is given by, 
\begin{align}
\mbox{sum d.o.f.}=2-\frac{2\lambda_N}{3} -\frac{\max(\lambda_N,2\lambda_D)}{3}.
\end{align}
Comparing with \eqref{eq:sum_rate_compact}, we see that the loss in d.o.f.~that must be incurred to incorporate secrecy constraints is given by,
\begin{align}
(\mbox{sum d.o.f.}) -(\mbox{sum s.d.o.f.})\triangleq \mbox{loss} = \begin{cases}
\lambda_N, &\mbox{if } \lambda_N \geq 2\lambda_D\\
\frac{2}{3}(2\lambda_N-\lambda_D), &\mbox{if } 2\lambda_D \geq \lambda_N \geq \lambda_D\\
\frac{1}{3}(\lambda_N+\lambda_D), &\mbox{if } \lambda_D \geq \lambda_N.
\end{cases}\label{eq:loss_due_to_security}
\end{align}
If we define $\alpha = \lambda_{D}/(\lambda_D+\lambda_N)$, we can rewrite \eqref{eq:loss_due_to_security} as follows,
\begin{align}
\mbox{loss} = (\lambda_{D}+ \lambda_{N})\times\begin{cases}
(1-\alpha), &\mbox{if } \alpha \leq \frac{1}{3}\\
\left(\frac{4}{3} - 2\alpha\right), &\mbox{if } \frac{1}{2} \geq \alpha \geq \frac{1}{3}\\
\frac{1}{3}, &\mbox{if } \alpha \geq \frac{1}{2}.
\end{cases}
\end{align}  
We show this loss as a function of $\alpha$ in Fig.~\ref{fig:cost_of_security}. Note that $\lambda_D+\lambda_N$ is the fraction of the time a user feeds back imperfect (delayed or none) $\CSIT$. If this fraction is fixed, increasing the fraction of the delayed $\CSIT$ decreases the penalty due to the security constraints, but only to a certain extent. When $\lambda_N \geq \lambda_D$, increasing the fraction of delayed $\CSIT$ leads to a decrease in the penalty due to the security constraints. However, once the fraction of the delayed $\CSIT$ (state $\mathsf{D}$) matches that of no $\CSIT$ ($\mathsf{N}$), that is, $\lambda_D \geq \lambda_N$, increasing the fraction of delayed $\CSIT$ further does not reduce the penalty any more.
\subsubsection*{Remark 6. [S.d.o.f.~characterization of individual CSIT states]}
As an additional relevant result, we also characterize the respective s.d.o.f.~regions for the $6$ individual CSIT states. To the best of our knowledge, the only CSIT states for which the s.d.o.f.~regions were previously known are: $\PP$ (with sum s.d.o.f.$=2$), $\DD$ (with sum s.d.o.f.$=1$), $\PN$ (with s.d.o.f.$=1$), and $\NN$ (with s.d.o.f.$=0$). For the remaining two CSIT states, i.e., $\PD$ and $\DN$, we establish the optimal s.d.o.f.~regions. In particular, for the $\PD$ CSIT state, we show in Appendix \ref{appendix:pd_alone}  that the s.d.o.f.~region is given by $d_{1}+d_{2}\leq 1$. For the $\DN$ state, we show in Appendix \ref{appendix:dn_alone} that the s.d.o.f.~region is given by $d_{1}+d_{2}\leq 1/2$. As the next remark shows, these complete set of results for the individual CSIT states confirm the synergistic benefits (or lack thereof) in various alternating CSIT scenarios. 

\subsubsection*{Remark 7. [Synergistic benefits]}
It was shown in \cite{ravi_alternating} that by coding across different states one can achieve higher sum d.o.f.~than by optimal encoding for each state separately and time sharing. A similar result holds true in our case as well. We illustrate this with the help of a few examples.

\emph{Example 1}. Consider a special case where only states $\PD$ and $\DP$ occur, each for half of the time. In our previous work,\cite{pmukherjee_ravi_ulukus_isit2014}, we showed that optimal sum s.d.o.f.~is $\frac{3}{2}$ in this case; see also \eqref{eq:sum_rate_compact} here. The best achievable scheme for the $\PD$ (or $\DP$) state alone was known to achieve a sum s.d.o.f.~of $1$. This was either by treating the $\PD$ state as a $\PN$ state and zero forcing, or by treating $\PD$ as a $\DD$ state. However a converse proof showing the optimality of $1$ sum s.d.o.f.~was not known. In Appendix~\ref{appendix:pd_alone}, we present a converse proof to show that the sum s.d.o.f.~of $1$ is indeed optimal for the $\PD$ state alone. Thus, by encoding for each state separately and time sharing between the $\PD$ and $\DP$ states, we can achieve only $1$ sum s.d.o.f., whereas joint encoding across the states achieves sum s.d.o.f.~of $\frac{3}{2}$. Thus, we have synergistic benefit of $50\%$ in this case.

\emph{Example 2}. Consider another special case with three states: $\PD$, $\DP$ and $\NN$ each occurring for one-third of the time. The optimal sum s.d.o.f.~is $\frac{4}{3}$. If we encode for each state separately and time share between them, we can achieve a sum s.d.o.f.~of $\frac{1}{3}\times 1 + \frac{1}{3}\times 1 + \frac{1}{3}\times 0 = \frac{2}{3} $, since the $\NN$ state does not provide any secrecy. If we encode across the $\PD$ and $\DP$ states optimally and then time share with the $\NN$ state, we can achieve $\frac{2}{3}\times \frac{3}{2} + \frac{1}{3}\times 0 = 1$ sum s.d.o.f. Thus, in this case too, we get synergistic benefit by coding across all the states together.

\emph{Example 3}. Now, assume we have the following three states: $\PN$, $\NP$ and $\DD$ each occurring for one-third of the time. The optimal sum s.d.o.f.~for this case is $\frac{4}{3}$. On the other hand, the optimal sum s.d.o.f.~of the $\PN$ state alone is $1$, \cite{aligned_image_sets_jafar}, and that of the $\DD$ state alone is also $1$,\cite{kobayashi_delayed_csit}. Thus, by separately encoding for each state and time sharing, we can achieve $\frac{1}{3}\times 1 + \frac{1}{3}\times 1 + \frac{1}{3}\times 1 = 1$ sum s.d.o.f. Note that the optimal sum s.d.o.f.~for $\PN$ and $\NP$ states, each occurring for half of the time, is also $1$, using \eqref{eq:sum_rate_compact}. Thus, by optimal encoding for $\PN$ and $\NP$ together and time sharing with the $\DD$ state also yields sum s.d.o.f.~of $1$. Therefore, there is synergistic benefit to be gained by coding across all the states together in this case too.

\emph{Example 4}. Consider the case where the two states, $\DD$ and $\NN$ occur for equal fractions of time. The optimal sum s.d.o.f.~of the $\DD$ state alone is $1$ \cite{kobayashi_delayed_csit}. The $\NN$ state, by itself does not provide any secrecy and its s.d.o.f.$\,=0$. Thus, by encoding for the individual states and time sharing, at most $1\times \frac{1}{2} + 0\times \frac{1}{2} = \frac{1}{2}$ sum s.d.o.f.~is achievable. However, by jointly encoding across both the $\DD$ and $\NN$ states, the optimal sum s.d.o.f.~of $1$ is achievable. Thus, we have synergistic benefit of $100\%$ in terms of sum s.d.o.f.~in this case.  

\emph{Example 5}. Finally, consider the case where the two states, $\DN$ and $\ND$ occur for equal fractions of time.  We show in  Appendix \ref{appendix:dn_alone} that  the optimal sum s.d.o.f.~for  $\DN$ state is $\frac{1}{2}$. Thus, by separately encoding across the individual states, only $\frac{1}{2}$ sum s.d.o.f.~is achievable. However, by jointly encoding across both the $\DN$ and $\DN$ states, the optimal sum s.d.o.f.~of $1$ is achievable. Thus, we have synergistic benefit of $100\%$ in terms of sum s.d.o.f.~in this case.       

\subsubsection*{Remark 7. [Lack of synergistic benefits]}
There are some situations where joint encoding across alternating states does not yield any benefit in terms of the s.d.o.f.~region. For example, consider a case with only $2$ states, $\PN$ and $\NP$, each occurring for half of the time. The optimal sum s.d.o.f.~for the $\PN$ state alone is $1$, which is achieved by zero forcing. The optimal sum s.d.o.f.~of both $\PN$ and $\NP$ states together is also $1$; thus, encoding for each state separately is optimal in this case. Indeed separable encoding for each individual state suffices to achieve the full s.d.o.f.~region as well. \emph{This result is perhaps surprising, since in the case with no security, we do get synergistic benefits of joint encoding across the $\PN$ and $\NP$ states. The optimal sum s.d.o.f.~with joint encoding is $\frac{3}{2}$, while that for each state alone is $1$,}\cite{ravi_alternating}. 

\section{Constituent Schemes}\label{constituent_schemes}
\begin{table}[h]
\large
\begin{center}
\begin{tabular}{ |c|c|c|c|c| }
\hline
\multicolumn{5}{ |c| }{Summary of Constituent Schemes (CS)} \\\hline
Sum s.d.o.f.~& CS Notation & CSIT States & Fractions of States & $(d_{1},d_{2})$\\ \hline
$2$  &   $S^{2}$  & $\PP$  & $1$ &  $(1,1)$\\\hline
\multirow{2}{*}{3/2}
 & $S_{1}^{3/2}$ & $\PD$, $\DP$ & $\left(\frac{1}{2},\frac{1}{2}\right)$ &   $\left(\frac{3}{4}, \frac{3}{4}\right)$\\
 & $S_{2}^{3/2}$ & $\PD$, $\DP$, $\PN, \NP$ & $\left(\frac{1}{4},\frac{1}{4}, \frac{1}{4}, \frac{1}{4}\right)$ &   $\left(\frac{3}{4}, \frac{3}{4}\right)$\\\hline
\multirow{2}{*}{4/3} 
 & $S_{1}^{4/3}$ & $\PD, \DP, \NN$ & $\left(\frac{1}{3},\frac{1}{3}, \frac{1}{3}\right)$ &   $\left(\frac{2}{3},\frac{2}{3}\right)$\\
 & $S_{2}^{4/3}$ & $\PN, \NP, \DD$ & $\left(\frac{1}{3},\frac{1}{3}, \frac{1}{3}\right)$ &   $\left(\frac{2}{3},\frac{2}{3}\right)$\\\hline
 \multirow{3}{*}{1}
 & $S_{1}^{1}$ & $\DD$ & $1$ &   $\left(\frac{1}{2}, \frac{1}{2}\right)$\\
 & $S_{2}^{1}$ & $\DD, \NN$ & $\left(\frac{1}{2},\frac{1}{2}\right)$ &   $\left(\frac{1}{2}, \frac{1}{2}\right)$\\
 & $S_{3}^{1}$ & $\DN, \ND$ & $\left(\frac{1}{2},\frac{1}{2}\right)$ &   $\left(\frac{1}{2}, \frac{1}{2}\right)$\\\hline
  \multirow{3}{*}{2/3}
 & $S_{1}^{2/3}$ & $\DD$ & $1$ &   $\left(\frac{2}{3}, 0\right)$\\
 & $S_{2}^{2/3}$ & $\DD, \NN$ & $\left(\frac{2}{3}, \frac{1}{3}\right)$ &   $\left(\frac{2}{3}, 0\right)$\\
 & $S_{3}^{2/3}$ & $\DN, \ND, \NN$ & $\left(\frac{1}{3}, \frac{1}{3},\frac{1}{3}\right)$ &   $\left(\frac{2}{3}, 0\right)$\\
\hline
 \end{tabular}
\vspace{8pt}
\caption{Constituent schemes.}\label{TableCSsummary}
\end{center}
\end{table}

Before we present the achievability of the s.d.o.f.~region, we first present the key constituent schemes that will be instrumental in the proof. We combine these schemes carefully and time share between them to achieve the s.d.o.f.~region. A summary of these constituent schemes is shown in Table~\ref{TableCSsummary}. Before we discuss the individual schemes we make the following remark that applies to all the schemes presented here. 

\subsection{A Note on the Achievable Security Guarantee}
Each scheme described in the following sections can be outlined as follows. We neglect the impact of noise at high SNR. Then, to achieve a certain s.d.o.f.~pair $(d_1,d_2)$, we send $n_1$ symbols $\ubar{u} = \left(u_1,\ldots,u_{n_1}\right)$ and $n_2$ symbols $\ubar{v}=\left(v_1,\ldots,v_{n_2}\right)$ intended for the first and second receivers, respectively, in $n_B$ slots, such that $d_1 = n_1/n_B$ and $d_2 = n_2/n_B$. Finally, we argue that the leakage of information symbols at the unintended receiver is $o(\log P)$. We  however want a stronger guarantee of security, namely,
\begin{align}
 \frac{1}{n} I(W_1;Z^n,\mathbf{H}^n) \leq \epsilon_n,\qquad
 \frac{1}{n} I(W_2;Y^n,\mathbf{H}^n)\leq \epsilon_n.
\end{align}      
To achieve this, we view the $n_B$ slots described in the scheme as a block and treat the equivalent channel from $\ubar{u}$ to $(\mathbf{Y},\mathbf{H})$ and $(\mathbf{Z},\mathbf{H})$ as a memoryless wiretap channel (with $(\mathbf{Y},\mathbf{H})$ being the legitimate receiver) by ignoring the $\CSI$ of the previous block. We do the same for the channel from $\ubar{v}$  to $(\mathbf{Z},\mathbf{H})$ and $(\mathbf{Y},\mathbf{H})$ (with $(\mathbf{Z},\mathbf{H})$ as the legitimate receiver). Note also that no information about $\mathbf{H}$ is used to create the codebooks for $\ubar{u}$ and $\ubar{v}$ in any of the schemes. More formally, the following secrecy rate pair is achievable for receivers 1 and 2, respectively, from \cite{Wyner}: 
\begin{align}
R_{1}=& I(\ubar{u};\mathbf{Y},\mathbf{H}) - I(\ubar{v};\mathbf{Z},\mathbf{H})
= I(\ubar{u};\mathbf{Y}|\mathbf{H}) - I(\ubar{v};\mathbf{Z}|\mathbf{H})\label{eq:r1}\\
R_{2}=& I(\ubar{v};\mathbf{Z},\mathbf{H}) - I(\ubar{u};\mathbf{Y},\mathbf{H})
= I(\ubar{v};\mathbf{Z}|\mathbf{H}) - I(\ubar{u};\mathbf{Y}|\mathbf{H}),\label{eq:r2}
\end{align}
where we noted that $\ubar{u}$ and $\ubar{v}$ are all independent of $\mathbf{H}$. Using the proposed scheme, $\ubar{u}$ (resp., $\ubar{v}$) can be reconstructed from $(\mathbf{Y},\mathbf{H})$ (resp., $(\mathbf{Z},\mathbf{H})$) to within a noise distortion. Thus, 
\begin{align}
I(\ubar{u};\mathbf{Y}|\mathbf{H})
=& n_1 \log P + o(\log P)\label{eq:indep1}\\
I(\ubar{v};\mathbf{Z}|\mathbf{H})=& n_2\log P + o(\log P) \label{eq:indep2}.
\end{align}
Also, for each scheme,
\begin{align}
I(\ubar{v};\mathbf{Y}|\mathbf{H})
=& o(\log P)\label{eq:leak1}\\
I(\ubar{u};\mathbf{Z}|\mathbf{H})=& o(\log P) \label{eq:leak2}.
\end{align}
Thus, from \eqref{eq:r1} and \eqref{eq:r2}, the achievable secure rates in each block are,
\begin{align}
R_1 =& n_1\log P + o(\log P)\\
R_2 =& n_2\log P + o(\log P).
\end{align}
Since our block contains $n_B$ channel uses, the effective secure rates are
\begin{align}
R_1 =& \frac{n_1}{n_B}\log P + o(\log P)\\
R_2 =& \frac{n_2}{n_B}\log P + o(\log P).
\end{align}
These rates clearly yield the required s.d.o.f.~pair $(d_1,d_2)$, while also conforming to our stringent security requirement.

In the following subsections, we now present the achievability of each scheme in detail.

Notation:  A particular sum s.d.o.f.~value can be achieved in various ways through alternation between different possible sets of $\CSIT$ states. To this end, we use the following notation: if there are $r$ schemes achieving a particular s.d.o.f.~value, we denote these schemes as: $S^{\text{sum s.d.o.f.}}_{1}, S^{\text{sum s.d.o.f.}}_{2}, \ldots, S^{\text{sum s.d.o.f.}}_{r}$. For example, in Table~\ref{TableCSsummary}, for achieving the sum s.d.o.f.~value of $1$, we present $r=3$ distinct schemes and these are denoted as $S^{1}_{1}, S^{1}_{2}$ and $S^{1}_{3}$.

Given a $1\times 2$ channel vector $\mathbf{H}(t)$, we denote by $\mathbf{H}(t)^{\perp}$, a $2\times 1$ beamforming vector that is orthogonal to the $1\times 2$ channel vector $\mathbf{H}(t)$; in other words, $\mathbf{H}(t)\mathbf{H}(t)^{\perp} =0$.  

\subsection{Scheme Achieving Sum s.d.o.f.~of 2} 
A sum s.d.o.f.~of $2$ is achievable only in the state $\PP$, that is, when the transmitter has perfect $\CSIT$ from both users. This is achievable using zero-forcing. The following scheme achieves a sum s.d.o.f.~of 2.
\subsubsection{Scheme $S^2$}

The scheme $S^2$ uses the state $\PP$ and achieves the rate pair $(d_1,d_2) = (1,1)$.
The scheme is as follows. We wish to send confidential symbols $u$ and $v$ to receivers $1$ and $2$, respectively, in one time slot, thus achieving a sum s.d.o.f.~of $2$. Since the transmitter knows both channel coefficients $\mathbf{H}_1$ and $\mathbf{H}_2$, it sends,
\begin{align}
\mathbf{X} = u\mathbf{H}_2^{\perp} + v\mathbf{H}_1^{\perp}, 
\end{align}
where, $\mathbf{H}_i(t)^{\perp}$ is a $2\times 1$ beamforming vector that is orthogonal to the $1\times 2$ channel vector $\mathbf{H}_i(t)$ for $i=1,2$. This is to ensure that the symbols do not leak to unintended receivers. For s.d.o.f.~calculations, we disregard the additive noise and the outputs at the receivers are:
\begin{align}
Y =& u \mathbf{H}_1\mathbf{H}_2^{\perp}\\
Z =& v \mathbf{H}_2\mathbf{H}_1^{\perp},
\end{align}
which allows both receivers to decode their respective messages. Also, since $u$ does not appear at all in $Z$, the confidentiality of $u$ is guaranteed. Similarly, the confidentiality of $v$ too is satisfied.

\subsection{Schemes Achieving Sum s.d.o.f.~of 3/2}
The following schemes achieve $\frac{3}{2}$ sum s.d.o.f.:

\subsubsection{Scheme $S_1^{3/2}$}

In this subsection, we present the scheme $S_1^{3/2}$ which uses the states $(\PD,\DP)$ with fractions $(\frac{1}{2},\frac{1}{2})$ to achieve rate pair $(d_1,d_2)=(\frac{3}{4},\frac{3}{4})$.

This scheme was presented in \cite{pmukherjee_ravi_ulukus_isit2014}. For the sake of completeness we reproduce the scheme here. We wish to send $3$ confidential symbols from the transmitter to each of the receivers in $4$ channel uses at high $P$ (that is negligible noise). Let us denote by $(u_1,u_2,u_3)$ and $(v_1,v_2,v_3)$ the confidential symbols intended for receivers $1$ and $2$, respectively. Also, in $2$ of the $4$ channel uses, the channel is in state $\textsf{PD}$; in the remaining $2$ uses, the channel is in state $\textsf{DP}$. The scheme is as follows:

1) At time \textbf{$t=1$}, $S(1) = \textsf{PD}$: As the transmitter knows $\mathbf{H}_1(1)$, it sends:
\begin{align}
\mathbf{X}(1) = [u_{1} \quad 0]^{T} + q\mathbf{H}_1(1)^{\perp},
\end{align}
where $\mathbf{H}_1(1)\mathbf{H}_1(1)^{\perp} =0$, and $q$ denotes an artificial noise distributed as $\mathcal{CN}(0, P)$. Here $\mathbf{H}_1(1)^{\perp}$ is a $2\times 1$ beamforming vector orthogonal to the $1\times 2$ channel vector $\mathbf{H}_{1}(1)$ of receiver $1$ that ensures that the artificial noise $q$ does not create interference at receiver $1$.  The receivers' outputs are:
\begin{align}
Y(1) &= h_{11}(1)u_1\\
Z(1) &= h_{21}(1)u_1 + q\mathbf{H}_2(1)\mathbf{H}_1(1)^{\perp} \stackrel{\Delta}{=} K.
\end{align}
Thus, receiver $1$ has observed $u_1$ while receiver $2$ gets a linear combination of $u_1$ and $q$, which we denote as $K$. Due to delayed $\CSIT$ from receiver $2$, the transmitter can reconstruct $K$ in the next channel use and use it for transmission.

2) At time \textbf{$t=2$}, $S(2) = \textsf{DP}$:  The transmitter knows $\mathbf{H}_2(2)$ and $K$. It sends
\begin{align}
\mathbf{X}(2) =\left[ v_1+K \quad v_2+K\right]^{T} + u_2\mathbf{H}_2(2)^{\perp}.
\end{align}
The received signals are:
\begin{align}
Y(2) =&\, h_{11}(2)v_1 + h_{12}(2)v_2 + (h_{11}(2)+h_{12}(2))K + u_2\mathbf{H}_1(2)\mathbf{H}_2(2)^{\perp}\\
=&\, L_1(v_1,v_2,K)  +  u_2\mathbf{H}_1(2)\mathbf{H}_2(2)^{\perp}\\
Z(2) =&\, h_{21}(2)v_1+h_{22}(2)v_2 + (h_{21}(2)+h_{22}(2))K \\ \stackrel{\Delta}{=}&\, L_2(v_1,v_2,K),
\end{align}
where we have defined $L_1(v_1,v_2,K)$ and $L_2(v_1,v_2,K)$ as linear combinations of $v_1, v_2$ and $K$ at receivers $1$ and $2$, respectively.

3) At time \textbf{$t=3$}, $S(3) = \textsf{DP}$:  The transmitter knows $\mathbf{H}_2(3)$ and $L_1(v_1,v_2,K)$ (via delayed $\CSIT$ from $t=2$). Using these, it transmits:
  \begin{align}
 \mathbf{X}(3) = \left[ L_1(v_1,v_2,K) \quad  0\right]^{T} + u_3\mathbf{H}_2(3)^{\perp},
 \end{align}
and  the channel outputs are:
 \begin{align}
 Y(3) &= h_{11}(3)L_1(v_1,v_2,K) + u_3\mathbf{H}_1(3)\mathbf{H}_2(3)^{\perp}\\
 Z(3) &= h_{21}(3)L_1(v_1,v_2,K).
 \end{align}
At the end of this step, note that, receiver $2$ can decode $v_1$ and $v_2$ by first eliminating $K$ using $Z(1)$ and $Z(3)$ to get a linear combination of $v_1$ and $v_2$, which it can then use with $Z(2)$ to solve for $v_1$ and $v_2$.

4) At time \textbf{$t=4$}, $S(4) = \textsf{PD}$: The transmitter knows $\mathbf{H}_1(4)$ and it sends
 \begin{align}
  \mathbf{X}(4) = \left[  L_{1}(v_1,v_2,K) \quad
  0
\right]^{T} + v_3\mathbf{H}_1(4)^{\perp},
 \end{align}
and the channel outputs are:
 \begin{align}
  Y(4) &= h_{11}(4)L_1(v_1,v_2,K) \\
  Z(4) &= h_{21}(4)L_1(v_1,v_2,K) + v_3 \mathbf{H}_2(4)\mathbf{H}_1(4)^{\perp}.
 \end{align}

\begin{figure*}[t]
\centering{\includegraphics[width=\linewidth]{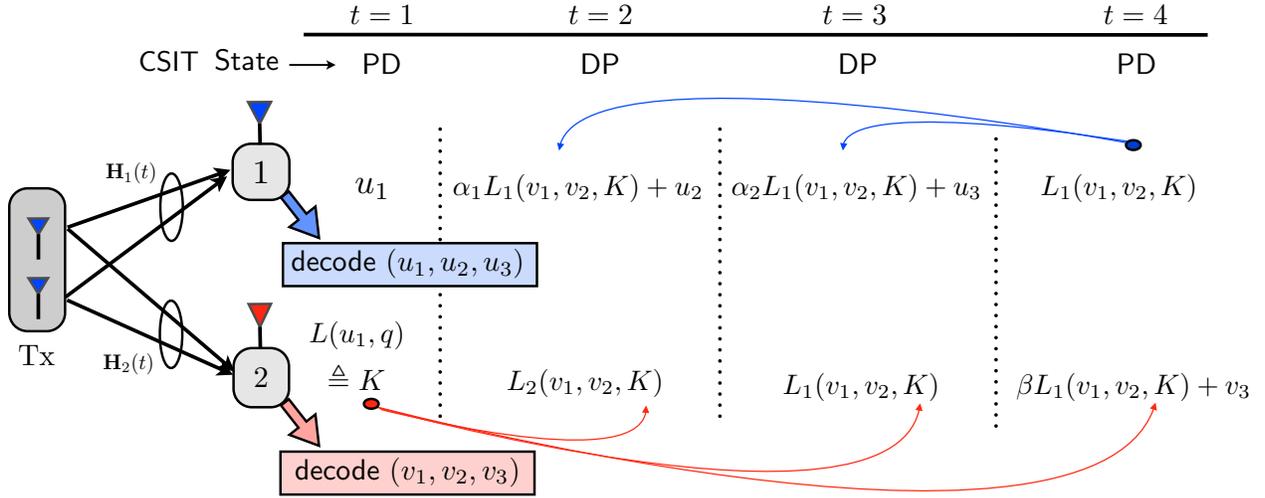}}
\caption{Achieving $\frac{3}{2}$ s.d.o.f.~using scheme $S_1^{3/2}$.}
\label{fig:achievability}
\end{figure*}

Thus, at the end of these four steps the outputs at the two receivers can be summarized (see Fig.~\ref{fig:achievability}) as:
 \begin{align}
 \mathbf{Y}\hspace{-2pt}= \hspace{-2pt}\left[ \hspace{-4pt}\begin{array}{c}
 u_1  \\
\hspace{-1pt}\alpha_{1}L_1(v_1,v_2,K)  +  u_2\\
\hspace{-1pt}\alpha_{2}L_1(v_1,v_2,K) + u_3 \\
\hspace{-1pt}L_1(v_1,v_2,K) \end{array} \hspace{-4pt}\right],\qquad
\mathbf{Z}\hspace{-1pt}= \hspace{-1pt}\left[ \hspace{-4pt}\begin{array}{c}K\\
\hspace{-1pt} L_2(v_1,v_2,K)\\
\hspace{-1pt} L_1(v_1,v_2,K)\\
\hspace{-1pt}\beta L_1(v_1,v_2,K) \hspace{-1pt} \hspace{-1pt}+ \hspace{-1pt}\hspace{-1pt}v_3
\end{array} \hspace{-4pt}\right].\nonumber
 \end{align}
Using $\mathbf{Y}$, receiver $1$ can decode all three symbols $(u_1,u_2,u_3)$ and using $\mathbf{Z}$, receiver $2$ can decode $(v_{1}, v_{2}, v_{3})$. Next we prove that the information leakage is only $o(\log P)$.

\emph{Security guarantees}:

We consider the four slots as a single block and the equivalent channel from $\ubar{u}=(u_1,u_2,u_3)$ to $(\mathbf{Y},\mathbf{H})$ and $(\mathbf{Z},\mathbf{H})$ as a memoryless channel by ignoring the $\CSI$ of the previous block. We do the same for the channel from $\ubar{v} = (v_1,v_2,v_3)$  to $(\mathbf{Y},\mathbf{H})$ and $(\mathbf{Z},\mathbf{H})$. Recall that all the random variables $\left\lbrace u_i,v_i,i=1,2,3\right\rbrace$ and $q$ are independent and distributed as $\mathcal{CN}(0,P)$. 

First, let us consider the confidentiality of the first user's symbols $\ubar{u}$. The information leakage at user 2 is:
\begin{align}
I(\ubar{u};\mathbf{Z}|\mathbf{H})
=& I(u_1,u_2,u_3;\mathbf{Z}|\mathbf{H})\\
=& I(u_1;\mathbf{Z}|\mathbf{H})\label{eq:user1_pddp_1}\\
\leq& I(u_1;K|\mathbf{H})\label{eq:user1_pddp_2} \\
=& I(u_1;h_{21}(1)u_1 + q\mathbf{H}_2(1)\mathbf{H}_1(1)^{\perp}|\mathbf{H})\\
=& h(h_{21}(1)u_1 + q\mathbf{H}_2(1)\mathbf{H}_1(1)^{\perp}|\mathbf{H}) - h(h_{21}(1)u_1 + q\mathbf{H}_2(1)\mathbf{H}_1(1)^{\perp}|u_1,\mathbf{H})\\
=& h(h_{21}(1)u_1 + q\mathbf{H}_2(1)\mathbf{H}_1(1)^{\perp}|\mathbf{H}) - h( q\mathbf{H}_2(1)\mathbf{H}_1(1)^{\perp}|\mathbf{H})\\
=& \left(\log P + o(\log P)\right) - \left(\log P + o(\log P)\right)\\
=& o(\log P),
\end{align}
where \eqref{eq:user1_pddp_1} follows from the fact that  $\mathbf{Z}$ does not have any term involving $(u_2, u_3)$, and \eqref{eq:user1_pddp_2} follows from the Markov chain $u_1 \rightarrow K \rightarrow \mathbf{Z}$.

For the second user's symbols, the information leakage at the first receiver is:
\begin{align}
I(\ubar{v};\mathbf{Y}|\mathbf{H})
=& I(v_1,v_2,v_3;\mathbf{Y}|\mathbf{H})\\
=& I(v_1,v_2;\mathbf{Y}|\mathbf{H})\label{eq:user2_pddp_1}\\
\leq& I(v_1,v_2;L_1(v_1,v_2,K)|\mathbf{H})\label{eq:user2_pddp_2}\\
=& h(L_1(v_1,v_2,K)|\mathbf{H})-h(L_1(v_1,v_2,K)|v_1,v_2,\mathbf{H})\\
\leq& \log P - h(K|v_1,v_2,\mathbf{H}) + o(\log P)\\
=& \log P - h(K|\mathbf{H}) + o(\log P)\\
=& \log P -\log P + o(\log P)\\
=& o(\log P),
\end{align}
where \eqref{eq:user2_pddp_1} follows since $v_3$ does not appear in $\mathbf{Y}$ and \eqref{eq:user2_pddp_2} follows from the Markov chain $(v_1,v_2)\rightarrow L_1(v_1,v_2,K)\rightarrow \mathbf{Y}$.

\subsubsection{Scheme $S_2^{3/2}$}
In this sub-section, we present the scheme $S_2^{3/2}$ which uses the states $(\PD,\DP,\PN,\NP)$ with fractions $(\frac{1}{4},\frac{1}{4},\frac{1}{4},\frac{1}{4})$ to achieve  $(d_1,d_2)=(\frac{3}{4},\frac{3}{4})$.

Let us consider the utilization of $\CSIT$ in the scheme $S_1^{3/2}$ stated above. In the first slot, delayed $\CSIT$ is required from the second user, since that knowledge allows the transmitter to reconstruct $K$ and use it in the second slot. Similarly, in the second time slot, delayed $\CSIT$ from the first user  is required so that the transmitter can reconstruct $L_1(v_1,v_2,K)$ to transmit in the third and fourth slots. However, in the third and fourth slots, the transmitter does not require any $\CSIT$ of the first and second users, respectively. Thus, the same scheme works with $\PN$ and $\NP$ states in the last two slots. Since it is essentially the same scheme interpreted in a different way, the security of the scheme follows from that of $S_1^{3/2}$.

\subsection{Schemes Achieving Sum s.d.o.f.~of 4/3}

\subsubsection{Scheme $S_1^{4/3}$}

In this sub-section, we present the scheme $S_1^{4/3}$ which uses the states $(\PD,\DP,\NN)$ for fractions $(\frac{1}{3},\frac{1}{3},\frac{1}{3})$ to achieve s.d.o.f.~pair $(d_1,d_2)=(\frac{2}{3},\frac{2}{3})$.

\begin{figure}
\includegraphics[width=0.9\linewidth]{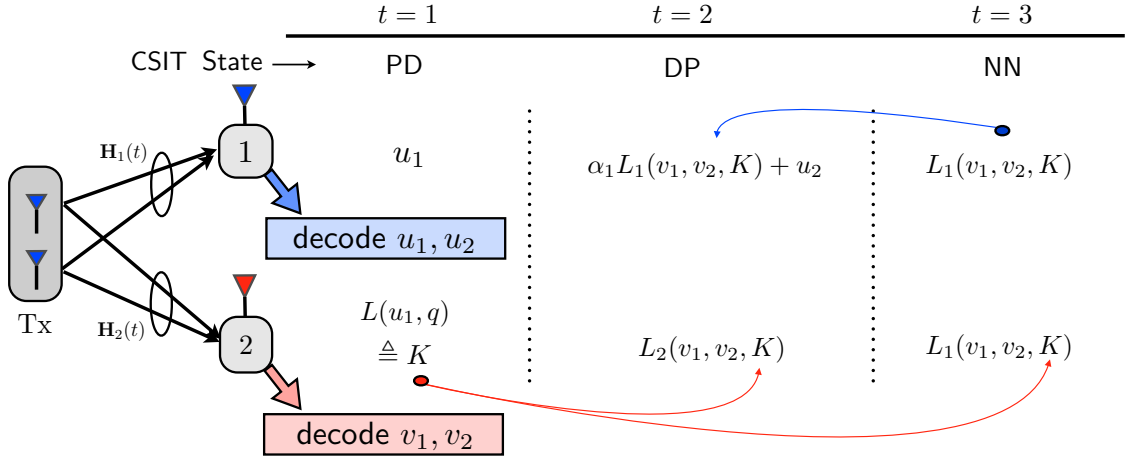}
\caption{Achieving sum s.d.o.f.~of $\frac{4}{3}$ using $S_1^{4/3}$.}
\label{fig:pd_dp_nn}
\end{figure}
We wish to send $2$ symbols to each user in $3$ time slots. Let $(u_1,u_2)$ and $(v_1,v_2)$ be the symbols intended for the first and second users, respectively. Fig.~\ref{fig:pd_dp_nn} shows the scheme. It is as follows:

1) At time \textbf{$t=1$}, $S(1) = \textsf{PD}$: As the transmitter knows $\mathbf{H}_1(1)$, it sends:
\begin{align}
\mathbf{X}(1) = [u_{1} \quad 0]^{T} + q\mathbf{H}_1(1)^{\perp},
\end{align}
where $\mathbf{H}_1(1)\mathbf{H}_1(1)^{\perp} =0$, and $q$ denotes an artificial noise distributed as $\mathcal{CN}(0, P)$. Here $\mathbf{H}_1(1)^{\perp}$ is a $2\times 1$ beamforming vector that ensures that the artificial noise $q$ does not create interference at receiver $1$. The receivers' outputs are:
\begin{align}
Y(1) &= h_{11}(1)u_1\\
Z(1) &= h_{21}(1)u_1 + q\mathbf{H}_2(1)\mathbf{H}_1(1)^{\perp} \stackrel{\Delta}{=} K.
\end{align}
Thus, receiver $1$ has observed $u_1$ while receiver $2$ gets a linear combination of $u_1$ and $q$, which we denote as $K$. Due to delayed $\CSIT$ from receiver $2$, the transmitter can reconstruct $K$ in the next channel use and use it for transmission.

2) At time \textbf{$t=2$}, $S(2) = \textsf{DP}$:  The transmitter knows $\mathbf{H}_2(2)$ and $K$. It sends
\begin{align}
\mathbf{X}(2) =\left[ v_1+K \quad v_2+K\right]^{T} + u_2\mathbf{H}_2(2)^{\perp}.
\end{align}
The received signals are:
\begin{align}
Y(2) =&\, h_{11}(2)v_1 + h_{12}(2)v_2 + (h_{11}(2)+h_{12}(2))K + u_2\mathbf{H}_1(2)\mathbf{H}_2(2)^{\perp}\\
=&\, L_1(v_1,v_2,K)  +  u_2\mathbf{H}_1(2)\mathbf{H}_2(2)^{\perp}\\
Z(2) =&\, h_{21}(2)v_1+h_{22}(2)v_2 + (h_{21}(2)+h_{22}(2))K \nonumber\\ \stackrel{\Delta}{=}&\, L_2(v_1,v_2,K),
\end{align}
where we have defined $L_1(v_1,v_2,K)$ and $L_2(v_1,v_2,K)$ as independent linear combinations of $v_{1}, v_{2}$ and $K$ at receivers $1$ and $2$, respectively.

3) At time $t=3$, $S(3)=\NN$: The transmitter transmits:
\begin{align}
\mathbf{X}(3) = \left[L_1(v_1,v_2,K) \quad 0\right]^T.
\end{align} 
The receivers get:
\begin{align}
Y(3) =& h_{11}(3)L_1(v_1,v_2,K)\\
Z(3) =& h_{21}(3)L_1(v_1,v_2,K).
\end{align}  

At the end of three slots, therefore, the received outputs can be summarized as:
 \begin{align}
 \mathbf{Y}= \hspace{-2pt}\left[ \begin{array}{c}
 u_1  \\
\alpha_{1}L_1(v_1,v_2,K)  +  u_2\\
L_1(v_1,v_2,K) \end{array} \right],\qquad
\mathbf{Z}= \left[ \begin{array}{c}K\\
L_2(v_1,v_2,K)\\
L_1(v_1,v_2,K)\\
\end{array} \right].\nonumber
 \end{align}
Using $\mathbf{Y}$, receiver $1$ can  decode $(u_1,u_2)$, while receiver $2$ can decode $(v_1,v_2)$ using $\mathbf{Z}$. The information leakage is only $o(\log P)$ as we show next. 

\emph{Security guarantees}:

The equivocation calculation follows similar to that of the scheme $S_1^{3/2}$. For the first user's symbols $\ubar{u}=(u_1,u_2)$, we have,
\begin{align}
I(\ubar{u};\mathbf{Z}|\mathbf{H})
=& I(u_1,u_2;\mathbf{Z}|\mathbf{H})\\
=& I(u_1;\mathbf{Z}|\mathbf{H})\label{eq:user1_pddpnn_1}\\
\leq& I(u_1;K|\mathbf{H})\label{eq:user1_pddpnn_2} \\
=& o(\log P),
\end{align}
where \eqref{eq:user1_pddpnn_1} follows from the fact that  $\mathbf{Z}$ does not have any term involving $u_2$, and \eqref{eq:user1_pddpnn_2} follows from the Markov chain $u_1 \rightarrow K \rightarrow \mathbf{Z}$.

For the second user's symbols, the information leakage at the first receiver is:
\begin{align}
I(\ubar{v};\mathbf{Y}|\mathbf{H})
\leq& I(v_1,v_2;L_1(v_1,v_2,K)|\mathbf{H})\label{eq:user2_pddpnn_1}\\
=& h(L_1(v_1,v_2,K)|\mathbf{H})-h(L_1(v_1,v_2,K)|v_1,v_2,\mathbf{H})\\
\leq& \log P - h(K|v_1,v_2,\mathbf{H}) + o(\log P)\\
=& \log P - h(K|\mathbf{H}) + o(\log P)\\
=& \log P -\log P + o(\log P)\\
=& o(\log P),
\end{align}
where \eqref{eq:user2_pddpnn_1} follows from the Markov chain $(v_1,v_2)\rightarrow L_1(v_1,v_2,K)\rightarrow \mathbf{Y}$.

\subsubsection{Scheme $S_2^{4/3}$}

We now present the scheme $S_2^{4/3}$ which uses the states $\PN,\NP,\DD$ with fractions $(\frac{1}{3},\frac{1}{3},\frac{1}{3})$ to achieve $(d_1,d_2)=(\frac{2}{3},\frac{2}{3})$. 

\begin{figure}
\includegraphics[width=\linewidth]{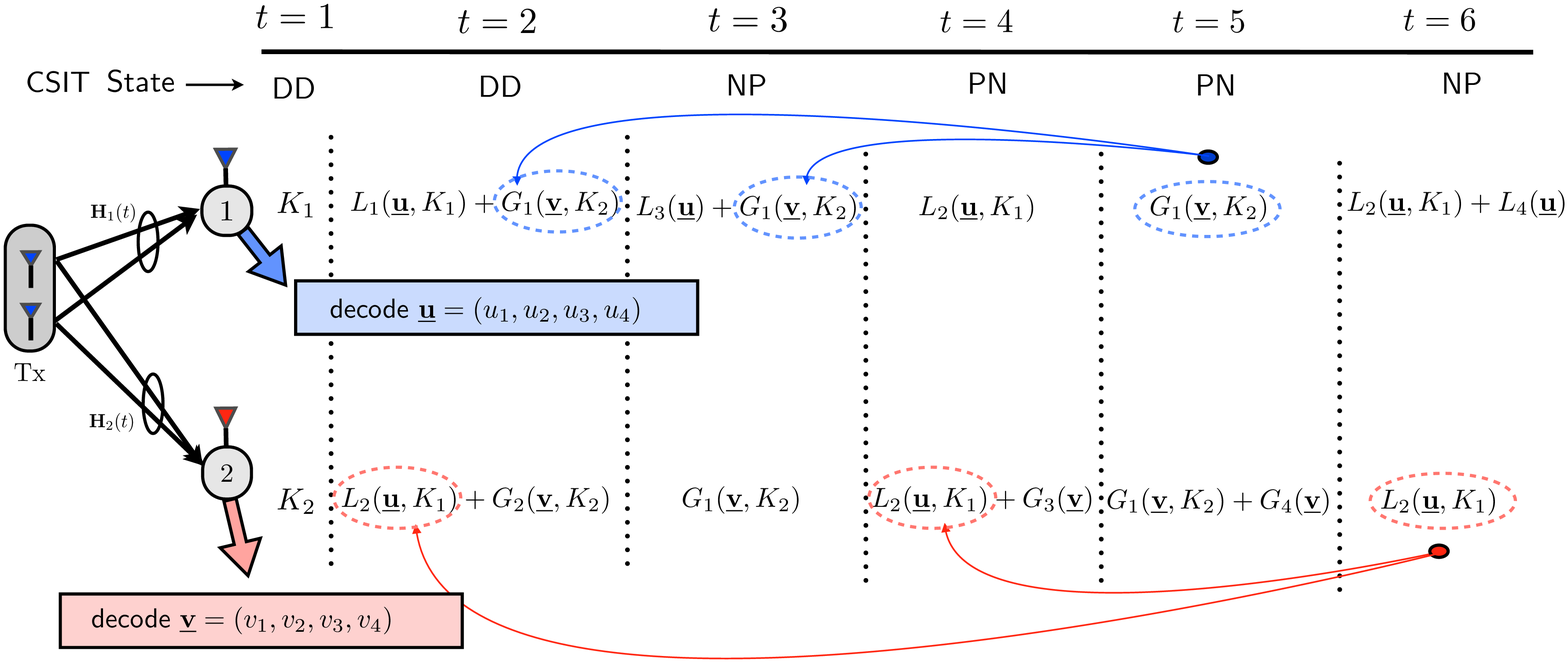}
\caption{Achieving sum s.d.o.f.~$\frac{4}{3}$ using $S_2^{4/3}$.}
\label{fig:pn_np_dd}
\end{figure}

In this case we will send $4$ symbols to each user in $6$ time slots. Let $\underline{\mathbf{u}}=(u_1,u_2,u_3,u_4)$ and $\ubar{v}=(v_1,v_2,v_3,v_4)$ be the symbols intended for the first and second users, respectively. Fig.~\ref{fig:pn_np_dd} shows the scheme. It is as follows:

1) At time $t=1$, $S(1) =\DD$: In this slot, the transmitter sends artificial noise symbols to create keys that can be used in later slots. The channel input is 
\begin{align}
\mathbf{X}(1) = \left[q_1 \quad q_2\right]^T,
\end{align}
where $q_1$ and $q_2$ are i.i.d.~as $\mathcal{CN}(0,P)$. The received signals are:
\begin{align}
Y(1) =& h_{11}(1)q_1 + h_{12}(1)q_2 \stackrel{\Delta}{=} K_1\\
Z(1) =& h_{21}(1)q_1 + h_{22}(1)q_2 \stackrel{\Delta}{=} K_2.
\end{align}
Due to delayed $\CSIT$, the transmitter learns $K_1$ and $K_2$ and uses them in the next time slots.
 
2) At time $t=2$, $S(2) =\DD$: In this slot, the transmitter sends:
\begin{align}
\mathbf{X}(2) = \left[u_1+u_2+v_3+v_4+K_1 \quad v_1+v_2+u_3+u_4+K_2 \right]^T.
\end{align} 
The received signals are:
\begin{align}
Y(2) &= h_{11}(2)(u_1+u_2+v_3+v_4+K_1) + h_{12}(2)(v_1+v_2+u_3+u_4+K_2) \\
&\stackrel{\Delta}{=} L_1(\ubar{u},K_1)+G_1(\ubar{v},K_2)\\
Z(2) &= h_{21}(2)(u_1+u_2+v_3+v_4+K_1) + h_{22}(2)(v_1+v_2+u_3+u_4+K_2) \\
&\stackrel{\Delta}{=} L_2(\ubar{u},K_1) + G_2(\ubar{v},K_2).
\end{align}
Note that since $K_1$ (or $K_2$) is known at the first (or second) receiver, it can be removed. The unintended symbols remain buried in the artificial noise, ensuring security. Also, if $G_1$ (or $L_2$) could be sent to the second (or first) receiver, it would provide a linear combination of the intended symbols  that is linearly independent of $G_2$ (or ${L_1}$). This is what we will do in the third and fourth  time slots.    

3) At time $t=3$, $S(3) =\NP$: In this state, the transmitter knows $\mathbf{H}_2$ perfectly. It sends,
\begin{align}
\mathbf{X}(3) = \left[G_1(\ubar{v},K_2)\quad 0\right]^T + L_3(\ubar{u})\mathbf{H}_2(3)^{\perp},
\end{align} 
where $L_3$ is linearly independent of both $L_1$ and $L_2$. The received signals are:
\begin{align}
Y(3) =& h_{11}(3)G_1(\ubar{v},K_2) + L_3(\ubar{u}) \mathbf{H}_1(3)\mathbf{H}_2(3)^{\perp} \\
Z(3) =& h_{21}(3)G_1(\ubar{v},K_2).
\end{align} 
    
4) At time $t=4$, $S(4) =\PN$: In this state, the transmitter knows $\mathbf{H}_1(4)$ perfectly. It sends,
\begin{align}
\mathbf{X}(4) = \left[L_2(\ubar{u},K_1)\quad 0\right]^T + G_3(\ubar{v})\mathbf{H}_1(4)^{\perp},
\end{align} 
where $G_3$ is linearly independent of both $G_1$ and $G_2$. The received signals are:
\begin{align}
Y(4) =& h_{11}(4)L_2(\ubar{u},K_1)   \\
Z(4)=& h_{21}(4)L_2(\ubar{u},K_1)+G_3(\ubar{v})\mathbf{H}_2(4)\mathbf{H}_1(4)^{\perp}.
\end{align} 
Now note that if we could supply $G_1$ and $L_2$ to the first and second receivers, respectively, both receivers will end up with $3$ linearly independent combinations of their intended symbols. Thus, in the next two slots, the transmitter will supply $G_1$ and $L_2$ to the first and second receivers, respectively, as well as send one more linearly independent combination of the intended information symbols to each receiver.    

5) At time $t=5$, $S(5) =\PN$: In this state, the transmitter knows $\mathbf{H}_1(5)$ perfectly. It sends,
\begin{align}
\mathbf{X}(5) = \left[G_1(\ubar{v},K_2) \quad 0\right]^T + G_4(\ubar{v})\mathbf{H}_1(5)^{\perp}.
\end{align}
The receivers receive:
\begin{align}
Y(5) &= h_{11}(5)G_1(\ubar{v},K_2)\\
Z(5) &= h_{21}(5)G_1(\ubar{v},K_2) + G_4(\ubar{v})\mathbf{H}_2(5)\mathbf{H}_1(5)^{\perp}.
\end{align}

6) At time $t=6$, $S(6) =\NP$: Now the transmitter knows $\mathbf{H}_2(6)$ perfectly, and it sends:
\begin{align}
\mathbf{X}(6) = \left[L_2(\ubar{u},K_1) \quad 0 \right] + L_4(\ubar{u})\mathbf{H}_2(6)^{\perp}.
\end{align}
The received signals are:
\begin{align}
Y(6) &= h_{11}(6)L_2(\ubar{u},K_1) + L_4(\ubar{u}) \mathbf{H}_1(6) \mathbf{H}_2(6)^{\perp}\\
Z(6) &= h_{21}(6)L_2(\ubar{u},K_1).
\end{align}

Let us summarize the received signals at each receiver after these 6 time slots:
\begin{align}
 \mathbf{Y}=\left[ \begin{array}{c}
K_1  \\
L_1(\ubar{u},K_1)+G_1(\ubar{v},K_2)\\
\alpha_1G_1(\ubar{v},K_2) + L_3(\ubar{u})\\
L_2(\ubar{u},K_1)\\
G_1(\ubar{v},K_2)\\
\alpha_2{L}_2(\ubar{u},K_1) + L_4(\ubar{u})
\end{array} \right], \qquad
\mathbf{Z}= \left[ \begin{array}{c} K_2\\
L_2(\ubar{u},K_1)+G_2(\ubar{v},K_2) \\
G_1(\ubar{v},K_2)\\
\beta_1L_2(\ubar{u},K_1)+G_3(\ubar{v})\\
\beta_2 G_1(\ubar{v},K_2) + G_4(\ubar{v})\\
L_2(\ubar{u},K_1)
\end{array} \right].\nonumber
 \end{align} 

The information symbols can now be decoded at the intended receivers from these observations. Also the leakage of information is only $o(\log P)$, as we prove next. 

\emph{Security guarantees}:

For the first user's symbols $\ubar{u}=(u_1,u_2,u_3,u_4)$, we have,
\begin{align}
I(\ubar{u};\mathbf{Z}|\mathbf{H})
\leq& I(\ubar{u};L_2(\ubar{u},K_1)|\mathbf{H})\label{eq:user1_pnnpdd_1} \\
=& h(L_2(\ubar{u},K_1)|\mathbf{H})-h(L_2(\ubar{u},K_1)|\ubar{u},\mathbf{H})\\
\leq& \log P - h(K_1|\ubar{u},\mathbf{H}) + o(\log P)\\
=& \log P - h(K_1|\mathbf{H}) + o(\log P)\\
=& \log P -\log P + o(\log P)\\
=& o(\log P),
\end{align}
where \eqref{eq:user1_pnnpdd_1} follows from the Markov chain $U \rightarrow L_2(\ubar{u},K_1) \rightarrow \mathbf{Z}$.

For the second user's symbols, the information leakage at the first receiver is:
\begin{align}
I(\ubar{v};\mathbf{Y}|\mathbf{H})
\leq& I(\ubar{v};G_1(\ubar{v},K_2)|\mathbf{H})\label{eq:user2_pnnpdd_1}\\
=& h(G_1(\ubar{v},K_2)|\mathbf{H})-h(G_1(\ubar{v},K_2)|\ubar{v},\mathbf{H})\\
\leq& \log P - h(K_2|\ubar{v},\mathbf{H}) + o(\log P)\\
=& \log P - h(K_2|\mathbf{H}) + o(\log P)\\
=& \log P -\log P + o(\log P)\\
=& o(\log P),
\end{align}
where \eqref{eq:user2_pddpnn_1} follows from the Markov chain $\ubar{v}\rightarrow G_1(\ubar{v},K_2)\rightarrow \mathbf{Y}$.

\subsection{Schemes Achieving Sum s.d.o.f.~of 1}

\subsubsection{Scheme $S_1^{1}$}

We first recap the scheme $S_1^{1}$ which uses the state $\DD$ to achieve $(d_1,d_2)=(\frac{1}{2},\frac{1}{2})$.   This scheme was presented in \cite{kobayashi_delayed_csit}. The scheme was used to transmit $2$ information symbols to each receiver in $4$ time slots. At $t=1$, the transmitter sends artificial noise symbols using both antennas. The received signals act as keys $K_1$ and $K_2$ for the respective users $1$ and $2$. Since there is delayed CSIT, the transmitter can reconstruct these keys and use them in the next slots. At $t=2$, the transmitter sends the two information symbols $(u_1,u_2)$ intended for the first receiver linearly combined with the first user's key. Thus, the first user can retrieve a linear combination of just its intended symbols. However, the second user gets a linear combination $L(u_1,u_2,K_1)$. Due to delayed $\CSIT$ however, the transmitter can reconstruct $L$. In the third slot, the roles of the receivers are reversed and the transmitter sends the second user's symbols $(v_1,v_2)$   linearly combined with the second user's key $K_2$. This allows the second user to retrieve a linear combination of just its information symbol, which however remain secure at the first user, which receives $G(v_1,v_2,K_2)$. In the fourth slot, the transmitter sends a linear combination of $L$ and $G$. Essentially this provides the first user with $L$, from which it can eliminate $K_1$ to get another independent linear combination of $(u_1,u_2)$. A similar situation takes place at the second user. Finally, each user has two linearly independent combinations of two symbols and thus can decode the information symbols intended for it. The information leakage is only $o(\log P)$, as shown in \cite{kobayashi_delayed_csit}.

\subsubsection{Scheme $S_{2}^{1}$}

In this sub-section, we present the scheme $S_2^{1}$ which uses the states $(\DD,\NN)$ with fractions $(\frac{1}{2},\frac{1}{2})$ to achieve $(d_1,d_2) = (\frac{1}{2},\frac{1}{2})$.

\begin{figure}
\centering
\includegraphics[width=0.85\linewidth]{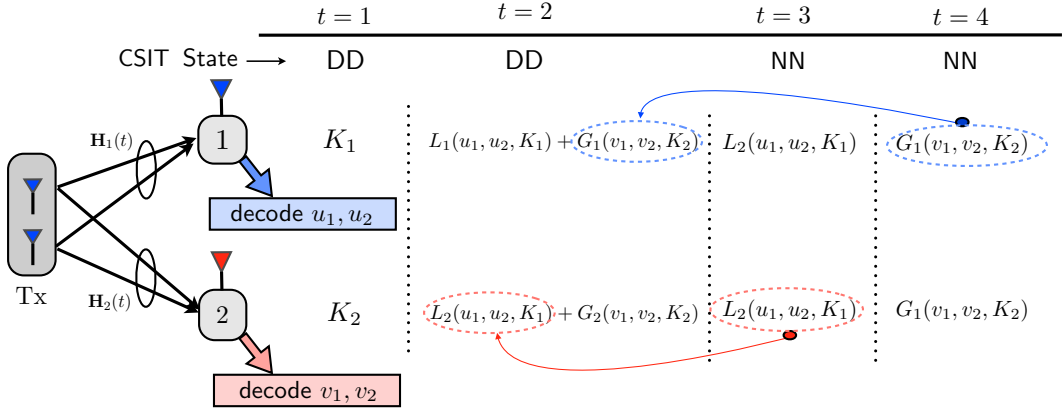}
\caption{Achieving sum s.d.o.f.~of $1$ using $S_2^{1}$.}\label{fig:dd_nn_1} 
\end{figure}

The scheme  $S_1^{1}$ requires delayed $\CSIT$ from at least one user for the first $3$ time slots. We need to modify this scheme to ensure that delayed $\CSIT$ is required only for $2$ of the $4$ time slots. Fig.~\ref{fig:dd_nn_1} shows the new scheme. It is as follows:

1) At time $t=1$, $S(1)=\DD$: The strategy in this slot is the same as in the scheme $S_1^1$. In this slot, the transmitter sends artificial noise symbols to create keys that can be used in later slots. The channel input is 
\begin{align}
\mathbf{X}(1) = \left[q_1 \quad q_2\right]^T,
\end{align}
where $q_1$ and $q_2$ are i.i.d.~as $\mathcal{CN}(0,P)$. The received signals are:
\begin{align}
Y(1) =& h_{11}(1)q_1 + h_{12}(1)q_2 \stackrel{\Delta}{=} K_1\\
Z(1) =& h_{21}(1)q_1 + h_{22}(1)q_2 \stackrel{\Delta}{=} K_2.
\end{align}
Due to delayed $\CSIT$, the transmitter learns $K_1$ and $K_2$ and uses them in the next time slots.   

2) At time $t=2$, $S(2)=\DD$: Instead of sending only the first user's symbols as in scheme $S_1^1$, the transmitter now sends linear combination of both users' symbols. It sends:
\begin{align}
\mathbf{X}(2) = \left[u_1+v_1+K_1 \quad u_2+v_2+K_2\right]^T.
\end{align}  
The received signals are:
\begin{align}
Y(2) =& h_{11}(u_1+v_1+K_1) + h_{12}(u_2+v_2+K_2)\\
\stackrel{\Delta}{=}& L_1(u_1,u_2,K_1) + G_1(v_1,v_2,K_2)\\
Z(2) =& h_{21}(u_1+v_1+K_1) + h_{22}(u_2+v_2+K_2)\\
\stackrel{\Delta}{=}& L_2(u_1,u_2,K_1) + G_2(v_1,v_2,K_2).
\end{align}
We notice that if $L_2$ and $G_1$ could be provided to both users, each user can get $2$ linear combinations of the symbols intended for it and hence decode both symbols. Hence, in the remaining two slots, we will transmit $L_2$ and $G_1$ to both users and this will not require any $\CSIT$ from any user.

3) At time $t=3$, $S(3)=\NN$: The transmitter does not have any $\CSIT$. It sends:
\begin{align}
\mathbf{X}(3) = \left[L_2(u_1,u_2,K_1)\quad 0\right]^T.
\end{align} 
The received signals are:
\begin{align}
Y(3) =& h_{11}(3)L_2(u_1,u_2,K_1)\\
Z(3) =& h_{21}(3)L_2(u_1,u_2,K_1).
\end{align}

4) At time $t=4$, $S(4)=\NN$: The transmitter sends:
\begin{align}
\mathbf{X}(4) = \left[G_1(v_1,v_2,K_2)\quad 0\right]^T.
\end{align}  
The received signals are:
\begin{align}
Y(4)=& h_{11}(4)G_1(v_1,v_2,K_2)\\
Z(4) =& h_{21}(4)G_1(v_1,v_2,K_2).
\end{align}

Thus, at the end of $4$ slots the received signals may be summarized as:
 \begin{align}
\mathbf{Y}= \left[ \begin{array}{c}
 K_1  \\
L_1(u_1,u_2,K_1) + G_1(v_1,v_2,K_2)\\
L_2(u_1,u_2,K_1)\\
G_1(v_1,v_2,K_2) \end{array} \right],\quad
\mathbf{Z}= \left[ \begin{array}{c}K_2\\
L_2(u_1,u_2,K_1) + G_2(v_1,v_2,K_2)\\
L_2(u_1,u_2,K_1)\\
G_1(v_1,v_2,K_2)
\end{array} \right].\nonumber
 \end{align}
Clearly, user $1$ can decode $(u_1,u_2)$ and user $2$ can get $(v_1,v_2)$. The information leakage is at most $o(\log P)$ as we show below.

\emph{Security guarantees}:

For the first user's symbols $\ubar{u}=(u_1,u_2)$, we have,
\begin{align}
I(\ubar{u};\mathbf{Z}|\mathbf{H})
\leq& I(\ubar{u};L_2(\ubar{u},K_1)|\mathbf{H})\label{eq:user1_ddnn_1} \\
=& h(L_2(\ubar{u},K_1)|\mathbf{H})-h(L_2(\ubar{u},K_1)|\ubar{u},\mathbf{H})\\
\leq& \log P - h(K_1|\ubar{u},\mathbf{H}) + o(\log P)\\
=& \log P - h(K_1|\mathbf{H}) + o(\log P)\\
=& \log P -\log P + o(\log P)\\
=& o(\log P),
\end{align}
where \eqref{eq:user1_ddnn_1} follows from the Markov chain $U \rightarrow L_2(\ubar{u},K_1) \rightarrow \mathbf{Z}$.

For the second user's symbols $\ubar{v} = (v_1,v_2)$, the information leakage at the first receiver is:
\begin{align}
I(\ubar{v};\mathbf{Y}|\mathbf{H})
\leq& I(\ubar{v};G_1(\ubar{v},K_2)|\mathbf{H})\label{eq:user2_ddnn_1}\\
=& h(G_1(\ubar{v},K_2)|\mathbf{H})-h(G_1(\ubar{v},K_2)|\ubar{v},\mathbf{H})\\
\leq& \log P - h(K_2|\ubar{v},\mathbf{H}) + o(\log P)\\
=& \log P - h(K_2|\mathbf{H}) + o(\log P)\\
=& \log P -\log P + o(\log P)\\
=& o(\log P),
\end{align}
where \eqref{eq:user2_ddnn_1} follows from the Markov chain $\ubar{v}\rightarrow G_1(\ubar{v},K_2)\rightarrow \mathbf{Y}$.

\subsubsection{Scheme $S_{3}^{1}$}

We next present a novel scheme $S_{3}^{1}$ which uses the states $(\DN,\ND)$ with fractions $(\frac{1}{2},\frac{1}{2})$ to achieve $(d_1,d_2)=(\frac{1}{2},\frac{1}{2})$. In particular, we present a scheme which achieves the s.d.o.f.~pair $(d_1, d_2)= \left(\frac{2n}{4n+1}, \frac{2n}{4n+1}\right)$ as a function of the block length $n$. Taking the limit $n\rightarrow \infty$ yields the s.d.o.f.~pair $\left(\frac{1}{2}, \frac{1}{2}\right)$.

\begin{figure}[t]
\centering
\includegraphics[width=0.7\linewidth]{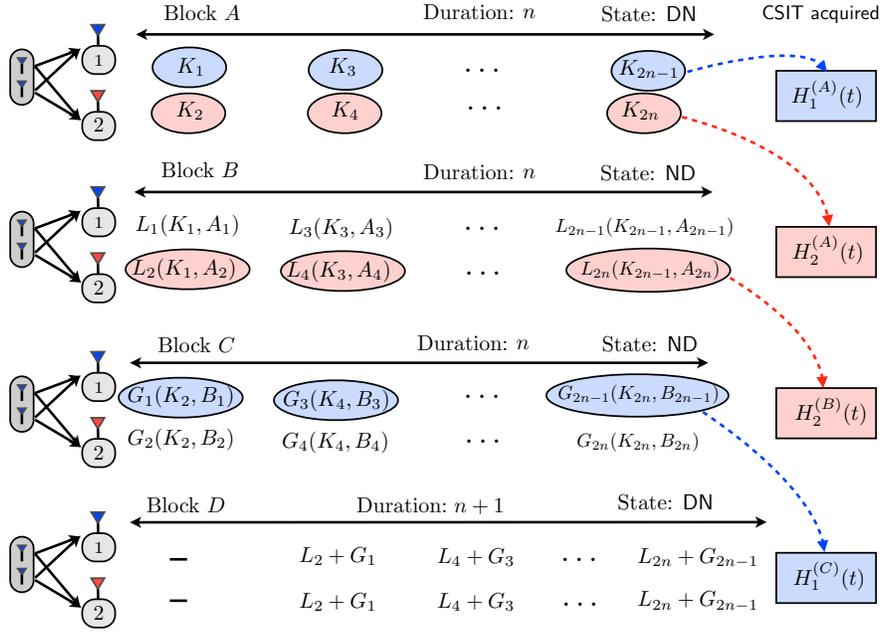}
\caption{Achieving sum s.d.o.f.~of $4n/(4n+1)$ using  scheme $S_3^1$.}
\label{fig:dn_nd_1}
\end{figure}

The scheme is shown in Fig.~\ref{fig:dn_nd_1}. Unlike all the other schemes in this paper where the optimal sum s.d.o.f.~can be achieved within a finite number of time slots, this scheme cannot achieve sum s.d.o.f.~of $1$ in a finite number of slots. Indeed, there does not exist a scheme that can achieve sum s.d.o.f.~of $1$ in finitely many slots. To see why, assume that there exists such a scheme with $n$ slots. In this scheme, states $\DN$ and $\ND$ occur for equal fractions of time; thus, $\lambda_D=\lambda_{N} =\frac{1}{2}$. Now, note that the delayed $\CSIT$ in the last slot cannot be used; thus, the scheme would work equally well if the last slot were $\NN$ instead of $\DN$ or $\ND$. However, changing the state in the last slot to $\NN$ would imply $\lambda_D < \frac{1}{2}$, which in turn implies that $d_1+d_2 < 1$ from \eqref{eq:sum_rate}. Thus, no scheme that uses only a finite number of slots can achieve a sum s.d.o.f.~of $1$. 

Here we provide an asymptotic scheme that achieves a sum s.d.o.f.~of $\frac{4n}{4n+1}$ in $n$ slots. As the number of slots $n\rightarrow\infty$, the sum s.d.o.f.~approaches $1$. We wish to send $2n$ symbols to each receiver in $4n+1$ time slots. The scheme involves transmission in $4$ blocks where the first $3$ blocks, say $A$, $B$ and $C$ each have $n$ time slots, while the last block $D$ has $n+1$ slots; thus, a total of $4n+1$ time slots are required in the scheme. The scheme is as follows:

1) In block $A$, $S(t)= \DN$: In each time slot $i$ in block $A$, the transmitter generates two artificial noise symbols and sends them using its two antennas. The receivers receive different linear combinations of the two artificial noise symbols $K_{2i-1}$ and $K_{2i}$ as shown in Fig.~\ref{fig:dn_nd_1}. Due to delayed $\CSIT$ from the first user, the transmitter can reconstruct each of $K_{2i-1}, i=1,\ldots$, by the end of block $A$. Thus, they can act as shared keys between the transmitter and the first receiver. However, since the second receiver does not feedback any $\CSIT$ (due to the fact that the state in the block is $\DN$), the transmitter cannot reconstruct the observations of the second receiver at the end of block $A$.   

2) In block $B$, $S(t) = \ND$: At the beginning of this slot, the transmitter has the keys $K_{2i-1}, i=1,\ldots,n$ shared with the first user. It uses these keys to send information intended for the first user. It creates $2n$ linearly independent combinations of the $2n$ symbols intended for the first receiver: $a_{1},\ldots,a_{2n}$. In slot $i$, it transmits 
\begin{align}
\mathbf{X}^{B}(i) = \left[a_{2i-1}+K_{2i-1}\quad a_{2i}+K_{2i-1}\right]^T.
\end{align}  
The first and second receivers receive linearly independent combinations $L_{2i-1}(A_{2i-1},K_{2i-1})$ and $L_{2i}(A_{2i},K_{2i-1})$ in slot $i$, where $A_{i}$ denotes the $i$th linear combination of the first user's symbols, as shown in Fig.~\ref{fig:dn_nd_1}. Since the state is $\ND$, the second user provides delayed $\CSIT$ to the transmitter. In the $i$th slot, the second user feeds back $\mathbf{H}_2^{A}(i)$, that is, the channel coefficients of the second user in slot $i$ within block $A$. \emph{Note that this is unlike any other achievable scheme we have encountered so far; in all other schemes, the receiver feeds back the channel coefficients of the current slot which appears as delayed $\CSIT$ at the beginning of the next slot}. Thus, at the end of slot $B$, the transmitter has all the channel coefficients of the second user from block $A$; thus, it can reconstruct the outputs of the second receiver in block $A$, $K_{2i}, i=1,\ldots,n$, which now act as shared keys between the transmitter and the second receiver.

3) In block $C$, $S(t) = \ND$: At the beginning of this slot, the transmitter has the keys $K_{2i}, i=1,\ldots, n$ shared with the second user. It uses these keys to send information securely to the second user. It creates $2n$ linearly independent combinations of the $2n$ symbols intended for the second receiver: $b_{1},\ldots,a_{2n}$. In slot $i$, it transmits 
\begin{align}
\mathbf{X}^{C}(i) = \left[b_{2i-1}+K_{2i-1}\quad b_{2i}+K_{2i-1}\right]^T.
\end{align}   
The first and second receivers receive linearly independent combinations $G_{2i-1}(B_{2i-1},K_{2i})$ and $G_{2i}(B_{2i},K_{2i})$ in slot $i$, where $B_{i}$ denotes the $i$th linear combination of the second user's symbols, as shown in Fig.~\ref{fig:dn_nd_1}. As $\CSIT$, in the $i$th slot, the second user feeds back the channel coefficients $\mathbf{H}^B_2(i)$, which allows the transmitter to reconstruct $L_{2i}(A_{2i},K_{2i-1})$. Note that now if $L_{2i}(A_{2i},K_{2i-1})$ and $G_{2i-1}(B_{2i-1},K_{2i})$ could be exchanged, each of the receivers would receive $2n$ linear combinations of the $2n$ symbols intended for it, thus, allowing both receivers to decode their own messages. However, $G_{2i-1}(B_{2i-1},K_{2i})$ is not known to the transmitter yet, since the first user has not fed back its channel in block $C$. This $\CSIT$ will be obtained in the next block.  

4) In block $D$, $S(t) = \ND$: The transmitter wishes to send the symbols $L_{2i}(A_{2i},K_{2i-1})+G_{2i-1}(B_{2i-1},K_{2i}), i=1,\ldots,n$, in this block. To do so, the transmitter does not transmit anything in the first slot in this block. It only acquires the channel coefficients $\mathbf{H}_1^C(i)$ from the first user who is supplying delayed $\CSIT$ in this block. In the $i$th slot, $i=1,\ldots,n$, the transmitter acquires the channel coefficients $\mathbf{H}_1^C(i)$ and transmits:
\begin{align}
\mathbf{X}^D (i) = \left[L_{2i-2}(A_{2i-2},K_{2i-3})+G_{2i-3}(B_{2i-3},K_{2i-2}) \quad 0\right]^T,\quad i= 2,\ldots,n+1.
\end{align}
The first user can now obtain $L_{2i-1}(A_{2i-1},K_{2i-1})$ and $L_{2i}(A_{2i},K_{2i-1})$ for every $i=1,\ldots,n$, while the second user obtains $G_{2i-1}(B_{2i-1},K_{2i})$ and $G_{2i}(B_{2i},K_{2i})$ for $i=1,\ldots,n$. Now by eliminating the respective keys, each user can decode the $2n$ symbols intended for it from the $2n$ linearly independent combinations available to it. Also the keys ensure the confidentiality, and the information leakage is only $o(\log P)$, as we show next.

\emph{Security guarantees}:

Let $\ubar{u}=(a_1,\ldots,a_{2n})$ and $\ubar{v}=(b_1,\ldots,b_{2n})$ be the symbols intended for users 1 and 2, respectively. The leakage of $\ubar{u}$ at user 2 is given by
\begin{align}
I(\ubar{u};\mathbf{Z}|\mathbf{H}) 
\leq& I(\ubar{u};\left\lbrace L_{2i}(A_{2i}, K_{2i-1})\right\rbrace_{i=1}^n |\mathbf{H})\label{eq:user1_dnnd}\\
=& h(\left\lbrace L_{2i}(A_{2i}, K_{2i-1})\right\rbrace_{i=1}^n|\mathbf{H}) - h(\left\lbrace L_{2i}(A_{2i}, K_{2i-1})\right\rbrace_{i=1}^n|\ubar{u},\mathbf{H}) \\
\leq& n\log P - h(\left\lbrace K_{2i-1}\right\rbrace_{i=1}^n|\mathbf{H}) + o(\log P)\\
=& n\log P -n\log P +o(\log P)\label{eq:indepedence_1}\\
=& o(\log P),
\end{align}
where \eqref{eq:user1_dnnd} follows due to the Markov chain $\ubar{u} \rightarrow \left\lbrace L_{2i}(A_{2i}, K_{2i-1})\right\rbrace_{i=1}^n \rightarrow \mathbf{Z}$, and \eqref{eq:indepedence_1} follows from the fact that  $\left\lbrace K_{2i-1} \right\rbrace_{i=1}^n $ are mutually independent and each is distributed as $\mathcal{N}(0,P)$.

Similarly, for the second user's symbols, the leakage at the first user is given by,
\begin{align}
I(\ubar{v};\mathbf{Y}|\mathbf{H}) 
\leq& I(\ubar{v};\left\lbrace G_{2i-1}(B_{2i-1}, K_{2i})\right\rbrace_{i=1}^n |\mathbf{H})\label{eq:user2_dnnd}\\
=& h(\left\lbrace G_{2i-1}(B_{2i-1}, K_{2i})\right\rbrace_{i=1}^n|\mathbf{H})- h(\left\lbrace G_{2i-1}(B_{2i-1}, K_{2i})\right\rbrace_{i=1}^n|\ubar{v},\mathbf{H}) \\
\leq& n\log P - h(\left\lbrace K_{2i}\right\rbrace_{i=1}^n|\mathbf{H}) + o(\log P)\\
=& n\log P -n\log P +o(\log P)\label{eq:indepedence_2}\\
=& o(\log P),
\end{align}
where \eqref{eq:user2_dnnd} follows due to the Markov chain $\ubar{v} \rightarrow \left\lbrace G_{2i-1}(B_{2i-1}, K_{2i})\right\rbrace_{i=1}^n \rightarrow \mathbf{Y}$, and \eqref{eq:indepedence_2} follows from the fact that  $\left\lbrace K_{2i} \right\rbrace_{i=1}^n $ are mutually independent and each is distributed as $\mathcal{N}(0,P)$.

\subsection{Schemes Achieving Sum s.d.o.f.~of 2/3}

\subsubsection{Scheme $S_1^{2/3}$}

The scheme $S_1^{2/3}$ uses the state $\DD$ to achieve $(d_1,d_2)=(\frac{2}{3},0)$. Such a scheme was presented in \cite{kobayashi_delayed_csit}. The scheme can be summarized as follows. At time $t=1$, the transmitter sends two artificial noise symbols using its two antennas. Each user receives a different linear combination of the noise symbols and they act as keys. Let $K_1$ and $K_2$ be the keys at receivers $1$ and $2$, respectively. Due to delayed $\CSIT$, the transmitter can reconstruct $K_1$. At time $t=2$, the transmitter sends the two symbols intended for the first receiver $(u_1,u_2)$, linearly combined with $K_1$. Receiver $1$ can remove $K_1$ from its received signal and get one linear combination of $(u_1,u_2)$ at the end of this slot. The second user receives a linear combination of $u_1, u_2$ and $K_1$, say $L(u_1,u_2,K_1)$; however, not knowing $K_1$, it cannot decode the information symbols. Due to delayed $\CSIT$, the transmitter learns $L$ and transmits it in $t=3$. The second receiver gets no new information but the first receiver can get a second linear combination of $(u_1,u_2)$ by eliminating $K_1$ from  $L$. This allows receiver $1$ to decode $(u_1,u_2)$, while the information leakage to receiver $2$ is $o(\log P)$.

\subsubsection{Scheme $S_2^{2/3}$}

The scheme $S_2^{2/3}$ uses the states $(\DD,\NN)$ with fractions $(\frac{2}{3},\frac{1}{3})$ to achieve $(d_1,d_2)=(\frac{2}{3},0)$. We note that in scheme $S_1^{2/3}$, the delayed $\CSIT$ in slot $t=3$ is not required. Thus, the scheme can work with the states  $(\DD,\NN)$ with fractions $(\frac{2}{3},\frac{1}{3})$, and we call this $S_2^{2/3}$.

\subsubsection{Scheme $S_3^{2/3}$}

Finally, the scheme $S_3^{2/3}$ uses the states $(\DN,\ND,\NN)$ with fractions $(\frac{1}{3},\frac{1}{3},\frac{1}{3})$ to achieve $(d_1,d_2)=(\frac{2}{3},0)$. We notice that instead of having $\DD$ state in the first two slots, it suffices to have $\DN$ in the first slot (since the transmitter does not need $K_2$) and $\ND$ in the second slot (since the transmitter only needs to reconstruct the second user's received signal $L$). Thus, it suffices to have the states $(\DN,\ND,\NN)$ with fractions $(\frac{1}{3},\frac{1}{3},\frac{1}{3})$ for the scheme to work, and we call this $S_3^{2/3}$.  

\section{Achievability}\label{achievability}
Now that we have all the required constituent schemes summarized in Table~\ref{TableCSsummary}, we proceed to show how these schemes can be combined to achieve the region stated in Theorem 1. We restate the region of Theorem 1 here for convenience:
\begin{align}
d_{1}&\leq \min\left(\frac{2+2\lambda_P-\lambda_{PP}}{3},1-\lambda_{NN}\right)\label{eq:single_user_rate1-A}\\
d_{2}&\leq \min \left(\frac{2+2\lambda_P-\lambda_{PP}}{3},1-\lambda_{NN}\right)\label{eq:single_user_rate2-A} \\
3d_1 + d_2 &\leq 2 + 2\lambda_{P}\label{eq:3d1+d2_bound1-A}\\
d_1 + 3d_2 &\leq 2 + 2\lambda_{P} \label{eq:3d1+d2_bound2-A}\\
d_{1}+ d_{2}&\leq 2(\lambda_{P}+ \lambda_{D}). \label{eq:sum_rate-A}
\end{align}

We classify this region into two cases:
\begin{itemize}
\item Case $A$: in which $d_{1}+d_{2}$ bound of \eqref{eq:sum_rate-A} is inactive. This corresponds to the condition 
\begin{align}
1+\lambda_{P}\leq 2\lambda_{P} + 2\lambda_{D},
\end{align}
which is equivalent to 
\begin{align}
\lambda_N &\leq \lambda_{D}.
\end{align}
\item Case $B$: in which $d_{1}+d_{2}$ bound of \eqref{eq:sum_rate-A} is active which corresponds to 
\begin{align}
\lambda_N &> \lambda_{D}.
\end{align}
\end{itemize}
In the next two sub-sections, we present the achievability for each of these cases separately.

\subsection{Achievability for Case $A$: $\lambda_{D}\geq \lambda_{N}$}
For Case $A$, the s.d.o.f.~region reduces to:
\begin{align}
d_{1}&\leq  \min \left(\frac{2+2\lambda_P-\lambda_{PP}}{3}, 1-\lambda_{NN}\right)\\
d_{2}&\leq  \min \left(\frac{2+2\lambda_P-\lambda_{PP}}{3},1-\lambda_{NN}\right)\\
3d_1 + d_2 &\leq 2 + 2\lambda_{P}\\
d_1 + 3d_2 &\leq 2 + 2\lambda_{P}.
\end{align}

Depending on which single user bound is active, we consider two cases:
\begin{enumerate}
\item $\frac{2+2\lambda_P-\lambda_{PP}}{3} \leq 1-\lambda_{NN}$, which is equivalent to the condition $\lambda_{DD}+2\lambda_{DN}\geq 2\lambda_{NN}$,
\item  $\frac{2+2\lambda_P-\lambda_{PP}}{3} \geq 1-\lambda_{NN}$, which is equivalent to the condition  $\lambda_{DD}+2\lambda_{DN}\leq 2\lambda_{NN}$.
\end{enumerate} 
As shown in Fig.~\ref{fig:case1}, due to symmetry, it suffices to achieve the points $P_1$ and $P_2$ in each case. 

\subsubsection{Achievability of Point $P_1$}
We first show the achievability of the point $P_1$ in both cases. To do so, let us consider the two cases one by one: 
\begin{enumerate}
\item $\lambda_{DD}+2\lambda_{DN}\geq 2\lambda_{NN}$: In this case, the single user bounds are:
\begin{align}
d_1 \leq& \frac{2+2\lambda_P-\lambda_{PP}}{3}\\
d_2 \leq& \frac{2+2\lambda_P-\lambda_{PP}}{3}.
\end{align}
\begin{figure}
\centering
\begin{subfigure}{0.45\textwidth}
	\includegraphics[height=135 pt]{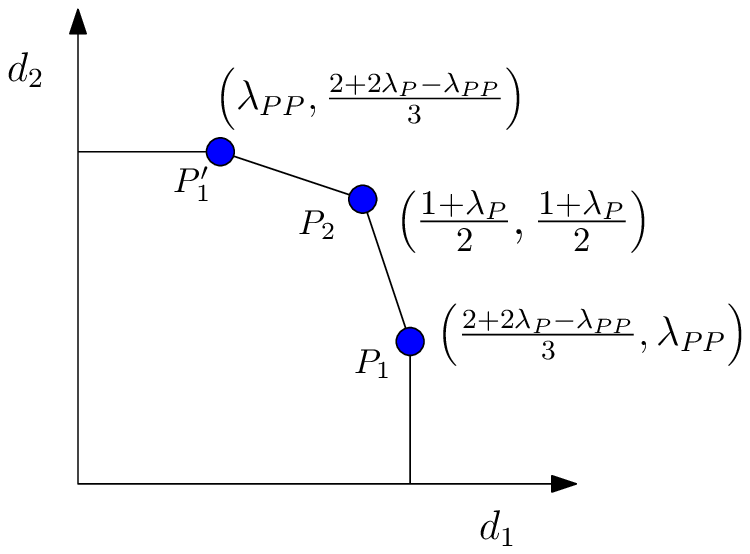}
	\caption{$\lambda_{DD} + 2\lambda_{DN}\geq 2\lambda_{NN}$.}
	\label{fig:case1A}
\end{subfigure}%
\begin{subfigure}{0.55\textwidth}
	\includegraphics[height=135 pt]{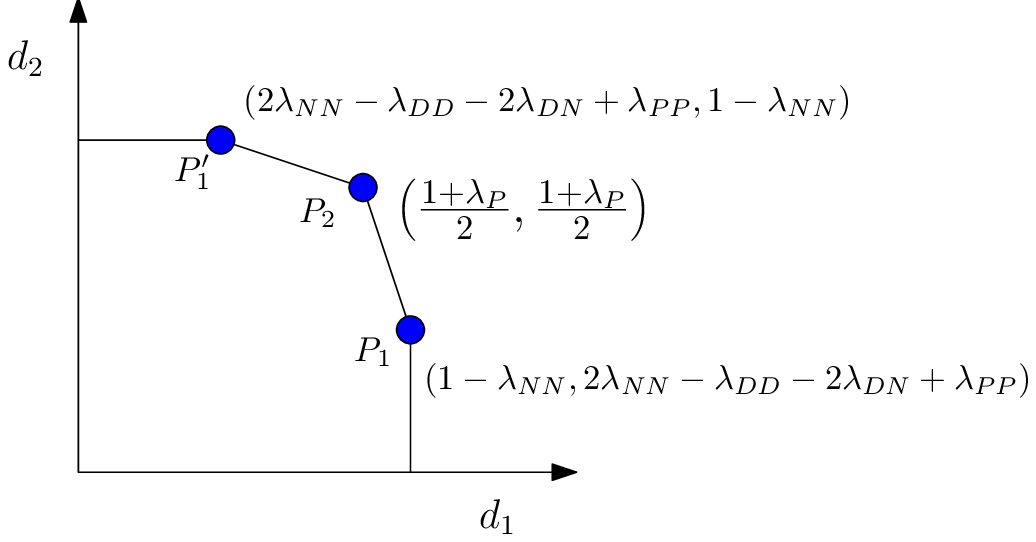}
	\caption{$\lambda_{DD}+2\lambda_{DN}\leq 2\lambda_{NN}$.}
	\label{fig:case1B}
\end{subfigure}
\caption{s.d.o.f.~regions in case $A$.}
\label{fig:case1}
\end{figure}

As seen in Fig.~\ref{fig:case1A}, the point $P_1$ is $\left(  \frac{2+ 2\lambda_{\mathsf{P}}- \lambda_{\PP}}{3},\lambda_{\PP}\right)$. To achieve this point, using the state $\PP$, we achieve $(1,1)$, with $\PD, \DP, \PN, \NP$, we achieve the pair $(1, 0)$ either through zero-forcing, or by transmitting artificial noise in a direction orthogonal to the first user's channel. For the states $(\DD, \NN)\sim (\frac{2}{3}, \frac{1}{3})$,  and $(\DN,\ND,\NN)\sim(\frac{1}{3},\frac{1}{3},\frac{1}{3})$, we achieve the pair $(\frac{2}{3}, 0)$ by using the schemes $S_2^{2/3}$ and $S_3^{2/3}$, respectively. Essentially, the $\NN$ state can be fully alternated with the $\DD$ state  and the $\DN$ and $\ND$ states to achieve $\frac{2}{3}$ s.d.o.f.~for user $1$.

Time sharing yields the following s.d.o.f.~pair:
\begin{align}
d_{2}&= \lambda_{PP}\\
d_{1}&= \lambda_{PP} + 2\lambda_{PD}+ 2\lambda_{PN}+  \underbrace{\frac{2}{3}}_{S_2^{2/3}}(\lambda_{DD}+2\lambda_{DN}+\lambda_{NN})\\
&=  2\lambda_{P} - \lambda_{PP} + \frac{2}{3}(\lambda_{DD}+ \lambda_{NN})\\
&= 2\lambda_{P}- \lambda_{PP} + \frac{2}{3}(1-2\lambda_{P}+\lambda_{PP})\\
&= \frac{2+ 2\lambda_{P}- \lambda_{PP}}{3}.
\end{align}

\item $\lambda_{DD} + 2\lambda_{DN}\leq 2\lambda_{NN}$:
In this case the single user bounds are:
\begin{align}
d_1 \leq& 1-\lambda_{NN}\\
d_2 \leq& 1-\lambda_{NN}.
\end{align}
Again, we wish to achieve the point $P_1$ in Fig.~\ref{fig:case1B}. The point $P_1$ is given by:
\begin{align}
P_1:(d_1,d_2) = (1-\lambda_{NN},\lambda_{PP} +  (2\lambda_{NN}-2\lambda_{DN}-\lambda_{DD}) ).
\end{align} 
Here we consider two further subcases
\begin{itemize}
\item $\lambda_{NN} \leq \lambda_{DD} + \lambda_{DN}$: In this case, to achieve the point $P_1$, we first use up the full $\DN$ and $\ND$ states with a part of the $\NN$ state using scheme $S_3^{2/3}$. We alternate the remaining $(\lambda_{NN}-\lambda_{DN})$ duration of $\NN$ state with the $\DD$ state using two schemes: $S_2^{2/3}$ and $S_2^1$. Note that in this subcase, $0\leq 2(\lambda_{DD} + \lambda_{DN}-\lambda_{NN})\leq \lambda_{DD}$. We use the state $\DD$  for duration $2(\lambda_{DD}+\lambda_{DN}-\lambda_{NN})$ and state $\NN$ for duration $(\lambda_{DD}+\lambda_{DN}-\lambda_{NN})$  together using scheme $S_2^{2/3}$ to achieve the s.d.o.f.~pair $\left(\frac{2}{3},0\right)$. The remaining $(2\lambda_{NN}-2\lambda_{DN}-\lambda_{DD})$ duration of the state $\NN$ is alternated with the remaining $(2\lambda_{NN}-2\lambda_{DN}-\lambda_{DD})$ duration of state $\DD$ using the scheme $S_{2}^{1}$ to achieve the s.d.o.f.~pair $\left(\frac{1}{2},\frac{1}{2}\right)$. The state $\PP$ allows us to achieve the s.d.o.f.~pair $(1,1)$ while the remaining states $\PD$, $\DP$, $\PN$, and $\NP$ each achieves $(1,0)$. Thus, by using time sharing, the s.d.o.f.~pair is:
\begin{align}
d_1 =& \lambda_{PP}+ 1\times 2\lambda_{PD} + 1\times 2\lambda_{PN} + \underbrace{\frac{2}{3}}_{S_3^{2/3}}\times 3\lambda_{DN} +  \underbrace{\frac{2}{3}}_{S_2^{2/3}}\times 3(\lambda_{DD}+\lambda_{DN}-\lambda_{NN})\nonumber\\&+ \underbrace{\frac{1}{2}}_{S_2^1}\times 2(2\lambda_{NN}-2\lambda_{DN}-\lambda_{DD})\\
=& 1-\lambda_{NN}\\
d_2 =&\lambda_{PP}+ \underbrace{\frac{1}{2}}_{S_2^1}\times 2(2\lambda_{NN}-2\lambda_{DN}-\lambda_{DD})\nonumber\\
=&\lambda_{PP} +  (2\lambda_{NN}-2\lambda_{DN}-\lambda_{DD}),
\end{align}
which is precisely the point $P_1$.
\item $\lambda_{NN} \geq \lambda_{DD} +\lambda_{DN}$:
In this case, the state $\NN$ cannot be completely used with the states $\DD$, $\DN$ and $\ND$. But we note that $\lambda_{D}\geq \lambda_{N}$ implies that $\lambda_{D}\geq \lambda_{NN}$. We first use up the $\DN$ and $\ND$ states by alternating with the $\NN$ state using scheme $S_3^{2/3}$. A portion $\lambda_{DD}$ of the remaining $(\lambda_{NN}-\lambda_{DN})$ duration of the $\NN$ state uses up the $\DD$ state in scheme $S_2^1$ achieving the pair $\left(\frac{1}{2},\frac{1}{2}\right)$. The remaining $(\lambda_{NN}-\lambda_{DN}-\lambda_{DD})$ portion of the $\NN$ state is used with the $\PD$ and $\DP$ states through the scheme $S_1^{4/3}$ to achieve the pair $\left(\frac{2}{3},\frac{2}{3}\right)$. For the remainder of the state $\PD$, $\DP$ and the states $\PN$, $\NP$, we can achieve the pair $(1,0)$, while $(1,1)$ is achieved in the $\PP$ state. By time sharing, we get
\begin{align}
d_1 =& \lambda_{PP} + 2\lambda_{PN} + \underbrace{\frac{2}{3}}_{S_3^{2/3}}\times 3\lambda_{DN}+\underbrace{\frac{2}{3}}_{S_1^{4/3}}\times 3(\lambda_{NN}-\lambda_{DN}-\lambda_{DD})\nonumber\\ &+ 2 (\lambda_{PD} - \lambda_{NN}+\lambda_{DN}+\lambda_{DD})+ \underbrace{\frac{1}{2}}_{S_2^1}\times 2\lambda_{DD} \\
=& 1- \lambda_{NN}\\
d_2 =& \lambda_{PP} + \underbrace{\frac{2}{3}}_{S_1^{4/3}}\times 3(\lambda_{NN}-\lambda_{DN}-\lambda_{DD})+\underbrace{\frac{1}{2}}_{S_2^1}\times 2\lambda_{DD}\\
=& \lambda_{PP} + 2\lambda_{NN} - 2\lambda_{DN}-\lambda_{DD},
\end{align}
which is again the point $P_1$.
\end{itemize}
\end{enumerate}

\subsubsection{Achieving the Sum s.d.o.f.~Achieving Point $P_{2}$}
The point $P_{2}$ corresponds to:
\begin{align}
P_{2}: (d_1, d_2)&= \left( \frac{1+\lambda_P}{2}, \frac{1+\lambda_P}{2}\right).
\end{align}
We rewrite the condition $\lambda_{D}\geq \lambda_{N}$ corresponding to case $A$ as:
\begin{align}
\lambda_{PD} + \lambda_{DD}&\geq \lambda_{PN}+ \lambda_{NN}. 
\end{align}
From this condition it is not immediately clear how the constituent schemes should be jointly utilized. Hence we break this condition into three mutually exclusive cases:
\begin{enumerate}
\item Sub-case $A1$: $\lambda_{PD}\geq \lambda_{PN}$ and $\lambda_{DD}\geq \lambda_{NN}$,
\item Sub-case $A2$: $\lambda_{PD}\geq \lambda_{PN}$ and $\lambda_{DD} \leq \lambda_{NN}$,
\item Sub-case $A3$: $\lambda_{PD}\leq \lambda_{PN}$ and $\lambda_{DD} \geq \lambda_{NN}$.
\end{enumerate}  
Now, we consider these three sub-cases one by one: \\

\noindent \textbf{Sub-case $A1$}: $\lambda_{PD}\geq \lambda_{PN}$ and $\lambda_{DD}\geq \lambda_{NN}$. In this sub-case, the original condition $\lambda_{D}\geq \lambda_{N}$ is automatically satisfied. For this sub-case, it is clear that the states $\PN$ and $\NP$ can be fully alternated along with the $\PD$ and $\DP$ using scheme $S_{2}^{3/2}$ to achieve $\frac{3}{2}$ s.d.o.f. The remaining fraction of time for $\PD$ (and $\DP$) is hence: $\lambda_{PD}- \lambda_{PN}$. The state $\NN$ can be fully utilized along with $\DD$ to achieve $1$ s.d.o.f.~using the scheme $S_2^1$. The $\DN$ and $\ND$ states are alternated with each other to achieve 1 s.d.o.f. Thus, we achieve the following sum s.d.o.f.:
\begin{align}
d_1+d_2&= \underbrace{2}_{S^2}\times\lambda_{PP} + \underbrace{\frac{3}{2}}_{S_{2}^{3/2}}\times (2\lambda_{PD}+ 2\lambda_{PN})  +  \underbrace{1}_{S_{2}^{1}}\times(\lambda_{DD}+ \lambda_{NN}) + 2\lambda_{DN}  \nonumber\\
&= 2\lambda_{PP} + 3\lambda_{PD} + 3\lambda_{PN} + \lambda_{DD}+ \lambda_{NN} + 2\lambda_{DN}\\
& = 1 + \lambda_{P}.
\end{align}

\noindent \textbf{Sub-case $A2$}: $\lambda_{PD}\geq \lambda_{PN}$, $\lambda_{DD} \leq \lambda_{NN}$. As in sub-case $A1$, we can fully alternate the $\PN$ and $\NP$ states with the $\PD$ and $\DP$ states using the scheme $S_2^{3/2}$ to achieve the s.d.o.f.~of $\frac{3}{2}$. Since $\lambda_{DD}\leq \lambda_{NN}$, we instead fully alternate the state $\DD$ along with $\NN$ using scheme $S_2^1$ to achieve a sum s.d.o.f.~of $1$. The remaining fraction of the $\NN$ state is $\lambda_{NN}-\lambda_{DD}$ which can be alternated with the remaining fraction of $(\PD, \DP)$, which is $\lambda_{PD}- \lambda_{PN}$ as long as 
$\lambda_{PD}- \lambda_{PN}\geq \lambda_{NN}- \lambda_{DD}$. This achieves $\frac{4}{3}$ sum s.d.o.f. Indeed, this is feasible as this is precisely the condition $\lambda_{D}\geq \lambda_{N}$. The $\DN$ and $\ND$ states are alternated with each other to achieve 1 s.d.o.f.
\begin{align}
d_1+d_2&= \underbrace{2}_{S^2}\times\lambda_{PP} + \underbrace{\frac{3}{2}}_{S_{2}^{3/2}}\times (4\lambda_{PN})  +  \underbrace{1}_{S_{2}^{1}}\times(2\lambda_{DD}) + 2\lambda_{DN}\nonumber\\
&\quad +   \underbrace{\frac{4}{3}}_{S_{1}^{4/3}}\times(3(\lambda_{NN}-\lambda_{DD})) + \underbrace{\frac{3}{2}}_{S_{1}^{3/2}}\times 2(\lambda_{PD}- \lambda_{PN}- \lambda_{NN}+ \lambda_{DD})\\
&= 2\lambda_{PP} + 6\lambda_{PN}+ 2\lambda_{DD} + 4\lambda_{NN}-4\lambda_{DD} + 3\lambda_{PD}+ 3\lambda_{DD}- 3\lambda_{PN}-3\lambda_{NN}+2\lambda_{DN}\\
&= 2\lambda_{PP}+ 3\lambda_{PD}+ 3\lambda_{PN} + \lambda_{DD}+ \lambda_{NN}+2\lambda_{DN}\\
& = 1 + \lambda_{P}.
\end{align}

\noindent \textbf{Sub-case $A3$}: $\lambda_{PD}\leq \lambda_{PN}$, $\lambda_{DD} \geq \lambda_{NN}$. Unlike the previous two sub-cases, here, we cannot fully alternate 
the $\PN$ and $\NP$ states with the $\PD$ and $\DP$ states. Instead, we fully use up the $\PD$ and $\DP$ states with a part of the $\PN$ and $\NP$ states using scheme $S_2^{3/2}$ to achieve the sum s.d.o.f.~of $\frac{3}{2}$. The remaining duration of $\PN$ (or the $\NP$) state is $\lambda_{PN}- \lambda_{PD}$. Now, we can also fully alternate the $\NN$ state with $\DD$ since $\lambda_{DD}\geq \lambda_{NN}$ using the scheme $S_2^1$ to achieve the sum s.d.o.f.~of $1$; and thus, the remaining fraction of $\DD$ state is $\lambda_{DD}-\lambda_{NN}$. We now alternate the remaining $\PN$ and $\NP$ states with the remaining $\DD$ state using the scheme $S_2^{4/3}$ to achieve the sum s.d.o.f.~of $\frac{4}{3}$. For this to be feasible, we require $\lambda_{DD}- \lambda_{NN} \geq \lambda_{PN}- \lambda_{PD}$ which is again precisely the condition $\lambda_{D}\geq \lambda_{N}$. The remaining $\DD$ state achieves sum s.d.o.f.~of $1$ using scheme $S_1^1$.  The $\DN$ and $\ND$ states are alternated with each other to achieve $1$ s.d.o.f.
\begin{align}
d_1+d_2&= \underbrace{2}_{S^2}\times\lambda_{PP} + \underbrace{\frac{3}{2}}_{S_{2}^{3/2}}\times (4\lambda_{PD})  +  \underbrace{1}_{S_{2}^{1}}\times(2\lambda_{NN}) \nonumber\\
&\quad +   \underbrace{\frac{4}{3}}_{S_{2}^{4/3}}\times(3(\lambda_{PN}-\lambda_{PD})) + \underbrace{1}_{S_{1}^{1}}\times (\lambda_{DD}-\lambda_{NN}-\lambda_{PN}+ \lambda_{PD})+2\lambda_{DN}\\
&= 2\lambda_{PP} + 6\lambda_{PD}+ 2\lambda_{NN} + 4\lambda_{PN}-4\lambda_{PD} + \lambda_{PD}+ \lambda_{DD}- \lambda_{PN}-\lambda_{NN}+2\lambda_{DN}\\
&= 2\lambda_{PP}+ 3\lambda_{PD}+ 3\lambda_{PN} + \lambda_{DD}+ \lambda_{NN}+2\lambda_{DN}\\
& = 1 + \lambda_{P}.
\end{align}

Hence, for Case A, i.e., when $\lambda_{D}\geq \lambda_{N}$, we have the complete characterization of the s.d.o.f.~region.

\subsection{Achievability for Case $B$: $\lambda_{N} > \lambda_{D}$}
In this case, the $3d_1+d_2/d_1+3d_2$ bounds are inactive at the symmetric sum rate point. However, these $3d_1+d_2/d_1+3d_2$ bounds play a role at other points in the region, in particular, when one of the users requires full secure rate, the $3d_1+d_2/d_1+3d_2$ bounds are relevant in some cases. Thus, these bounds are still partially relevant. Based on whether the $3d_1+d_2/d_1+3d_2$ bounds are partially relevant or completely irrelevant, we divide our achievability into two broad cases:
\begin{enumerate}
\item $3d_1+d_2$ bounds are partially relevant, at the point where one user requires full secret rate,
\item $3d_1+d_2$ bounds are completely irrelevant to the region.
\end{enumerate}
Now let us investigate each of these two cases individually.
\subsubsection{When $3d_1+d_2$ Bounds are Partially Relevant}
\begin{figure}
\centering
\begin{subfigure}{0.45\textwidth}
	\includegraphics[height=120 pt]{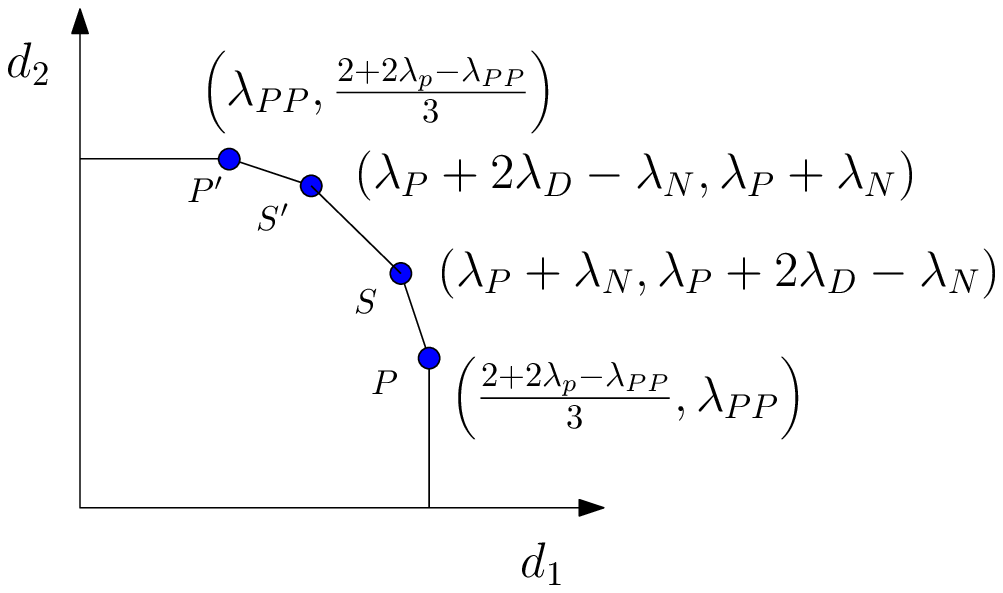}
	\caption{$\lambda_{DD} + 2\lambda_{DN}\geq 2\lambda_{NN}$.}
	\label{fig:case2A}
\end{subfigure}%
\begin{subfigure}{0.55\textwidth}
	\includegraphics[height=120 pt]{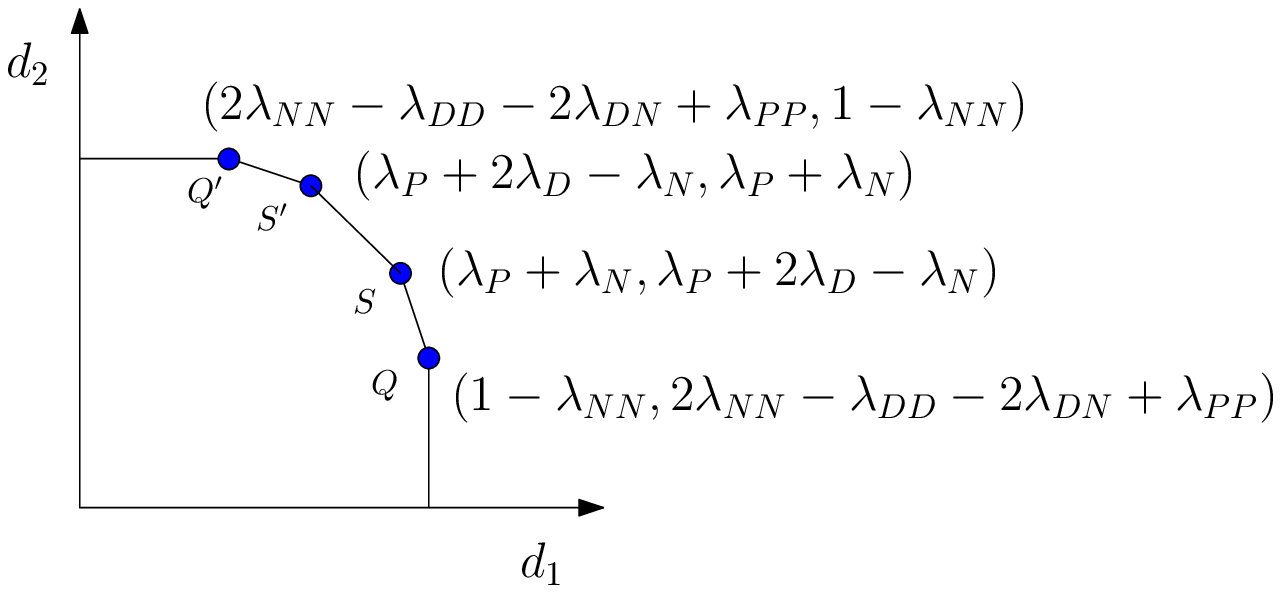}
	\caption{$\lambda_{DD}+2\lambda_{DN}\leq 2\lambda_{NN}$.}
	\label{fig:case2B}
\end{subfigure}
\caption{s.d.o.f.~regions in case $B$ when $3d_1+d_2$ and  $d_1+3d_2$ bounds are partially active.}
\label{fig:case2}
\end{figure}
This case happens when the intersection of the lines defined by the $3d_1+d_2$ bound and the single user bound is inside the region defined by the lines $d_1=0$, $d_2=0$, single user bounds and the $d_1+d_2$ bound. We note that this depends on which of the single user bounds is active, giving rise to two cases, as shown in Fig.~\ref{fig:case2}:
\begin{itemize}
\item $1-\lambda_{NN} \geq \frac{2+2\lambda_P-\lambda_{PP}}{3}$, in which case, the $3d_1+d_2$ bounds are always relevant, since $\lambda_{PP} \leq 2(\lambda_P+\lambda_D) -\frac{2+2\lambda_P-\lambda_{PP}}{3}$. In this case, when one user requires full rate, it suffices to achieve extremal point given by:
\begin{align}
P:(d_1,d_2) =& \left(\frac{2+2\lambda_P -\lambda_{PP}}{3},\lambda_{PP}\right),
\end{align}  
\item $1-\lambda_{NN} \leq \frac{2+2\lambda_P-\lambda_{PP}}{3}$, in which case, the $3d_1+d_2$ bounds are relevant as long as $\lambda_{NN}\leq \lambda_D$. We will need to show the achievability of one of the extremal points when one of the users requires full rate, given by:
\begin{align}
Q:(d_1,d_2) = (1-\lambda_{NN},\lambda_{PP} +  (2\lambda_{NN}-2\lambda_{DN}-\lambda_{DD}) ).
\end{align}
\end{itemize}
However, we note that in both cases, the extremal points that achieve the sum rate are defined by the intersection of the lines $3d_1+d_2=2+2\lambda_P$ and $d_1+d_2 = 2(\lambda_P + \lambda_D)$. These points are symmetric with respect to the line $d_1=d_2$ and it suffices to show the achievability of either one of them. As shown in the figures, it suffices to achieve the point 
\begin{align}
S:(d_1,d_2) = (\lambda_P+\lambda_N,\lambda_P+2\lambda_D-\lambda_N).
\end{align}
Thus, to show the achievability of the full region, we need to show how the points $P$, $Q$ and $S$ are achieved in their relevant cases. We will begin with point $S$ since it remains unaffected by which of the single user bounds is active.

\vspace{5pt}
\noindent\textbf{The sum rate point S:}

Now we are effectively operating under the constraint $\lambda_{NN}\leq \lambda_D \leq \lambda_N$, and wish to achieve the point $(\lambda_P+\lambda_N,\lambda_P+2\lambda_D-\lambda_N)$. From this condition it is not immediately clear how the constituent schemes should be jointly utilized. Hence we focus on the second half of the inequality, which simplifies to $\lambda_{PD} + \lambda_{DD}\leq \lambda_{PN} +\lambda_{NN}$, and break this condition into three mutually exclusive cases:
\begin{itemize}
\item Sub-case $B1$: $\lambda_{PD}\leq \lambda_{PN}$ and $\lambda_{DD}\leq \lambda_{NN}$,
\item Sub-case $B2$: $\lambda_{PD}\geq \lambda_{PN}$ and $\lambda_{DD} \leq \lambda_{NN}$,
\item Sub-case $B3$: $\lambda_{PD}\leq \lambda_{PN}$ and $\lambda_{DD} \geq \lambda_{NN} $.
\end{itemize}
Now let us consider each case one by one:

\noindent \textbf{Sub-case $B1$}: $\lambda_{PD}\leq \lambda_{PN}$ and $\lambda_{DD}\leq \lambda_{NN}$: 
In this case, the full $\DD$ state will be used up with a part of the $\NN$ state using scheme $S_2^1$ to achieve the rate pair $\left(\frac{1}{2},\frac{1}{2}\right)$. The duration of the remaining $\NN$ state is $(\lambda_{NN}-\lambda_{DD})$. Now if $\lambda_{NN}-\lambda_{DD} \leq \lambda_{DN}$, this remaining $\NN$ state can be fully used up with the $\DN$ and $\ND$ states using scheme $S_3^{2/3}$ achieving the pair $(\frac{2}{3},0)$. The remaining $\DN$ and $\ND$ states achieve the pair $(\frac{1}{2},\frac{1}{2})$ using the scheme $S_3^1$. The $\PD$  and $\DP$ states are fully alternated with the $\PN$ and $\NP$ states using scheme $S_2^{3/2}$ to achieve the pair $\left(\frac{3}{4},\frac{3}{4}\right)$.  The remaining $\PN$ and $\NP$ states achieve the pair $(1,0)$. The rate pair achieved then is
\begin{align}
d_1 =& \lambda_{PP} + \underbrace{\frac{3}{4}}_{S_2^{3/2}}\times 4\lambda_{PD} + \underbrace{\frac{1}{2}}_{S_2^1}\times 2\lambda_{DD}+1\times 2(\lambda_{PN}-\lambda_{PD})+ \underbrace{\frac{2}{3}}_{S_3^{2/3}}\times 3(\lambda_{NN}-\lambda_{DD})\nonumber\\&+ \underbrace{\frac{1}{2}}_{S_3^1}\times 2(\lambda_{DN}-\lambda_{NN}+\lambda_{DD})\nonumber\\
=& \lambda_{PP} + \lambda_{PD} + \lambda_{DN} +\lambda_{NN}+2\lambda_{PN}\nonumber\\
=& \lambda_{P}+\lambda_{N}\\
d_2 =&  \lambda_{PP} +\underbrace{\frac{3}{4}}_{S_2^{3/2}}\times 4\lambda_{PD} + \underbrace{\frac{1}{2}}_{S_2^1}\times 2\lambda_{DD}+ \underbrace{\frac{1}{2}}_{S_3^1}\times 2(\lambda_{DN}-\lambda_{NN}+\lambda_{DD})\nonumber\\
=& \lambda_{PP} + 3\lambda_{PD}+2\lambda_{DD} +\lambda_{DN} -\lambda_{NN}\nonumber\\
=& \lambda_P+2\lambda_D-\lambda_N.
\end{align} 

If on the other hand, $\lambda_{NN}-\lambda_{DD} \geq \lambda_{DN}$, the remaining state $\NN$ cannot be fully alternated with the states $\DN$ and $\ND$. However, $\lambda_{NN} \leq \lambda_{DN}+\lambda_{DD}+\lambda_{PD}$ from our original condition. Therefore, the full $\DN$ and $\ND$ states are alternated with a part of the $\NN$ state using scheme $S_3^{2/3}$ achieving the pair $(\frac{2}{3},0)$. The remaining duration of the $\NN$ state is $(\lambda_{NN}-\lambda_{DD}-\lambda_{DN})$, which can be fully alternated with the $\PD$ and $\DP$ states using the scheme $S_1^{4/3}$ achieving the pair $(\frac{2}{3},\frac{2}{3})$. The remaining $\PD$ and $\DP$ states can be alternated with the $\PN$ and $\NP$ states using scheme $S_2^{3/2}$ achieving the point $(\frac{3}{4},\frac{3}{4})$. The rest of the $\PN$ and $\NP$ states achieve the point $(1,0)$. Thus, we have,
\begin{align}
d_1 =& \lambda_{PP} + \underbrace{\frac{1}{2}}_{S_2^1}\times 2\lambda_{DD} + \underbrace{\frac{2}{3}}_{S_3^{2/3}}\times 3\lambda_{DN} + \underbrace{\frac{2}{3}}_{S_1^{4/3}} \times 3(\lambda_{NN}-\lambda_{DN}-\lambda_{DD})\nonumber\\
&+ \underbrace{\frac{3}{4}}_{S_2^{3/2}}\times 4(\lambda_{PD} -(\lambda_{NN}-\lambda_{DN}-\lambda_{DD}))+ 1\times 2(\lambda_{PN} - \lambda_{PD} + (\lambda_{NN}-\lambda_{DN}-\lambda_{DD}))\nonumber\\
=& \lambda_{P} + \lambda_{N}\\
d_2 =& \lambda_{PP} + \underbrace{\frac{1}{2}}_{S_2^1}\times 2\lambda_{DD}  + \underbrace{\frac{2}{3}}_{S_1^{4/3}} \times 3(\lambda_{NN}-\lambda_{DN}-\lambda_{DD})
+\underbrace{\frac{3}{4}}_{S_2^{3/2}}\times 4(\lambda_{PD} -(\lambda_{NN}-\lambda_{DN}-\lambda_{DD})) \nonumber\\
=& \lambda_{P} + 2\lambda_{D}-\lambda_{N}.
\end{align} 

\noindent \textbf{Sub-case $B2$}: $\lambda_{PD}\geq \lambda_{PN}$ and $\lambda_{DD} \leq \lambda_{NN}$: In this case, since $\lambda_{NN} \geq \lambda_{DD}$, the entire $\DD$ state is alternated with a portion of the $\NN$ state using scheme $S_2^1$ to achieve the s.d.o.f.~pair $\left(\frac{1}{2},\frac{1}{2}\right)$. The remaining duration of the $\NN$ state is $\lambda_{NN}-\lambda_{DD}$. Now if $\lambda_{NN}-\lambda_{DD}\leq \lambda_{PD}$, the remaining $\NN$ state is used with a part of the $\PD$ and $\DP$ states in scheme $S_1^{4/3}$ achieving the pair $\left(\frac{2}{3},\frac{2}{3}\right)$. The remaining portion of the $\PD$ and $\DP$ states can then be utilized with the $\PN$ and $\NP$ states using scheme $S_2^{3/2}$ achieving the pair $\left(\frac{3}{4},\frac{3}{4}\right)$. The remaining $\PN$ and $\NP$ states are utilized to just achieve the rate pair $\left(1,0\right)$. The $\DN$ and $\ND$ states are used to achieve the pair $(\frac{1}{2},\frac{1}{2})$ using the scheme $S_3^1$.
Thus, we have,
\begin{align}
d_1 =& \lambda_{PP} + \underbrace{\frac{1}{2}}_{S_2^1} \times \left(2\lambda_{DD}\right) + \underbrace{\frac{2}{3}}_{S_1^{4/3}} \times \left(3(\lambda_{NN}-\lambda_{DD})\right) +\underbrace{\frac{3}{4}}_{S_2^{3/2}}\times \left(4\left(\lambda_{PD} - (\lambda_{NN} -\lambda_{DD})\right)\right) \nonumber\\
&+ 1 \times \left(2\lambda_{PN} - 2\left(\lambda_{PD} - (\lambda_{NN} -\lambda_{DD})\right)\right)+\frac{1}{2}\times 2\lambda_{DN}\\
=& \lambda_P + \lambda_N\\
d_2 =& \lambda_{PP} + \underbrace{\frac{1}{2}}_{S_2^1} \times \left(2\lambda_{DD}\right) + \underbrace{\frac{2}{3}}_{S_1^{4/3}} \times \left(3(\lambda_{NN}-\lambda_{DD})\right) +\underbrace{\frac{3}{4}}_{S_2^{3/2}}\times \left(4\left(\lambda_{PD}-(\lambda_{NN} -\lambda_{DD})\right)\right)\nonumber\\ &+\frac{1}{2}\times 2\lambda_{DN}\\
=& \lambda_{PP} +2\lambda_{DD} +3 \lambda_{PD} -\lambda_{NN}+\lambda_{DN}\\
=& \lambda_P +2\lambda_D -\lambda_N.
\end{align} 
If on the other hand, $\lambda_{NN}-\lambda_{DD}\geq \lambda_{PD}$, the full $\PD$ and $\DP$ states will be used up with a part of the remaining $\NN$ state using scheme $S_1^{4/3}$ achieving the pair $(\frac{2}{3},\frac{2}{3})$. The remaining duration of the $\NN$ state is $\lambda_{NN}-\lambda_{DD}-\lambda_{PD}$, which is less than $\lambda_{DN}$ from our original condition. Therefore, this remaining $\NN$ state can be fully utilized with the $\DN$ and $\ND$ states using scheme $S_3^{2/3}$ to achieve the pair $(\frac{2}{3},0)$. The remaining $\DN$ and $\ND$ states achieve the pair $(\frac{1}{2},\frac{1}{2})$, while the $\PN$ and $\NP$ states achieve the pair $(1,0)$. Thus, we have,
\begin{align}
d_1 =& \lambda_{PP} + \underbrace{\frac{1}{2}}_{S_2^1} \times 2\lambda_{DD} + \underbrace{\frac{2}{3}}_{S_1^{4/3}}\times 3 \lambda_{PD} + \underbrace{\frac{2}{3}}_{S_3^{2/3}}\times 3(\lambda_{NN}-\lambda_{DD}-\lambda_{PD})\nonumber\\&+\underbrace{\frac{1}{2}}_{S_3^1} \times 2(\lambda_{DN}+\lambda_{DD}+\lambda_{PD}-\lambda_{NN}) +1\times 2\lambda_{PN}\\
=&\lambda_{PP} + \lambda_{PD}+2\lambda_{PN} +\lambda_{DN} +\lambda_{NN}\\
=& \lambda_{P}+\lambda_{N}\\
d_2 =&  \lambda_{PP} + \underbrace{\frac{1}{2}}_{S_2^1} \times 2\lambda_{DD} +\underbrace{\frac{2}{3}}_{S_1^{4/3}}\times 3 \lambda_{PD} + \underbrace{\frac{1}{2}}_{S_3^1} \times 2(\lambda_{DN}+\lambda_{DD}+\lambda_{PD}-\lambda_{NN})\nonumber\\
=& \lambda_{PP}+2\lambda_{DD} + 3\lambda_{PD} +\lambda_{DN} -\lambda_{NN}\\
=& \lambda_P+2\lambda_D-\lambda_N.
\end{align}      
 
\noindent \textbf{Sub-case $B3$}: $\lambda_{PD}\leq \lambda_{PN}$ and $\lambda_{DD} \geq \lambda_{NN} $: To achieve the sum rate point, we should alternate the entire $\PD$ and $\DP$ states with part of the $\PN$ and $\NP$ states  using the scheme $S_2^{3/2}$. Also the entire $\NN$ state should be alternated with the $\DD$ state using the scheme $S_2^1$. The remaining $\DD$ state can then be fully utilized with a part of the remaining $\PN$ and $\NP$ states using scheme $S_2^{4/3}$, since, $\lambda_{DD}-\lambda_{NN} \leq \lambda_{PN} - \lambda_{PD}$. The remaining $\PN$ and $\NP$ states will be exploited to achieve the s.d.o.f.~pair $(1,0)$. The $\DN$ and $\ND$ states together achieve the pair $(\frac{1}{2},\frac{1}{2})$. Thus, we have,
 \begin{align}
 d_1 =& \lambda_{PP} + \underbrace{\frac{3}{4}}_{S_2^{3/2}}\times \left(4\lambda_{PD}\right) + \underbrace{\frac{1}{2}}_{S_2^1}\times \left(2\lambda_{NN}\right) + \underbrace{\frac{2}{3}}_{S_2^{4/3}}\times \left(3(\lambda_{DD}-\lambda_{NN})\right)+\frac{1}{2}\times 2\lambda_{DN}\nonumber\\ &+ 1\times\left(2(\lambda_{PN}-\lambda_{PD})- 2(\lambda_{DD}-\lambda_{NN})\right)\\
 =& \lambda_{PP} + \lambda_{PD} + 2\lambda_{PN} +\lambda_{NN}+\lambda_{DN}\\
 =& \lambda_{P} +\lambda_{N}\\
 d_2 =& \lambda_{PP} + \underbrace{\frac{3}{4}}_{S_2^{3/2}}\times \left(4\lambda_{PD}\right) + \underbrace{\frac{1}{2}}_{S_2^1}\times \left(2\lambda_{NN}\right) + \underbrace{\frac{2}{3}}_{S_2^{4/3}}\times \left(3(\lambda_{DD}-\lambda_{NN})+\frac{1}{2}\times 2\lambda_{DN}\right)\\
 =&\lambda_{PP} +3\lambda_{PD} +2\lambda_{DD} -\lambda_{NN}+\lambda_{DN}\\
 =& \lambda_P +2\lambda_{D} - \lambda_{N}.
 \end{align}

\noindent\textbf{The points $P$ and $Q$:}
\begin{itemize}
\item Point $P$: Recall that we need to achieve the point $P:\left(\frac{2+2\lambda_P -\lambda_{PP}}{3},\lambda_{PP}\right)$ when $1-\lambda_{NN} \geq \frac{2+2\lambda_P-\lambda_{PP}}{3}$, a condition that simplifies to $\lambda_{DD}+2\lambda_{DN}\geq 2\lambda_{NN}$. To achieve this point, using the state $\PP$, we achieve $(1,1)$, with $\PD, \DP, \PN, \NP$, we achieve the pair $(1, 0)$. For the states $(\DD, \NN)\sim (\frac{2}{3}, \frac{1}{3})$,  and $(\DN,\ND,\NN)\sim(\frac{1}{3},\frac{1}{3},\frac{1}{3})$, we achieve the pair $(\frac{2}{3}, 0)$ by using the schemes $S_2^{2/3}$ and $S_3^{2/3}$, respectively. Essentially, the $\NN$ state is used up with the $\DD$ state  and the $\DN$ and $\ND$ states to achieve $\frac{2}{3}$ s.d.o.f.~for user $1$.

Time sharing yields the following s.d.o.f.~pair:
\begin{align}
d_{2}&= \lambda_{PP}\\
d_{1}&= \lambda_{PP} + 2\lambda_{PD}+ 2\lambda_{PN}+  \underbrace{\frac{2}{3}}_{S_2^{2/3}}(\lambda_{DD}+2\lambda_{DN}+\lambda_{NN})\\
&=  2\lambda_{P} - \lambda_{PP} + \frac{2}{3}(\lambda_{DD} +2\lambda_{DN}+\lambda_{NN})\\
&= 2\lambda_{P}- \lambda_{PP} + \frac{2}{3}(1-2\lambda_{P}+\lambda_{PP})\\
&= \frac{2+ 2\lambda_{P}- \lambda_{PP}}{3}.
\end{align}

\item Point $Q$: We need to achieve the point $Q:(1-\lambda_{NN},\lambda_{PP} +  (2\lambda_{NN}-2\lambda_{DN}-\lambda_{DD}) )$ when $1-\lambda_{NN} \leq \frac{2+2\lambda_P-\lambda_{PP}}{3}$, or equivalently, when $\lambda_{DD}+2\lambda_{DN}\leq \lambda_{NN}$ and under the added constraint $\lambda_{NN}\leq \lambda_D$. Here, we consider two further subcases:
\begin{itemize}
\item $\lambda_{NN} \leq \lambda_{DD} + \lambda_{DN}$: In this case, to achieve the point $Q$, we first use up the full $\DN$ and $\ND$ states with a part of the $\NN$ state using scheme $S_3^{2/3}$. We alternate the remaining $(\lambda_{NN}-\lambda_{DN})$ duration of $\NN$ state with the $\DD$ state using two schemes: $S_2^{2/3}$ and $S_2^1$. Note that in this case, $0\leq 2(\lambda_{DD} + \lambda_{DN}-\lambda_{NN})\leq \lambda_{DD}$. We use the state $\DD$  for duration $2(\lambda_{DD}+\lambda_{DN}-\lambda_{NN})$ and state $\NN$ for duration $(\lambda_{DD}+\lambda_{DN}-\lambda_{NN})$  together using scheme $S_2^{2/3}$ to achieve the s.d.o.f.~pair $\left(\frac{2}{3},0\right)$. The remaining $(2\lambda_{NN}-2\lambda_{DN}-\lambda_{DD})$ duration of the state $\NN$ is alternated with the remaining $(2\lambda_{NN}-2\lambda_{DN}-\lambda_{DD})$ duration of state $\DD$ using the scheme $S_{2}^{1}$ to achieve the s.d.o.f.~pair $\left(\frac{1}{2},\frac{1}{2}\right)$. The state $\PP$ allows us to achieve the s.d.o.f.~pair $(1,1)$ while the remaining states $\PD$, $\DP$, $\PN$, and $\NP$ each achieves $(1,0)$. Thus, by using time sharing, the s.d.o.f.~pair is:
\begin{align}
d_1 =& \lambda_{PP}+ 1\times 2\lambda_{PD} + 1\times 2\lambda_{PN} + \underbrace{\frac{2}{3}}_{S_3^{2/3}}\times 3\lambda_{DN} +  \underbrace{\frac{2}{3}}_{S_2^{2/3}}\times 3(\lambda_{DD}+\lambda_{DN}-\lambda_{NN})\\&+ \underbrace{\frac{1}{2}}_{S_2^1}\times 2(2\lambda_{NN}-2\lambda_{DN}-\lambda_{DD})\\
=& 1-\lambda_{NN}\\
d_2 =&\lambda_{PP}+ \underbrace{\frac{1}{2}}_{S_2^1}\times 2(2\lambda_{NN}-2\lambda_{DN}-\lambda_{DD})\nonumber\\
=&\lambda_{PP} +  (2\lambda_{NN}-2\lambda_{DN}-\lambda_{DD}), 
\end{align}
which is precisely the point $Q$.
\item $\lambda_{NN} \geq \lambda_{DD} +\lambda_{DN}$:
In this case, the state $\NN$ cannot be completely used with the states $\DD$, $\DN$ and $\ND$. But we note that $\lambda_{D}\geq \lambda_{NN}$. We first use up the $\DN$ and $\ND$ states by alternating with the $\NN$ state using scheme $S_3^{2/3}$. A portion $\lambda_{DD}$ of the remaining $(\lambda_{NN}-\lambda_{DN})$ duration of the $\NN$ state uses up the $\DD$ state in scheme $S_2^1$ achieving the pair $\left(\frac{1}{2},\frac{1}{2}\right)$. The remaining $(\lambda_{NN}-\lambda_{DN}-\lambda_{DD})$ portion of the $\NN$ state is used with the $\PD$ and $\DP$ states through the scheme $S_1^{4/3}$ to achieve the pair $\left(\frac{2}{3},\frac{2}{3}\right)$. For the remainder of the state $\PD$, $\DP$ and the states $\PN$, $\NP$, we can achieve the pair $(1,0)$, while $(1,1)$ is achieved in the $\PP$ state. By time sharing, we get
\begin{align}
d_1 =& \lambda_{PP} + 2\lambda_{PN} + \underbrace{\frac{2}{3}}_{S_3^{2/3}}\times 3\lambda_{DN}+\underbrace{\frac{2}{3}}_{S_1^{4/3}}\times 3(\lambda_{NN}-\lambda_{DN}-\lambda_{DD})\nonumber\\ &+ 2 (\lambda_{PD} - \lambda_{NN}+\lambda_{DN}+\lambda_{DD})+ \underbrace{\frac{1}{2}}_{S_2^1}\times 2\lambda_{DD} \\
=& 1- \lambda_{NN}\\
d_2 =& \lambda_{PP} + \underbrace{\frac{2}{3}}_{S_1^{4/3}}\times 3(\lambda_{NN}-\lambda_{DN}-\lambda_{DD})+\underbrace{\frac{1}{2}}_{S_2^1}\times 2\lambda_{DD}\\
=& \lambda_{PP} + 2\lambda_{NN} - 2\lambda_{DN}-\lambda_{DD},
\end{align}
which is again the point $Q$.
\end{itemize} 
Thus, we have achieved the point $Q$ as well.
\end{itemize}
This completes the achievability of the full region when the $3d_1+d_2$ bounds are relevant.
\subsubsection{When $3d_1+d_2$ Bounds are Irrelevant}
\begin{figure}
\centering
\begin{subfigure}{0.5\textwidth}
	\includegraphics[height=135 pt]{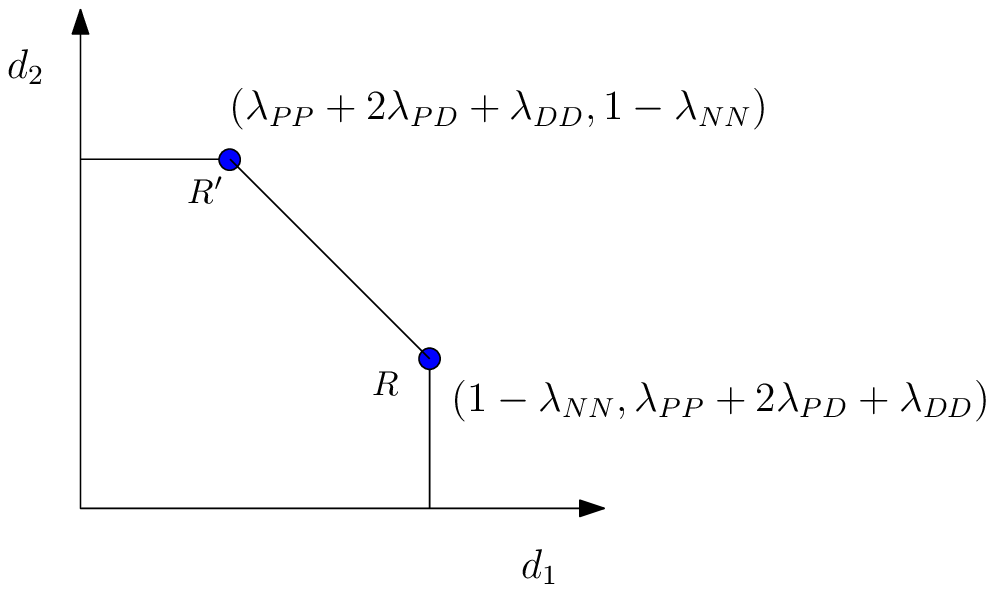}
	\caption{$\lambda_{DN}+\lambda_{PN}\neq 0$.}
	\label{fig:case3A}
\end{subfigure}%
\begin{subfigure}{0.5\textwidth}
	\includegraphics[height=135 pt]{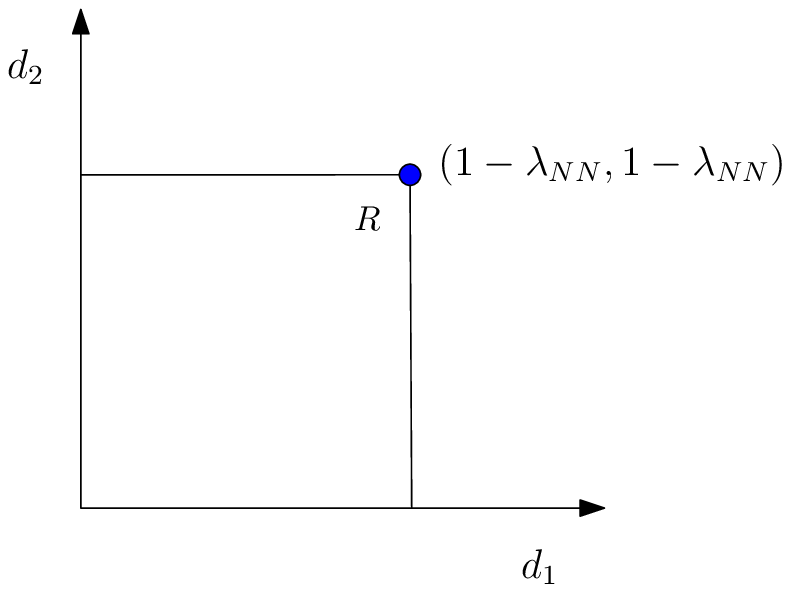}
	\caption{$\lambda_{DN}=\lambda_{PN}= 0$.}
	\label{fig:case3B}
\end{subfigure}
\caption{s.d.o.f.~regions in case $B$, when $3d_1+d_2$ and $d_1+3d_2$ bounds are completely irrelevant.}
\label{fig:case3}
\end{figure}

This case occurs when $\lambda_{NN}\geq \lambda_D$. In this case, the single user bounds are
\begin{align}
d_1\leq& 1-\lambda_{NN}\\
d_2\leq& 1-\lambda_{NN},
\end{align} 
and as shown in Fig.~\ref{fig:case3A} the only point to achieve is given by:
\begin{align}
R:(d_1,d_2) = (1-\lambda_{NN},\lambda_{PP}+2\lambda_{PD}+\lambda_{DD}).
\end{align} 
Note that $\lambda_{PP}+2\lambda_{PD}+\lambda_{DD} \leq 1-\lambda_{NN}$ with equality if and only if $\lambda_{PN}=\lambda_{DN}=0$. Thus, it suffices to achieve the point $R$ which goes to the degenerate point $(1-\lambda_{NN},1-\lambda_{NN})$ when $\lambda_{PN}=\lambda_{DN}=0$, as shown in Fig.~\ref{fig:case3B}.

To achieve this point, we alternate part of the $\NN$ state with the $\DD$ state using scheme $S_2^1$ to achieve the pair $\left(\frac{1}{2},\frac{1}{2}\right)$, and  with the $\PD$ and $\DP$ states using the scheme $S_{1}^{4/3}$ to achieve the pair $\left(\frac{2}{3},\frac{2}{3}\right)$ and with the $\DN$ and $\ND$ states using the scheme $S_3^{2/3}$ to achieve the pair $(\frac{2}{3},0)$. The remaining $\NN$ state is left unused. The $\PN$ and $\NP$ states, if available, is used to achieve the s.d.o.f.~pair $\left(1,0\right)$. Thus, we have,
\begin{align}
d_1 =& \lambda_{PP} + \underbrace{\frac{1}{2}}_{S_2^1}\times \left(2\lambda_{DD}\right) + \underbrace{\frac{2}{3}}_{S_1^{4/3}}\times \left(3\lambda_{PD}\right) + \underbrace{\frac{2}{3}}_{S_3^{2/3}}\times 3\lambda_{DN}+ 1\times 2\lambda_{PN}\nonumber\\
=& 1 - \lambda_{NN}\\
d_2 =& \lambda_{PP} + \underbrace{\frac{1}{2}}_{S_2^1}\times \left(2\lambda_{DD}\right) + \underbrace{\frac{2}{3}}_{S_1^{4/3}}\times \left(3\lambda_{PD}\right)\\
=& \lambda_{PP} + \lambda_{DD} +2\lambda_{PD}\\
=& 1-\lambda_{NN} \mbox{ if } \lambda_{PN} =\lambda_{DN}=0.
\end{align}
This completes the proof of the achievability.
\section{Proof of the Converse}\label{converse}
\subsection{Local Statistical Equivalence Property and Associated Lemma}
We introduce a property of the channel which we call \emph{local statistical equivalence}. Let us focus on the channel output of receiver $2$ corresponding to the state $\PD$ and $\DD$ at time $t$:
\begin{align}
Z_{pd}(t) &= \mathbf{H}_{2, pd}(t)\mathbf{X}_{pd}(t) + N_{2,pd}(t)\\
Z_{dd}(t) &= \mathbf{H}_{2, dd}(t)\mathbf{X}_{dd}(t) + N_{2,dd}(t).
\end{align}
Now consider $(\tilde{\mathbf{H}}_{2, pd}(t),\tilde{\mathbf{H}}_{2, dd}(t))$, $(\tilde{N}_{2,pd}(t),\tilde{N}_{2,dd}(t))$, which are i.i.d.~as $(\mathbf{H}_{2, pd}(t),\mathbf{H}_{2,dd}(t))$ and $(N_{2,pd}(t),N_{2,dd}(t))$, respectively. Using these random variables, we define  artificial channel outputs as:
\begin{align}
\tilde{Z}_{pd}(t) &= \tilde{\mathbf{H}}_{2, pd}(t)\mathbf{X}_{pd}(t) + \tilde{N}_{2,pd}(t)\\
\tilde{Z}_{dd}(t) &= \tilde{\mathbf{H}}_{2, dd}(t)\mathbf{X}_{dd}(t) + \tilde{N}_{2,dd}(t).
\end{align}
Let $\Omega = (\mathbf{H}^n, \tilde{\mathbf{H}}^n)$. Now the \emph{local statistical equivalence} property is the following:
\begin{align}
h(Z_{pd}(t),Z_{dd}(t)|Z_{pd}^{t-1},Z_{dd}^{t-1},\Omega) = h(\tilde{Z}_{pd}(t),\tilde{Z}_{dd}(t)|Z_{pd}^{t-1},Z_{dd}^{t-1},\Omega).
\end{align}
This property shows that if we consider the outputs of a receiver for such states in which it supplies delayed $\CSIT$, then the entropy of the channel outputs conditioned on the past outputs is the same as that of another artificial receiver whose channel is distributed identically as the original receiver. Note that in an alternating $\CSIT$ setting, we focus on only the states in which the receiver provides delayed $\CSIT$; hence we call it \textit{local}. The original and artificial receivers have \emph{statistically equivalent} channels in the sense that the conditional differential entropies of the outputs at the real and the artificial receivers given the past outputs are equal. The proof of this property is given in Appendix~\ref{appendix-local-equiv}. We next present the following lemma which together with the local statistical equivalence property is instrumental in the converse proofs. 

\begin{Lem}\label{bounds-lemma}
For our channel model, with $\CSIT$ alternating among the states $\DD$, $\PD$ and $\DP$ we have:
\begin{align}
h(Z^n|\Omega) \stackrel{.}{\geq}& \;h(Y_{pd}^n, Y_{dd}^n|Z^n,\Omega)\label{Lemma:1}\\
2h(Z^n|\Omega) \stackrel{.}{\geq}&\; h(Y_{pd}^n,Y_{dd}^n|\Omega)\label{Lemma:2}\\
h(Y^n|\Omega) \stackrel{.}{\geq}& \; h(Z_{dp}^n, Z_{dd}^n|Y^n,\Omega)\label{Lemma:3}\\
2h(Y^n|\Omega) \stackrel{.}{\geq}& \; h(Z_{dp}^n, Z_{dd}^n|\Omega),\label{Lemma:4}
\end{align}
where $a \stackrel{.}{\geq} b$ denotes $\lim\limits_{P \rightarrow \infty} \frac{a}{\log P} \geq \lim\limits_{P \rightarrow \infty} \frac{b}{\log P}$.
\end{Lem}
This lemma is proved in Appendix~\ref{appendix-bounds-lemma}.

In the following sections, we use the local statistical equivalence property along with Lemma \ref{bounds-lemma} to prove the bounds on individual d.o.f.~$d_{1}$ and $d_{2}$, the sum d.o.f.~$(d_{1}+d_{2})$ and the weighted d.o.f.~$3d_{1}+d_{2}$ and $d_{1}+3d_{2}$. 
\subsection{The Single User Bounds}
We recall the single user bounds in \eqref{eq:single_user_rate1}-\eqref{eq:single_user_rate2}:
\begin{align}
d_1 &\leq \min\left(\frac{2+2\lambda_{P}-\lambda_{PP}}{3},1-\lambda_{NN}\right)\\
d_2 &\leq \min\left(\frac{2+2\lambda_{P}-\lambda_{PP}}{3},1-\lambda_{NN}\right). 
\end{align} 

\subsubsection{Proof of $d_{i}\leq \frac{2+2\lambda_P - \lambda_{PP}}{3}$, $i=1,2$}
In this section, we prove the following single-user bounds:
\begin{align}
d_1 &\leq \frac{2+2\lambda_P - \lambda_{PP}}{3} = \frac{2+2\lambda_{P}+2\lambda_{PD}+2\lambda_{PN}}{3}\label{eq:single_user1}\\
d_2 &\leq \frac{2+2\lambda_P - \lambda_{PP}}{3}
= \frac{2+2\lambda_{P}+2\lambda_{PD}+2\lambda_{PN}}{3}.\label{eq:single_user2}
\end{align}  
To do so, we enhance the transmitter in the following way: 
\begin{itemize}
\item First, if in any state, the transmitter has perfect CSIT from any of the users, we provide perfect CSI for the other user too, that is, the states $\PP, \PD, \DP, \PN, \NP$ are all enhanced to the state $\PP$. 
\item Next, we enhance all the remaining states, (i.e., $\DD, \DN, \ND, \NN$) to $\DD$. 
\end{itemize}
The enhanced channel has two states: $\PP$ occurring for $\lambda_{pp} = \lambda_{PP} + 2 \lambda_{PD}+2\lambda_{PN}$ (using symmetry of the alternation), and $\DD$ occurring for the remaining fraction of the time. Now, we have the following lemma for such a channel with only $\PP$ and $\DD$ states.

\begin{Lem}\label{lem:single_user_lemma1}
Consider the two-user MISO BCCM with only two states: $\PP$ and $\DD$ occurring for $\lambda_{pp}$ and $\lambda_{dd}$ fractions of time, respectively, such that $\lambda_{pp}+\lambda_{dd}=1$. Then,
\begin{align}
d_1 &\leq \frac{2+\lambda_{pp}}{3}\\
d_2 &\leq \frac{2+\lambda_{pp}}{3}.
\end{align}
\end{Lem}
The proof of this lemma is provided in Appendix~\ref{appendix:single_user_lemma1}.

Now using $\lambda_{pp} = \lambda_{PP} + 2 \lambda_{PD}+2\lambda_{PN}$ in Lemma \ref{lem:single_user_lemma1}, we get the bounds in \eqref{eq:single_user1}-\eqref{eq:single_user2}. 

\subsubsection{Proof of $d_{i}\leq 1-\lambda_{NN}$, $i=1,2$}
In this section, we prove the following single user bounds:
\begin{align}
d_1 &\leq 1-\lambda_{NN}\label{eq:single_user21}\\
d_2 &\leq 1-\lambda_{NN}\label{eq:single_user22}.
\end{align}
To prove these, we again enhance the transmitter, but in a different way. We provide the transmitter with perfect $\CSIT$ in every state except the $\NN$ state, that is, every state except the $\NN$ state is enhanced to the $\PP$ state. Thus, we end up with a system with two states: $\PP$ occurring for $1-\lambda_{NN}$ fraction of the time and $\NN$ occurring for $\lambda_{NN}$ fraction of the time. Note that since there is no delayed $\CSIT$ in the enhanced system, there is no feedback. For such a system we have the following lemma.

\begin{Lem}\label{lem:single_user_bound2}
For the two-user MISO BCCM with only two states: $\PP$ and $\NN$ occurring for $1-\lambda_{nn}$ and $\lambda_{nn}$ fractions of time, respectively, and no feedback,
\begin{align}
d_1 &\leq 1-\lambda_{nn}\\
d_2 &\leq 1-\lambda_{nn}.
\end{align}
\end{Lem}
The proof of this lemma is provided in Appendix~\ref{appendix:single_user_bound2}.

Using $\lambda_{nn} = \lambda_{NN}$ in Lemma \ref{lem:single_user_bound2}, we get the bounds in \eqref{eq:single_user21}-\eqref{eq:single_user22}.

Combining the bounds in \eqref{eq:single_user1}-\eqref{eq:single_user2} and \eqref{eq:single_user21}-\eqref{eq:single_user22}, we have the bounds in \eqref{eq:single_user_rate1}-\eqref{eq:single_user_rate2}.

\subsection{Proof of $d_{1}+d_{2}$ Bound}
Recall the sum s.d.o.f.~bound from \eqref{eq:sum_rate}:
\begin{align}
d_1+d_2 \leq 2(\lambda_P+\lambda_D).
\end{align}
The original system model has nine possible states, namely, $\PP$, $\DD$, $\NN$, $\DP$, $\PD$, $\PN$, $\NP$, $\DN$, and $\ND$. We enhance the transmitter in the following way: whenever in any state, the transmitter receives delayed CSI of a channel, we provide perfect CSI of the channel to the transmitter; in other words, we convert each $\mathsf{D}$ state to a $\mathsf{P}$ state. This clearly does not decrease the secrecy capacity (and thus, the s.d.o.f.~region). Also note that the enhanced system does not have any delayed $\CSIT$, and hence no feedback. Now the enhanced system has only four states: $\PP$,  $\PN$, $\NP$, $\NN$, occurring for $\lambda_{pp}=\lambda_{PP}+\lambda_{DD}+\lambda_{DP}+\lambda_{PD}$, $\lambda_{pn} = \lambda_{PN}+\lambda_{DN}$, $\lambda_{np} = \lambda_{NP}+\lambda_{ND}$ and $\lambda_{nn}= \lambda_{NN}$ fractions of time, respectively. For such a system with four states we have the following lemma:

\begin{Lem}\label{lem:four_state_sum_rate}
Consider the two-user MISO BCCM with only four of the nine states: $\PP$, $\PN$, $\NP$ and $\NN$ occurring for $\lambda_{pp}$, $\lambda_{pn}$, $\lambda_{np}$ and $\lambda_{nn}$ fractions of the time, with  $\lambda_{pp}+\lambda_{pn}+\lambda_{np}+\lambda_{nn}=1$. Also, assume there is no feedback. Then,
\begin{align}
d_1+d_2 \leq 2\lambda_{pp}+ \lambda_{pn}+\lambda_{np}.
\end{align}  
\end{Lem}
Proof of this lemma is presented in Appendix~\ref{appendix:four_state_sum_rate}.

Thus, using $\lambda_{pp}=\lambda_{PP}+\lambda_{DD}+\lambda_{DP}+\lambda_{PD}$, $\lambda_{pn} = \lambda_{PN}+\lambda_{DN}$, $\lambda_{np} = \lambda_{NP}+\lambda_{ND}$ and $\lambda_{nn}= \lambda_{NN}$ in Lemma \ref{lem:four_state_sum_rate}, we have,
\begin{align}
d_1+d_2 &\leq 2(\lambda_{PP}+\lambda_{DP}+\lambda_{PD}+\lambda_{DD}) + \lambda_{PN}+\lambda_{DN} +  \lambda_{NP}+\lambda_{ND} \\
&= 2(\lambda_P+\lambda_D),\label{eq:symmetry}
\end{align} 
where \eqref{eq:symmetry} follows due to the assumed symmetry: $\lambda_{PD}=\lambda_{DP}$, and this completes the proof of the bound on $d_{1}+d_{2}$.

\subsection{Proof of $3d_1+d_2$ and $d_1+3d_2$ Bounds}

In this section, we prove the following bounds from \eqref{eq:3d1+d2_bound1}-\eqref{eq:3d1+d2_bound2}:
\begin{align}
3d_1+d_2 &\leq 2+2\lambda_{PP}+ 2\lambda_{PD}+2\lambda_{PN}\\
d_1+3d_2 &\leq 2+2\lambda_{PP}+2\lambda_{PD}+2\lambda_{PN}.
\end{align}
To do so, we enhance the system in the following way: Whenever in any state, the transmitter has no CSIT from a user, we provide the transmitter delayed CSIT of that user's channel; in other words, we enhance each $\mathsf{N}$ state to a $\mathsf{D}$ state. After this enhancement, we are left with only four states, namely $
\PP$, $\PD$, $\DP$ and $\DD$ occurring for $\lambda_{pp}=\lambda_{PP}$, $\lambda_{pd}=\lambda_{PD}+\lambda_{PN}$, $\lambda_{dp}=\lambda_{DP}+\lambda_{NP}$ and $\lambda_{dd}=\lambda_{DD}+\lambda_{DN}+\lambda_{ND}+\lambda_{NN}$ fractions of the time, respectively. We have the following lemma for such a system with four states:

\begin{Lem}\label{lem:pd_dp_bound}
Consider the two-user MISO BCCM with only four of the nine states: $\PP$, $\PD$, $\DP$ and $\DD$ occurring for $\lambda_{pp}$, $\lambda_{pd}$, $\lambda_{dp}$ and $\lambda_{dd}$ fractions of the time, with $\lambda_{pd}=\lambda_{dp}$ and $\lambda_{pp}+\lambda_{pd}+\lambda_{dp}+\lambda_{dd}=1$. Then,
\begin{align}
3d_1+d_2 \leq 2+2\lambda_{pp}+2\lambda_{pd}\\
d_1+3d_2 \leq 2+2\lambda_{pp}+2\lambda_{pd}.
\end{align}  
\end{Lem}
We provide a proof for this lemma in Appendix~\ref{appendix:pd_dp_bound}. 

Using $\lambda_{pp}=\lambda_{PP}$, $\lambda_{pd}=\lambda_{PD}+\lambda_{PN}$, $\lambda_{dp}=\lambda_{DP}+\lambda_{NP}$ and $\lambda_{dd}=\lambda_{DD}+\lambda_{DN}+\lambda_{ND}+\lambda_{NN}$ in Lemma~\ref{lem:pd_dp_bound}, and symmetry of the alternating states, we have,
\begin{align}
3d_1+d_2 &\leq 2+2\lambda_{PP}+2\lambda_{PD} + 2\lambda_{PN}\\
&= 2 +2\lambda_{P}\\
d_1+3d_2 &\leq 2+2\lambda_{PP}+2\lambda_{PD}+2\lambda_{PN}\\
&= 2+2\lambda_P,
\end{align}
which completes the proofs for the bounds on $3d_{1}+d_2$ and $d_{1}+3d_{2}$.

\section{Conclusions}
In this paper, we studied the two-user MISO broadcast channel with confidential messages (BCCM) and characterized its secure degrees of freedom (s.d.o.f.) region with alternating channel state information at the transmitter ($\CSIT$). The converse proofs for the s.d.o.f.~region presented in the paper are based on novel arguments such as local statistical equivalence property and enhancing the system model in different ways, where each carefully chosen enhancement strictly improves the quality of $\CSIT$ in a certain manner.  For each such enhanced system, we invoke the local statistical equivalence property and incorporate the confidentiality constraints and obtain corresponding upper bounds on the individual $(d_{1}, d_{2})$, sum $(d_{1}+d_{2})$ and weighted $(3d_{1} +d_{2}, d_{1}+3d_{2})$ s.d.o.f.  

To establish the achievability of the s.d.o.f.~region, several constituent schemes are developed, where each scheme by itself only operates over a subset of $9$ states. The achievability of the optimal s.d.o.f.~region is then established by time-sharing between the core constituent schemes. The core constituent schemes not only serve the purpose of establishing the s.d.o.f.~region but also highlight the synergies across multiple $\CSIT$ states which can be exploited to achieve higher s.d.o.f.~in comparison to their individually optimal s.d.o.f.~values. Besides highlighting the synergistic benefits of alternating $\CSIT$ for secrecy, the optimal s.d.o.f.~region also quantifies the information theoretic minimal $\CSIT$ required from each user to attain a certain s.d.o.f.~value. In addition, we also quantify the loss in d.o.f., as a function of the overall $\CSIT$ quality, which must be incurred for incorporating confidentiality constraints. 


\begin{appendices}
\section{Proof of Local Statistical Equivalence} \label{appendix-local-equiv}
In this section, we prove the \emph{local statistical equivalence} property:
\begin{align}
h(Z_{pd}(t),Z_{dd}(t)|Z_{pd}^{t-1},Z_{dd}^{t-1},\Omega) = h(\tilde{Z}_{pd}(t),\tilde{Z}_{dd}(t)|Z_{pd}^{t-1},Z_{dd}^{t-1},\Omega).
\end{align}
To this end, first denote the common distribution of $(\mathbf{H}_{2, pd}(t),\mathbf{H}_{2, dd}(t))$, $(\tilde{\mathbf{H}}_{2, pd}(t),\tilde{\mathbf{H}}_{2, dd}(t))$ by $F$. Let $\Omega = \left\lbrace \mathbf{H}_1(t), \mathbf{H}_2(t), \tilde{\mathbf{H}}_1(t),\tilde{\mathbf{H}}_2(t), t=1,\ldots,n\right\rbrace $ be the set of all channel vectors upto and including time $n$.
 Also, let $\Omega_t = \Omega \backslash \left\lbrace \mathbf{H}_{2,pd}(t),\tilde{\mathbf{H}}_{2,pd}(t),\mathbf{H}_{2,dd}(t),\tilde{\mathbf{H}}_{2,dd}(t) \right\rbrace $.
We have,
\begin{align}
&h(Z_{pd}(t),Z_{dd}(t)|Z_{pd}^{t-1},Z_{dd}^{t-1}\Omega)\nonumber\\=&\mathbb{E}_F\left[h(Z_{pd}(t), Z_{dd}(t)|Z_{pd}^{t-1},Z_{dd}^{t-1},\Omega_t,\tilde{\mathbf{H}}_{2,pd}(t),\tilde{\mathbf{H}}_{2,dd}(t),\mathbf{H}_{2,pd}(t)=\mathbf{h}(t),\mathbf{H}_{2,dd}(t)=\mathbf{g}(t)) \right]\\
=&\mathbb{E}_F\left[h(\mathbf{h}(t)\mathbf{X}_{pd}(t) +N_{2,pd}(t),\mathbf{g}(t)\mathbf{X}_{dd}(t) +N_{2,dd}(t)|Z_{pd}^{t-1},Z_{dd}^{t-1},\Omega_t) \right]\label{eq:delayed_csit2}\\
=&\mathbb{E}_F\left[h(\mathbf{h}(t)\mathbf{X}_{pd}(t) +\tilde{N}_{2,pd}(t),\mathbf{g}(t)\mathbf{X}_{dd}(t) +\tilde{N}_{2,dd}(t)|Z_{pd}^{t-1},Z_{dd}^{t-1},\Omega_t) \right]\label{eq:noise_independence} \\
=&\mathbb{E}_F\left[h(\mathbf{h}(t)\mathbf{X}_{pd}(t) +\tilde{N}_{2,pd}(t),\mathbf{g}(t)\mathbf{X}_{dd}(t) +\tilde{N}_{2,dd}(t)|Z_{pd}^{t-1},Z_{pd}^{t-1},\Omega_t, \tilde{\mathbf{H}}_{2,pd}(t) = \mathbf{h}(t),\right.\nonumber\\ &\hspace{255 pt}\left.\tilde{\mathbf{H}}_{2,dd}(t) = \mathbf{g}(t)) \right]\label{eq:delayed_csit3}\\
=&\mathbb{E}_F\left[h(\tilde{Z}_{pd}(t),\tilde{Z}_{dd}(t)|Z_{pd}^{t-1},Z_{dd}^{t-1},\Omega_t, \mathbf{H}_{2,pd}(t),\mathbf{H}_{2,dd}(t),\tilde{\mathbf{H}}_{2,pd}(t) = \mathbf{h}(t),\tilde{\mathbf{H}}_{2,dd}(t) = \mathbf{g}(t)) \right]\label{eq:delayed_csit4}\\
=& h(\tilde{Z}_{pd}(t),\tilde{Z}_{dd}(t)|Z_{pd}^{t-1},Z_{dd}^{t-1},\Omega),
\end{align}
where \eqref{eq:delayed_csit2} follows because $(\mathbf{X}_{pd}(t),\mathbf{X}_{dd}(t))$ does not depend on $\left(\mathbf{H}_{2,pd}(t),\tilde{\mathbf{H}}_{2,pd}(t),\mathbf{H}_{2,dd}(t),\right. \\ \left.\tilde{\mathbf{H}}_{2,dd}(t)\vphantom{\tilde{\mathbf{H}}_{2,pd}(t)}\right)$, \eqref{eq:noise_independence} follows since the additive noises $(N_{2,pd}(t),N_{2,dd}(t))$ and $(\tilde{N}_{2,pd}(t),\tilde{N}_{2,dd}(t))$ are i.i.d.~and independent of all other random variables, \eqref{eq:delayed_csit3}-\eqref{eq:delayed_csit4} follow since $(\mathbf{H}_{2,pd}(t),\mathbf{H}_{2,dd}(t))$ and  $(\tilde{\mathbf{H}}_{2,pd}(t),\tilde{\mathbf{H}}_{2,dd}(t))$ have the same distribution $F$ and the fact that $(\mathbf{X}_{pd}(t),\mathbf{X}_{dd}(t))$ does not depend on $(\mathbf{H}_{2,pd}(t), \tilde{\mathbf{H}}_{2,pd}(t),\mathbf{H}_{2,pd}(t), \tilde{\mathbf{H}}_{2,dd}(t))$.  

\section{Proof of Lemma~\ref{bounds-lemma}} \label{appendix-bounds-lemma}
We consider the scenario in which there are only three $\CSIT$ states, namely $\DD, \PD$ and $\DP$. For such a specific alternating $\CSIT$ model, we define the channel outputs as:
\begin{align}
Z^{n}&\triangleq  \left(Z_{dd}^{n}, Z_{pd}^{n}, Z_{dp}^{n}\right)\nonumber\\
Y^{n}&\triangleq  \left(Y_{dd}^{n}, Y_{pd}^{n}, Y_{dp}^{n}\right)\nonumber.
\end{align}
Also let $\Omega$ denote the set of all channel vectors upto and including time $n$, that is, in other words, $\Omega = \left\lbrace \mathbf{H}_1(t), \mathbf{H}_2(t), \tilde{\mathbf{H}}_1(t),\tilde{\mathbf{H}}_2(t), t=1,\ldots,n\right\rbrace $.
We wish to prove that with $\CSIT$ alternating among the states $\DD$, $\PD$ and $\DP$ we have:
\begin{align}
h(Z^n|\Omega) \stackrel{.}{\geq}& \;h(Y_{pd}^n, Y_{dd}^n|Z^n,\Omega)\label{Lemma:11}\\
2h(Z^n|\Omega) \stackrel{.}{\geq}&\; h(Y_{pd}^n,Y_{dd}^n|\Omega)\label{Lemma:22}\\
h(Y^n|\Omega) \stackrel{.}{\geq}& \; h(Z_{dp}^n, Z_{dd}^n|Y^n,\Omega)\label{Lemma:33}\\
2h(Y^n|\Omega) \stackrel{.}{\geq}& \; h(Z_{dp}^n, Z_{dd}^n|\Omega).\label{Lemma:44}
\end{align}
First we note that due to symmetry, it suffices to prove (\ref{Lemma:11}) and (\ref{Lemma:22}). We proceed as follows:
\begin{align}
h(Z^n|\Omega) =& h(Z_{pd}^n,\zdd|\Omega) + h(Z_{dp}^n|Z_{pd}^n,\zdd \Omega)\\
=& \sum_{t=1}^{n} h(Z_{pd}(t), Z_{dd}(t)|Z_{pd}^{t-1},Z_{dd}^{t-1},\Omega) + h(Z_{dp}^n|Z_{pd}^n,\zdd, \Omega).\label{eq:eqn1}
\end{align}
Using the \emph{local statistical equivalence} property, we get,
\begin{align}
h(Z^n|\Omega) = \sum_{t=1}^{n} h(\tilde{Z}_{pd}(t),\tilde{Z}_{dd}(t)|Z_{pd}^{t-1},Z_{dd}^{t-1},\Omega) + h(Z_{dp}^n|Z_{pd}^n,\zdd, \Omega). \label{eq:eqn2}
\end{align}
Adding \eqref{eq:eqn1} and \eqref{eq:eqn2}, and lower bounding, we get,
\begin{align}
2h(Z^n|\Omega)
\geq& \sum_{t=1}^{n} h(Z_{pd}(t),Z_{dd}(t), \tilde{Z}_{pd}(t),\tilde{Z}_{dd}(t)|Z_{pd}^{t-1},Z_{dd}^{t-1},\Omega) +  2h(Z_{dp}^n|Z_{pd}^n,\zdd \Omega)\nonumber\\
\geq& \sum_{t=1}^{n} h(Z_{pd}(t),Z_{dd}(t), \tilde{Z}_{pd}(t),\tilde{Z}_{dd}(t)|Z_{pd}^{t-1},Z_{dd}^{t-1},\Omega) + h(Z_{dp}^n|Z_{pd}^n,\zdd, \Omega)\nonumber\\ &+ no(\log P)\label{eq:lemma1_eqn1}\\
=& \sum_{t=1}^{n}  h(Z_{pd}(t),Z_{dd}(t), \tilde{Z}_{pd}(t),\tilde{Z}_{dd}(t),Y_{pd}(t),Y_{dd}(t)|Z_{pd}^{t-1},Z_{dd}^{t-1},\Omega)\nonumber\\ &-  \sum_{t=1}^{n} h(Y_{pd}(t), Y_{dd}(t)|Z_{pd}(t),Z_{dd}(t), \tilde{Z}_{pd}(t),\tilde{Z}_{dd}(t),Z_{pd}^{t-1},Z_{dd}^{t-1}\Omega)  \nonumber \\&+ h(Z_{dp}^n|Z_{pd}^n,\zdd, \Omega) + no(\log P) \label{eq:lemma1_eqn2}\\
\geq&  \sum_{t=1}^{n}  h(Z_{pd}(t),Z_{dd}(t), Y_{pd}(t), Y_{dd}(t)|Z_{pd}^{t-1},Z_{dd}^{t-1},\Omega)  + h(Z_{dp}^n|Z_{pd}^n,\zdd, \Omega)\nonumber\\&+ no(\log P)\\
\geq&  \sum_{t=1}^{n}  h(Z_{pd}(t),Z_{dd}(t), Y_{dd}(t) Y_{pd}(t)|Z_{pd}^{t-1},Z_{dd}^{t-1},Y_{pd}^{t-1},Y_{dd}^{t-1}\Omega) \nonumber\\
&+ h(Z_{dp}^n|Z_{pd}^n,Y_{pd}^n,Y_{dd}^n,\zdd, \Omega) + no(\log P) \\
=& h(Z_{pd}^n,Z_{dd}^n, Y_{pd}^n, Y_{dd}^n| \Omega) +  h(Z_{dp}^n|Z_{pd}^n,Y_{pd}^n,\zdd,Y_{dd}^n \Omega) + no(\log P) \\
=& h(Z^n, Y_{pd}^n, Y_{dd}^n|\Omega) + no(\log P),  \label{eq:eqn3}
\end{align}
where \eqref{eq:lemma1_eqn1} follows by noting that 
\begin{align}
h(Z_{dp}^n|Z_{pd}^n,Z_{dd}^n,\Omega) \geq h(Z_{dp}^n|Z_{pd}^n,Z_{dd}^n,\mathbf{X}^n,\Omega) = no(\log P)
\end{align}
and \eqref{eq:lemma1_eqn2} follows since given $(Z_{pd}(t), \tilde{Z}_{pd}(t),Z_{dd}(t), \tilde{Z}_{dd}(t))$, one can reconstruct $(\mathbf{X}_{pd}(t),\mathbf{X}_{dd}(t))$ and hence $(Y_{pd}(t),Y_{dd}(t))$ within noise distortion, implying that
\begin{align}
h(Y_{pd}(t), Y_{dd}(t)|Z_{pd}(t),Z_{dd}(t) \tilde{Z}_{pd}(t),\tilde{Z}_{dd}(t),Z_{pd}^{t-1},\Omega ) \leq no(\log P).
\end{align}

Now both \eqref{Lemma:11} and \eqref{Lemma:22} can be derived from \eqref{eq:eqn3}. We simply expand the right hand side of \eqref{eq:eqn3} in two ways:
\begin{align}
2h(Z^n|\Omega) \geq&  h(Z^n, Y_{pd}^n, Y_{dd}^n|\Omega) + no(\log P) \\
=& h(Z^n|\Omega) + h(Y_{pd}^n, Y_{dd}^n|Z^n,\Omega) + no(\log P),
\end{align}
which implies  $h(Z^n|\Omega) \stackrel{.}{\geq} h(Y_{pd}^n, Y_{dd}^n|Z^n,\Omega)$, which is exactly \eqref{Lemma:11}.
Alternatively from  \eqref{eq:eqn3}, we also have
\begin{align}
2h(Z^n|\Omega) \geq& h(Y_{pd}^n, Y_{dd}^n|\Omega) + h(Z^n|Y_{pd}^n,Y_{dd}^n \Omega) + no(\log P) \\
\geq& h(Y_{pd}^n, Y_{dd}^n| \Omega) + no(\log P),
\end{align}
which implies $2h(Z^n|\Omega) \stackrel{.}{\geq} h(Y_{pd}^n, Y_{dd}^n|\Omega)$, thus proving the relation in \eqref{Lemma:22}. This completes the proof of Lemma~\ref{bounds-lemma}.

\section{Proofs of Lemmas \ref{lem:single_user_lemma1}-\ref{lem:pd_dp_bound}}
\subsection{Proof of Lemma \ref{lem:single_user_lemma1}}\label{appendix:single_user_lemma1}
Recall that we wish to prove that for the two-user MISO BC with only two states: $\PP$ and $\DD$ occurring for $\lambda_{pp}$ and $\lambda_{dd}$ fractions of time, respectively, such that $\lambda_{pp}+\lambda_{dd}=1$,
\begin{align}
d_1 \leq& \frac{2+\lambda_{pp}}{3}, \quad d_2 \leq \frac{2+\lambda_{pp}}{3}.
\end{align}

To do so, we proceed as follows:
\begin{align}
nR_1 &\leq I(W_1;Y_{pp}^n, Y_{dd}^n| \Omega)+no(n)\label{eq:i1}\\
&=  I(W_1;Y_{dd}^n|\Omega) + I(W_1;Y_{pp}^n|Y_{dd}^n,\Omega)+no(n)\\
&\leq n\lambda_{pp}\log P + I(W_1;Y_{dd}^n|\Omega)+no(n)\label{in-between}\\
&\leq n\lambda_{pp}\log P + I(W_1;Y_{dd}^n, Z_{dd}^n| \Omega)+no(n)\\
&\leq n\lambda_{pp}\log P + I(W_1;Y_{dd}^n| Z_{dd}^n, \Omega) + no(\log P)+no(n)\\
&\leq n\lambda_{pp}\log P + h(Y_{dd}^n|Z_{dd}^n,\Omega) + no(\log P)+no(n)\label{eq:i2}\\
&\leq n\lambda_{pp}\log P + h(Z_{dd}^n|\Omega) + no(\log P)+no(n),\label{eq:bound_single_user1} 
\end{align}
where \eqref{eq:i1} follows from decodability of $W_{1}$ at receiver $1$ and Fano's inequality, \eqref{eq:i2} follows from confidentiality constraint of message $W_{1}$ at receiver $2$, and \eqref{eq:bound_single_user1} follows from application of Lemma \ref{bounds-lemma}.

Starting from (\ref{in-between}), we also have
\begin{align}
nR_1 &\leq n\lambda_{pp}\log P + I(W_1;Y_{dd}^n|\Omega)+no(n)\\
&\leq n\lambda_{pp}\log P + I(W_1;Y_{dd}^n|\Omega) - I(W_1;Z_{dd}^n|\Omega) + no(\log P)+no(n)\label{eq:j1}\\
&\leq n\lambda_{pp}\log P + h(Y_{dd}^n|\Omega) - h(Y_{dd}^n|W_1,\Omega) - h(Z_{dd}^n|\Omega) + h(Z_{dd}^n|W_1, \Omega)+ no(\log P)\nonumber\\&\hspace{7 pt}+no(n)\\   
&\leq n\lambda_{pp}\log P + h(Y_{dd}^n|\Omega) - \frac{1}{2}h(Z_{dd}^n|W_1, \Omega) - h(Z_{dd}^n|\Omega) + h(Z_{dd}^n|W_1,\Omega)+ no(\log P) \nonumber\\&\hspace{7 pt}+no(n)\label{eq:j2}\\
&\leq n\lambda_{pp}\log P + h(Y_{dd}^n|\Omega) + \frac{1}{2}h(Z_{dd}^n|W_1,\Omega) - h(Z_{dd}^n|\Omega) + no(\log P)+no(n)\\  
&\leq  n\lambda_{pp}\log P + h(Y_{dd}^n|\Omega) + \frac{1}{2}h(Z_{dd}^n|\Omega) - h(Z_{dd}^n|\Omega) + no(\log P)+no(n)\label{eq:j3}\\  
&= n\lambda_{pp}\log P + h(Y_{dd}^n|\Omega) -\frac{1}{2}h(Z_{dd}^n|\Omega) + no(\log P)+no(n)\\   
&\leq n\lambda_{pp}\log P + n\lambda_{dd}\log P -\frac{1}{2}h(Z_{dd}^n|\Omega) + no(\log P)+no(n),\label{eq:bound_single_user2}  
\end{align}
where \eqref{eq:j1} follows from confidentiality constraint of message $W_{1}$ at receiver $2$, \eqref{eq:j2} follows from application of Lemma \ref{bounds-lemma}, and \eqref{eq:j3} follows from the fact that conditioning reduces differential entropy. 

Eliminating $h(Z_{dd}^{n}|\Omega)$ from the bounds \eqref{eq:bound_single_user2} and \eqref{eq:bound_single_user1}, we have,
\begin{align}
3nR_1 \leq& (3n\lambda_{pp}+ 2n\lambda_{dd})\log P + no(\log P)+no(n)\\
=& (2 + \lambda_{pp})n\log P + no(\log P).
\end{align}
Now first dividing by $n$ and letting $n\rightarrow \infty$, then dividing by $\log P$ and letting $P \rightarrow \infty$, we get,
\begin{align}
d_1 \leq& \frac{2 + \lambda_{pp}}{3}.
\end{align}
By symmetry, we get the same single user bound for user 2, completing the proof of Lemma~\ref{lem:single_user_lemma1}.
\subsection{Proof of Lemma \ref{lem:single_user_bound2}}\label{appendix:single_user_bound2} 
We want to show that for the two-user MISO BC with only two states: $\PP$ and $\NN$ occurring for $1-\lambda_{nn}$ and $\lambda_{nn}$ fractions of time, respectively,
\begin{align}
d_1 \leq& 1-\lambda_{nn}\\
d_2 \leq& 1-\lambda_{nn}.
\end{align}

To prove this, we note that since there is no feedback, the secrecy capacity depends only on the marginal distributions of channel outputs given the input distribution; \cite{csiszar}. Since the transmitter does not have channel knowledge of any of the users in the state $\NN$, our system with outputs
\begin{align}
Y^n =& (Y_{pp}^n, Y_{nn}^n)\\
Z^n =& (Z_{pp}^n, Z_{nn}^n) 
\end{align}
has the same secrecy capacity of a new system with outputs given by
\begin{align}
Y^n =& (Y_{pp}^n, Y_{nn}^n)\\
Z^n =& (Z_{pp}^n, Y_{nn}^n). 
\end{align}
Thus, from the secrecy requirement, we get,
\begin{align}
I(W_1;Y_{nn}^n) = I(W_1;Z_{nn}^n) \leq I(W_1;Z^n) \leq n o(\log P). \label{eq:int_step1}
\end{align}
Then we have,
\begin{align}
nR_1 \leq& I(W_1;Y_{pp}^n, Y_{nn}^n) + n o(n)\\
=& I(W_1;Y_{nn}^n) + I(W_1;Y_{pp}^n|Y_{nn}^n)+  no(n)\\
\leq& I(W_1;Y_{pp}^n|Y_{nn}^n) +no(\log P) + no(n)\label{eq:k1}\\
\leq& h(Y_{pp}^n|Y_{nn}^n) + no(\log P) + no(n)\label{eq:k2}\\
\leq& h(Y_{pp}^n) + no(\log P) + no(n)\label{eq:k3}\\
\leq& n(1-\lambda_{nn})\log P +no(\log P) + no(n),
\end{align}
where, \eqref{eq:k1} follows from equation \eqref{eq:int_step1}, \eqref{eq:k2} follows since $h(Y_{pp}^n|Y_{nn}^n,W_1) \geq\\ h(Y_{pp}^n|Y_{nn}^n,W_1,\mathbf{X}^n) \geq o(\log P)$, and \eqref{eq:k3} follows since conditioning reduces differential entropy.

Dividing by $n$, and letting $n\rightarrow \infty$, we get,
\begin{align}
R_1 \leq (1-\lambda_{nn})\log P + o(\log P).
\end{align}
Dividing by $\log P$ and letting $P \rightarrow \infty$, we have,
\begin{align}
d_1 \leq 1-\lambda_{nn}.
\end{align} 
By symmetry, we also have,
\begin{align}
d_2\leq 1-\lambda_{nn}.
\end{align}
This completes the proof of Lemma~\ref{lem:single_user_bound2}.

\subsection{Proof of Lemma~\ref{lem:four_state_sum_rate}}\label{appendix:four_state_sum_rate}
We wish to prove that for the two-user MISO BC with no feedback and only four of the nine states: $\PP$, $\PN$, $\NP$ and $\NN$ occurring for $\lambda_{pp}$, $\lambda_{pn}$, $\lambda_{np}$ and $\lambda_{nn}$ fractions of the time, with  $\lambda_{pp}+\lambda_{pn}+\lambda_{np}+\lambda_{nn}=1$,
\begin{align}
d_1+d_2 \leq 2\lambda_{pp}+ \lambda_{pn}+\lambda_{np}.
\end{align}
To that end, for each of the two receivers, we introduce another statistically equivalent receiver. At receiver $1$, we introduce a virtual receiver $\tilde{1}$, with channel output denoted by $\tilde{Y}$, while the channel output at the virtual receiver $\tilde{2}$ at receiver $2$ is denoted by $\tilde{Z}$. Since the secrecy capacity without feedback depends only on the marginals \cite{csiszar}, without loss of generality, we can assume that the channels in the state $\NN$ are the same for all receivers. The outputs at each of the receivers are
\begin{align}
 Y^n =& (\ypp,\ypn,\ynp,\ynn)\\
 Z^n =& (\zpp,\zpn,\znp,\ynn)\\
 \tilde{Y}^n =& (\ypp,\ypn,\ytildenp,\ynn)\\
\tilde{Z}^n =& (\zpp,\ztildepn,\znp,\ynn),
\end{align}
where
\begin{align}
\tilde{Y}_{np}(t) =& \tilde{\mathbf{H}}_{1,np}(t)\mathbf{X}_{np}(t) + \tilde{N}_{1,np}(t)\\ 
\tilde{Z}_{pn}(t) =& \tilde{\mathbf{H}}_{2,pn}(t)\mathbf{X}_{pn}(t) + \tilde{N}_{2,pn}(t), 
\end{align}
such that $\tilde{\mathbf{H}}_{1,np}$, $\tilde{\mathbf{H}}_{2,pn}$ are i.i.d.~with the same distribution as $\mathbf{H}_{1,np}$, $\mathbf{H}_{2,pn}$, respectively, and $\tilde{N}_{1,np}$, $\tilde{N}_{2,pn}$ are i.i.d.~with same distribution as $N_{1,np}$, $N_{2,pn}$. We upper bound the first receiver's rate as
\begin{align}
nR_1 \leq& I(W_1;\ypp,\ypn,\ynp,\ynn|\Omega) + no(n)\\
=& I(W_1,\ypn,\ynp,\ynn|\Omega) + I(W_1,\ypp|\ypn,\ynp,\ynn,\Omega)\\
\leq& n\lambda_{pp}\log P + I(W_1,\ypn,\ynp,\ynn|\Omega)\\
=& n\lambda_{pp}\log P+ I(W_1;\ypn\ynn|\Omega)+I(W_1;\ynp|\ypn\ynn,\Omega)+no(n)\\
=& n\lambda_{pp}\log P+I(W_1;\ypn,\ynn|\Omega) + h(\ynp|\ypn,\ynn,\Omega) - h(\ynp|\ypn,\ynn, W_1,\Omega)+no(n)\\
\leq& n(\lambda_{pp}+\lambda_{np})\log P + I(W_1;\ypn,\ynn|\Omega) - h(\ynp|\ynn,\ypn,W_1,\Omega)+no(n)+ no(\log P) \label{eq:gaussian}\\
\leq& n(\lambda_{pp}+\lambda_{np})\log P + I(W_1;\ypn,\ynn,\zpn,\ztildepn,\znp,W_2|\Omega)\nonumber\\ &- h(\ynp|\ypn,\ynn,W_1,\Omega)+no(n)+ no(\log P)\\
=& n(\lambda_{pp}+\lambda_{np})\log P + I(W_1;\ypn,\ztildepn|\ynn,\zpn,\znp,W_2,\Omega)\nonumber\\ &- h(\ynp|\ypn,\ynn,W_1,\Omega)+no(n)+ no(\log P) \label{eq:secrecy}\\
=& n(\lambda_{pp}+\lambda_{np})\log P + h(\ypn,\ztildepn|\ynn,\zpn,\znp,W_2,\Omega) \nonumber\\&- h(\ypn,\ztildepn|\zpn,\ynn,\znp,W_1,W_2,\Omega) - h(\ynp|\ypn,\ynn,W_1,\Omega)+no(n)+ no(\log P)\\ 
\leq& n(\lambda_{pp}+\lambda_{np})\log P + h(\ypn,\ztildepn|\zpn,\znp,\ynn,W_2,\Omega)\nonumber\\ &-  h(\ynp|\ypn,\ynn,W_1,\Omega) +no(n) + no(\log P)\label{eq:lower_bound}\\
=&  n(\lambda_{pp}+\lambda_{np})\log P + h(\ztildepn|\zpn,\znp,\ynn,W_2,\Omega) + h(\ypn|\zpn,\ztildepn,\znp,\ynn, W_2,\Omega)  \nonumber\\ &-h(\ynp|\ypn,\ynn,W_1,\Omega) +no(n) + no(\log P)\\
\leq& n(\lambda_{pp}+\lambda_{np})\log P + h(\ztildepn|\znp,\ynn,W_2,\Omega) -  h(\ynp|\ypn,\ynn,W_1,\Omega)\nonumber\\&+no(n) + no(\log P)\label{eq:reconstruction}\\
=& n(\lambda_{pp}+\lambda_{np})\log P + h(\zpn|\znp,\ynn,W_2,\Omega) -  h(\ynp|\ypn,\ynn,W_1,\Omega)\nonumber\\ &+no(n) + no(\log P),\label{eq:r1_bound}
\end{align}
where \eqref{eq:secrecy} follows since,
\begin{align}
I(W_1;\zpn,\znp,\ynn,W_2|\Omega)\leq& I(W_1;\zpp,\zpn,\znp,\ynn,W_2|\Omega)\label{eq:sec_eq}\\
=& I(W_1,\zpp, \zpn,\znp,\ynn|\Omega) + I(W_1;W_2|\zpp,\zpn,\znp,\ynn,\Omega)\\
=& n o(\log P) + I(W_1;W_2|\zpp,\zpn,\znp,\ynn,\Omega) \label{eq: secrecy_condition}\\
\leq& n o(\log P) + H(W_2|\zpp,\zpn,\znp,\ynn,\Omega) \\
\leq& no(\log P)+ no(n), \label{eq:decodability} 
\end{align}
where, \eqref{eq: secrecy_condition} and \eqref{eq:decodability} follow from the secrecy and decodability requirements, respectively. In addition, \eqref{eq:lower_bound}  follows since $h(\ypn,\ztildepn|\zpn,\znp,\ynn,W_1,W_2,\Omega) \geq o(\log P)$, \eqref{eq:reconstruction} follows since given $\zpn$ and $\ztildepn$, one can reconstruct $\mathbf{X}_{pn}^n$ and hence $\ypn$ to within noise distortion, and \eqref{eq:r1_bound} follows due to the statistical equivalence of receivers $2$ and $\tilde{2}$ in the state $\PN$.

Similarly, by symmetry, we have,
\begin{align}
nR_2 \leq&  n(\lambda_{pp}+\lambda_{pn}) \log P + h(\ynp|\ypn,\ynn, W_1,\Omega)\nonumber\\ &-  h(\zpn|\znp,\ynn,W_2,\Omega) +no(n) + no(\log P).\label{eq:r2_bound}
\end{align}
Adding \eqref{eq:r1_bound} and \eqref{eq:r2_bound}, we have,
\begin{align}
n(R_1+R_2) \leq  n(2\lambda_{pp}+\lambda_{pn}+\lambda_{np}) \log P + 2no(n) + o(\log P).
\end{align}
First dividing by $n\log(P)$ and letting $n \rightarrow \infty$, and then letting $P \rightarrow \infty$, we obtain,
\begin{align}
d_1+d_2 \leq 2\lambda_{pp}+\lambda_{pn}+\lambda_{np}.
\end{align}
This completes the proof of Lemma~\ref{lem:four_state_sum_rate}.

\subsection{Proof of Lemma~\ref{lem:pd_dp_bound}}\label{appendix:pd_dp_bound}
We want to show that for the two-user MISO BC with only four of the nine states: $\PP$, $\PD$, $\DP$ and $\DD$ occurring for $\lambda_{pp}$, $\lambda_{pd}$, $\lambda_{dp}$ and $\lambda_{dd}$ fractions of the time, with $\lambda_{pd}=\lambda_{dp}$ and $\lambda_{pp}+\lambda_{pd}+\lambda_{dp}+\lambda_{dd}=1$,
\begin{align}
3d_1+d_2 \leq& 2+2\lambda_{pp}+2\lambda_{pd}\\
d_1+3d_2 \leq& 2+2\lambda_{pp}+2\lambda_{pd}.
\end{align} 

To do so, for each of the two receivers, we introduce another statistically equivalent receiver. At receiver $1$, we introduce a virtual receiver $\tilde{1}$, with channel output denoted by $\tilde{Y}$, while the channel output at the virtual receiver $\tilde{2}$ at receiver $2$ is denoted by $\tilde{Z}$. Since the capacity depends on the marginals, without loss of generality, we can assume that the channels in the state $\NN$ are the same for all receivers. The outputs at each of the receivers can be written as
\begin{align}
 Y^n =& (\ypp,\ypd,\ydp,\ynn)\\
 Z^n =& (\zpp,\zpd,\zdp,\ynn)\\
 \tilde{Y}^n =& (\ypp,\ypd,\ytildedp,\ynn)\\
\tilde{Z}^n =& (\zpp,\ztildepd,\zdp,\ynn),
\end{align}
where
\begin{align}
\tilde{Y}_{dp}(t) =& \tilde{\mathbf{H}}_{1,dp}(t)\mathbf{X}_{dp}(t) + \tilde{N}_{1,dp}(t)\\ 
\tilde{Z}_{pd}(t) =& \tilde{\mathbf{H}}_{2,pd}(t)\mathbf{X}_{pd}(t) + \tilde{N}_{2,pd}(t), 
\end{align}
such that $\tilde{\mathbf{H}}_{1,dp}$, $\tilde{\mathbf{H}}_{2,pd}$ are i.i.d.~with the same distribution as $\mathbf{H}_{1,dp}$, $\mathbf{H}_{2,pd}$, respectively, and $\tilde{N}_{1,dp}$, $\tilde{N}_{2,pd}$ are i.i.d.~with same distribution as $N_{1,dp}$, $N_{2,pd}$.
We consider a special case with  only four states $\PP$, $\PD$, $\DP$ and $\DD$. Aided by Lemma \ref{bounds-lemma}, we proceed to prove Lemma \ref{lem:pd_dp_bound}, as follows: 
\begin{align}
nR_1 \leq& I(W_1;Y^n |\Omega)+no(n)\\
\leq& I(W_1;Y^n |\Omega) - I(W_1;Z_{dp}^n\zdd|\Omega) + no(\log P) + no(n)\label{eq:l1}\\
\leq& h(Y^n|\Omega) - \frac{1}{2}h(Z_{dp}^n, Z_{dd}^n|W_1,\Omega) - h(Z_{dp}^n,\zdd|\Omega)+ h(Z_{dp}^n, Z_{dd}^n|W_1,\Omega) \nonumber\\& + no(\log P) + no(n)\label{eq:R1_bound_conditioning}\\
=&  h(Y^n|\Omega) + \frac{1}{2}h(Z_{dp}^n,\zdd|W_1,\Omega)- h(Z_{dp}^n,\zdd|\Omega) + no(\log P) + no(n)\\
\leq& h(Y^n|\Omega) + \frac{1}{2}h(Z_{dp}^n,\zdd|\Omega) - h(Z_{dp}^n,\zdd|\Omega)+ no(\log P) + no(n)\label{eq:l2}\\
=& h(Y^n|\Omega) - \frac{1}{2}h(Z_{dp}^n,\zdd|\Omega) + no(\log P) + no(n)\\
\leq& n\log P - \frac{1}{2}h(Z_{dp}^n,\zdd|\Omega)+no(\log P) + no(n), \label{eq:R1_bound1}
\end{align}
where \eqref{eq:l1} follows from the security constraints, \eqref{eq:R1_bound_conditioning} follows from a conditioned version of Lemma \ref{bounds-lemma} (conditioned on $W_{1}$), and \eqref{eq:l2} follows, since conditioning reduces differential entropy.

We also have the following bounds for user $1$:
\begin{align}
nR_1 \leq& I(W_1;Y^n| W_{2}, \Omega)+ no(n)\\
\leq& I(W_1;Y^n,Z^n| W_2, \Omega) + no(n)\\
=& I(W_1;Y^n|Z^n, W_2, \Omega) + no(\log P) + no(n)\label{eq:m1}\\
\leq& h(Y^n|Z^n,W_2,\Omega) + no(\log P) + no(n)\\
=& h(Y_{pd}^n, Y_{dp}^n,Y_{dd}^n|Z^n, W_2, \Omega)+h(Y_{pp}^n|Y_{pd}^n, Y_{dp}^n,Y_{dd}^n, Z^n, W_2, \Omega) +no(\log P) + no(n)\\
\leq& n\lambda_{pp}\log P+ h(Y_{dp}^n|Z^n, W_2,\Omega) + h(Y_{pd}^n, Y_{dd}^n|Z^n,W_2, \Omega)+no(\log P) + no(n)\\
\leq& n(\lambda_{pp}+\lambda_{dp})\log P + h(Y_{pd}^n, Y_{dd}^n|Z^n,W_2, \Omega)+no(\log P) + no(n)\\
\leq& n(\lambda_{pp}+\lambda_{dp})\log P + h(Z^n|W_2, \Omega)+no(\log P) + no(n), \label{eq:using_cor_lemma_1}
\end{align}
where \eqref{eq:m1} follows since,
\begin{align}
I(W_1;Z^n|W_2,\Omega) \leq& I(W_1;Z^n,W_2|\Omega)\label{eq:sec_eq2} \\
=& I(W_1;Z^n|\Omega) + I(W_1;W_2|Z^n,\Omega)\\
\leq& no(\log P) + H(W_2|Z^n,\Omega)\\
\leq& no(\log P) + no(n),
\end{align}
using the security and reliability constraints. In addition, \eqref{eq:using_cor_lemma_1} follows from the conditional version of Lemma \ref{bounds-lemma} (conditioned on $W_{2}$). 

For receiver $2$, we have
\begin{align}
nR_2 \leq& I(W_2;Z^n|\Omega) + no(n)\\
=& h(Z^n| \Omega) - h(Z^n|W_2, \Omega)  + no(n)\\
=& h(Z_{pp}^n|Z_{pd}^n, Z_{dp}^n,\zdd,\Omega )+ h(Z_{pd}^n, Z_{dp}^n,\zdd|\Omega) - h(Z^n|W_2,\Omega) + no(n)\\
\leq& n\lambda_{pp}\log P + h(Z_{pd}^n | \Omega) + h(Z_{dp}^n,\zdd|\Omega)  - h(Z^n|W_2,\Omega) + no(n)\\
\leq& n(\lambda_{pp}+\lambda_{dp})\log P + h(Z_{dp}^n,\zdd|\Omega)  - h(Z^n|W_2,\Omega)+ no(n). \label{eq:R2_bound}
\end{align}
In summary, from \eqref{eq:R1_bound1}, \eqref{eq:using_cor_lemma_1} and \eqref{eq:R2_bound}, we have,
\begin{align}
nR_{1}\leq& n\log P - \frac{1}{2}h(Z_{dp}^n,\zdd|\Omega)+no(\log P) + no(n),\\
nR_{1} \leq& n(\lambda_{pp}+\lambda_{dp})\log P + h(Z^n|W_2, \Omega)+no(\log P) + no(n), \\
nR_{2} \leq& n(\lambda_{pp}+\lambda_{dp})\log P + h(Z_{dp}^n,\zdd|\Omega)  - h(Z^n|W_2,\Omega)+ no(n).
\end{align}
Eliminating $h(Z_{dp}^{n}, Z_{dd}^{n}|\Omega)$ and $h(Z^n|W_2, \Omega)$ from these inequalities and taking the limit $n \rightarrow \infty$, we arrive at
\begin{align}
3R_1+R_2 \leq (2+2\lambda_{pp}+2\lambda_{dp})\log P + o(\log P).
\end{align}
Dividing by $\log P$ and taking the limit $P \rightarrow \infty$, we get the required result  
\begin{align}
3d_1 +d_2 \leq 2+2\lambda_{pp}+2\lambda_{dp}.
\end{align}

\section{Proof of the s.d.o.f.~Region for $\PD$ State}\label{appendix:pd_alone}
In this section, we present the proof for the s.d.o.f.~region of the fixed $\PD$ state (perfect $\CSIT$ from user $1$ and delayed $\CSIT$ from user $2$). The s.d.o.f.~region in this case is given by all non-negative pairs $(d_1,d_2)$ satisfying,
\begin{align}
d_{1}+d_{2}\leq 1.\label{eq:pd_region}
\end{align}
To prove this claim, we first provide a proof of the converse and then two achievable schemes that are sufficient to achieve the full region.
\subsection{Converse}
To this end, we create a virtual receiver with output $\tilde{Z}^n$ with a channel that is statistically equivalent to user $2$. The channel output $\tilde{Z}$ is given by
\begin{align}
\tilde{Z}(t) = \tilde{\mathbf{H}}_2(t)\mathbf{X}(t) + \tilde{N}_2(t),
\end{align}  
where $\tilde{\mathbf{H}}_2$ and $\tilde{N}_2$ are i.i.d.~as $\mathbf{H}_2$ and $N_2$, respectively. Then, the local statistical equivalence property implies that
\begin{align}
h(Z(t)|Z^{t-1},W_2,\Omega)=h(\tilde{Z}(t)|Z^{t-1},W_2,\Omega),\label{eq:prop}
\end{align}
where $\Omega$ is the set of all channel coefficients upto and including time $n$. Let us now bound the rate of user $1$:
\begin{align}
nR_1 \leq& I(W_1;Y^n|W_2,\Omega) + n o(n)\\
\leq& I(W_1;Y^n,Z^n|W_2,\Omega)  +no(n)\\
=& I(W_1;Y^n|Z^n,W_2,\Omega)+no(\log P) +no(n)\label{eq:n1}\\
\leq& I(W_1;Y^n,\tilde{Z}^n|Z^n,W_2,\Omega)+no(\log P) +no(n)\\
=& h(Y^n,\tilde{Z}^n|Z^n,W_2,\Omega) - h(Y^n,\tilde{Z}^n|Z^n,W_1,W_2,\Omega) + no(\log P) +no(n)\\
\leq& h(Y^n,\tilde{Z}^n|Z^n,W_2,\Omega)+no(\log P) +no(n)\\
=& h(\tilde{Z}^n|Z^n,W_2,\Omega)+h(Y^n|Z^n,\tilde{Z}^n,W_2,\Omega)+no(\log P) +no(n)\\
\leq&  h(\tilde{Z}^n|Z^n,W_2,\Omega) + no(\log P) +no(n)\label{eq:using_lemma}\\
=& \sum_{t=1}^n h(\tilde{Z}(t)|\tilde{Z}^{t-1},Z^n,W_2,\Omega) + no(\log P) +no(n)\\
\leq& \sum_{t=1}^n h(\tilde{Z}(t)|Z^{t-1},W_2,\Omega) + no(\log P) +no(n)\\
=& \sum_{t=1}^n h({Z}(t)|Z^{t-1},W_2,\Omega) + no(\log P) +no(n)\label{eq:using_prop}\\
=& h(Z^n|W_2,\Omega)+no(\log P) +no(n), \label{eq:user1_rate}
\end{align} 
where \eqref{eq:n1} follows since $I(W_1;Z^n|W_2,\Omega) \leq no(\log P)$ from \eqref{eq:sec_eq2}, \eqref{eq:using_lemma} follows due to the fact that given $Z^n$ and $\tilde{Z}^n$, it is possible to reconstruct $\mathbf{X}^n$ and hence $Y^n$ to within noise distortion, and \eqref{eq:using_prop} follows from \eqref{eq:prop}.

For the second user, we have,
\begin{align}
nR_2 \leq& I(W_2;Z^n|\Omega) + no(n)\\
=& h(Z^n|\Omega) - h(Z^n|W_2,\Omega) + no(n)\\
\leq& n\log P - h(Z^n|W_2,\Omega) +no(n). \label{eq:user2_rate}
\end{align} 
Adding \eqref{eq:user1_rate} and \eqref{eq:user2_rate}, we have,
\begin{align}
n(R_1+R_2) \leq n\log P + no(\log P) + no(n).
\end{align} 
Dividing by $n$ and letting $n\rightarrow \infty$,
\begin{align}
R_1+R_2 \leq \log P + o(\log P).
\end{align}
Now dividing by $\log P$ and letting $P\rightarrow \infty$,
\begin{align}
d_1+d_2 \leq 1.
\end{align}
This completes the proof of the converse for the case of $\PD$ state alone. 

\subsection{Achievable Schemes}
Note that it is sufficient to achieve only two points: a) $(d_1,d_2) = (1,0)$ and b) 
$(d_1,d_2)=(0,1)$. The achievability of these corner points follow in straightforward manner from existing arguments as follows:  sending message to user $1$ by superimposing it with artificial noise in a direction orthogonal to user $1$'s channel to achieve the pair $(1,0)$; and sending the message to user $2$ in a direction orthogonal to user $1$'s channel to achieve the pair $(0,1)$. This  completes the proof of the achievability of the region in \eqref{eq:pd_region}. 


\section{Proof of the s.d.o.f.~Region for $\DN$ State}\label{appendix:dn_alone}
For the MISO BCCM with the fixed state $\DN$ (delayed CSIT from the first user and no CSIT from the second user), the s.d.o.f.~region is given by the set of all non-negative pairs $(d_1,d_2)$ satisfying,
\begin{align}
d_1+d_2 \leq \frac{1}{2}.\label{eq:dn_region}
\end{align} 
To prove this claim, we first provide a proof of the converse and then two achievable schemes that are sufficient to achieve the full region.

\subsection{Converse}
We first create a virtual receiver with output $\tilde{Y}^n$ with a statistically equivalent channel as user $1$. The channel output $\tilde{Y}(t)$ is given by
\begin{align}
\tilde{Y}(t) = \tilde{\mathbf{H}}_1(t)\mathbf{X}(t) + \tilde{N}_1(t),
\end{align}  
where $\tilde{\mathbf{H}}_1$ and $\tilde{N}_1$ are i.i.d.~as $\mathbf{H}_1$ and $N_1$, respectively. Then, the local statistical equivalence property implies that
\begin{align}
h(Y(t)|Y^{t-1},W_1,\Omega)=h(\tilde{Y}(t)|Y^{t-1},W_1,\Omega),\label{eq:prop2}
\end{align}
where $\Omega$ is the set of all channel coefficients upto and including time $n$. Similar to the proof of Lemma~\ref{bounds-lemma}, Appendix \ref{appendix-bounds-lemma}, it can be readily shown that,
\begin{align}
2h(Y^n|W_1,\Omega) \geq h(Z^n|W_1,\Omega) + o(\log P). \label{eq:prop3}
\end{align} 
Then, for the first user, we have,
\begin{align}
nR_1 \leq& I(W_1;Y^n|\Omega) - I(W_1;Z^n|\Omega) + no(n) + no(\log P)\\
=& h(Y^n|\Omega) - h(Y^n|W_1,\Omega) - h(Z^n|\Omega) + h(Z^n|W_1,\Omega)\\
\leq&  h(Y^n|\Omega) -\frac{1}{2} h(Z^n|W_1,\Omega) - h(Z^n|\Omega) + h(Z^n|W_1,\Omega)\label{dn:bound1}\\
=& h(Y^n|\Omega) +\frac{1}{2} h(Z^n|W_1,\Omega) - h(Z^n|\Omega)\\
\leq& h(Y^n|\Omega) +\frac{1}{2} h(Z^n|\Omega) - h(Z^n|\Omega)\\
=& h(Y^n|\Omega) -\frac{1}{2} h(Z^n|\Omega),\label{eq:dn_user1}
\end{align}
where (\ref{dn:bound1}) follows from (\ref{eq:prop3}).
For the second user,
\begin{align}
nR_2 \leq& I(W_2;Z^n|\Omega) -  I(W_2;Y^n|\Omega) + no(n) + no(\log P)\\
=& h(Z^n|\Omega) - h(Y^n|\Omega) + \left(h(Y^n|W_2,\Omega) - h(Z^n|W_2,\Omega)\right) + no(n) + no(\log P).\label{eq:dn_user2}
\end{align}
Adding \eqref{eq:dn_user1} and \eqref{eq:dn_user2}, we obtain,
\begin{align}
n(R_1+R_2) \leq& \frac{1}{2} h(Z^n|\Omega) +   \left(h(Y^n|W_2,\Omega) - h(Z^n|W_2,\Omega)\right) + no(n)+no(\log P)\\
\leq& \frac{n}{2}\log P + \left(h(Y^n|W_2,\Omega) - h(Z^n|W_2,\Omega)\right)+no(n)+no(\log P).\label{dn:verylast}
\end{align}
Thus, in order to obtain $d_1+d_2\leq 1/2$, it suffices to  show that $\left(h(Y^n|W_2,\Omega) - h(Z^n|W_2,\Omega)\right)\leq no(\log P)$, where the transmitter has delayed CSIT from user $1$ and no CSIT from user $2$. To this end, we invoke a recent result  in \cite[(39)-(66)]{aligned_image_sets_jafar},  which showed that the maximum of $h(Y^n|W_2,\Omega) - h(Z^n|W_2,\Omega)$ is less than $no(\log P)$, under the assumption of perfect CSIT from user $1$ and no CSIT from user $2$. Hence, the same upper bound on the maximum value also holds under a weaker assumption of delayed CSIT from user $1$. Thus, using the fact that 
\begin{align}
\left(h(Y^n|W_2,\Omega) - h(Z^n|W_2,\Omega)\right)\leq no(\log P),
\end{align}
and substituting in (\ref{dn:verylast}), we have,
\begin{align}
n(R_1+R_2) \leq \frac{n}{2}\log P + no(n)+no(\log P).
\end{align}
Dividing by $n$ and letting $n\rightarrow \infty$, we get,
\begin{align}
R_1+R_2 \leq \frac{1}{2}\log P + o(\log P).
\end{align}  
Dividing by $\log P$ and letting $P\rightarrow \infty$ yields 
\begin{align}
d_1+d_2 \leq \frac{1}{2}.
\end{align}
This completes the proof of the converse.

\subsection{Achievable Schemes}
To prove the achievability of the s.d.o.f.~region in \eqref{eq:dn_region}, it suffices to consider only the two points: a) $(d_1,d_2)= \left(\frac{1}{2},0\right)$ and b) $(d_1,d_2)= \left(0,\frac{1}{2}\right)$. Every other point in the region can be obtained by time-sharing. A scheme for achieving $(d_1,d_2)= \left(\frac{1}{2},0\right)$ was presented in \cite{kobayashi_delayed_csit}. We include it here for completeness. 

\subsubsection{Scheme Achieving $(d_1,d_2)= \left(\frac{1}{2},0\right)$:} 
We wish to send $1$ symbol $u$ securely to the first user in $2$ time slots. This can be done as follows:

1) At time \textbf{$t=1$}: The transmitter does not have any channel knowledge. It sends:
\begin{align}
\mathbf{X}(1) = [q_1 \quad q_2]^{T},
\end{align}
where $q_1$ and $q_2$ denote independent artificial noise symbols distributed as $\mathcal{CN}(0, P)$. Both receivers receive linear combinations of the two symbols $q_1$ and $q_2$. The receivers' outputs are:
\begin{align}
Y(1) &= h_{11}(1)q_1 + h_{12}(1)q_2 \stackrel{\Delta}{=} L_1(q_1,q_2)\\
Z(1) &= h_{21}(1)q_1 + h_{22}(1)q_2.
\end{align}
Due to delayed $\CSIT$ from receiver $1$, the transmitter can reconstruct $L_1(q_1,q_2)$ in the next time slot and use it for transmission.

2) At time \textbf{$t=2$}:  The transmitter sends:
\begin{align}
\mathbf{X}(2) =\left[ u \quad L_1(q_1,q_2)\right]^{T}.
\end{align}
The received signals are:
\begin{align}
Y(2) =& h_{11}(2)u + h_{12}(2)L_1(q_1,q_2)\\ 
Z(2) =& h_{21}(2)u + h_{22}(2)L_1(q_1,q_2).
\end{align}
Since the receivers have full channel knowledge, receiver $1$ can recover $u$ by eliminating $L_1(q_1,q_2)$ from Y(1) and Y(2). On the other hand, the information leakage to the second user is given by,
\begin{align}
I(u;Z(1), Z(2)|\Omega) =& h(Z(1),Z(2)|\Omega) - h(Z(1),Z(2)|u,\Omega)\\
\leq& 2\log P - h(h_{21}(1)q_1 + h_{22}(1)q_2,h_{11}(1)q_1 + h_{12}(1)q_2|\Omega)\\
=& 2\log P - 2\log P + o(\log P)\\
=& o(\log P). 
\end{align} 

\subsubsection{Scheme Achieving $(d_1,d_2)= \left(0,\frac{1}{2}\right)$:}
In this scheme, we wish to send $1$ symbol $u$ securely to the second user in $2$ time slots. This can be done as follows:

1) At time \textbf{$t=1$}: The transmitter does not have any channel knowledge. It sends:
\begin{align}
\mathbf{X}(1) = [u \quad q_1]^{T},
\end{align}
where $q$ denotes an independent artificial noise symbol distributed as $\mathcal{CN}(0, P)$. Both receivers receive linear combinations of the two symbols $u$ and $q$. The receivers' outputs are:
\begin{align}
Y(1) &= h_{11}(1)u + h_{12}(1)q \stackrel{\Delta}{=} L(u,q)\\
Z(1) &= h_{21}(1)u + h_{22}(1)q \stackrel{\Delta}{=} G(u,q) .
\end{align}
Due to delayed $\CSIT$ from receiver $1$, the transmitter can reconstruct $L(u,q)$ in the next times lot and use it for transmission.

2) At time \textbf{$t=2$}:  The transmitter sends:
\begin{align}
\mathbf{X}(2) =\left[L(u,q) \quad 0\right]^{T}.
\end{align}
The received signals are:
\begin{align}
Y(2) =& h_{11}(2)L(u,q)\\ 
Z(2) =& h_{21}(2)L(u,q).
\end{align}
Since the receivers have full channel knowledge, receiver $2$ can recover $u$ by eliminating $q$ from $L(u,q)$ and $G(u,q)$. On the other hand, the information leakage to the first user is given by,
\begin{align}
I(u;Y(1), Y(2)|\Omega) =& I(u;L(u,q)|\Omega)\\
=& h(L(u,q)|\Omega) - h(L(u,q)|u,\Omega)\\
\leq& \log P - \log P +o(\log P)\\
=& o(\log P). 
\end{align} 

This completes the proof of achievability. 

\end{appendices}
\bibliographystyle{unsrt}
\bibliography{IEEEabrv,references}
\end{document}